\documentclass[onecolumn,12pt]{article}
\pdfoutput=1
\usepackage{jheppub}
\usepackage{ifpdf}
\usepackage[bb=boondox]{mathalfa}
\graphicspath{ {figures/} }
\usepackage{amsfonts}
\usepackage{amsmath,braket}
\usepackage{amssymb}
\usepackage{epsfig}
\usepackage{tensor}
\usepackage{pdfpages}
\usepackage{bbm}
\usepackage{graphicx,epstopdf}
\usepackage[numbers]{natbib} 
\usepackage[makeroom]{cancel}
\usepackage{hyperref}
\usepackage{array}
\usepackage[export]{adjustbox}

\usepackage[normalem]{ulem}
\usepackage{romannum}

\usepackage{ulem}

\numberwithin{equation}{section}									% equation numbering by section

\usepackage{comment}

\newcommand{\de}{\partial}
\newcommand{\be}{\begin{equation}}
	\newcommand{\ba}{\begin{eqnarray}}
		\newcommand{\ea}{\end{eqnarray}}
	\newcommand{\ee}{\end{equation}}

\newcommand{\s}{\sqrt}

\newcommand{\no}{\nonumber \\}
\newcommand{\la}{\langle}
\newcommand{\lb}{\rangle}
\newcommand{\bea}{\begin{eqnarray}}
	\newcommand{\eea}{\end{eqnarray}}
\newcommand{\bes}{\begin{equation*}}
	\newcommand{\beas}{\begin{eqnarray*}}
		\newcommand{\eeas}{\end{eqnarray*}}
	\newcommand{\bas}{\begin{array*}}
		\newcommand{\eas}{\end{array*}}
	\newcommand{\ees}{\end{equation*}}
\newcommand{\nn}{\nonumber}

\newcommand{\ep}{\epsilon}

\newcommand{\ra}{\rangle}

\def\nn{\nonumber}
\def\inf{\infty}

\newcommand{\eg}{{\it e.g.,}\ }
\newcommand{\ie}{{\it i.e.,}\ }
\newcommand{\viz}{{\it viz,}\ }
\newcommand{\mt}[1]{\textrm{\tiny #1}}
\renewcommand{\(}{\left(}
\renewcommand{\)}{\right)}

\newcommand{\GN}{G_\mt{N}}

\newcommand{\TE}{T_{\mt{E}}}

\newcommand{\tP}{t_{\mt{P}}}
\newcommand{\TP}{T_{\mt{P}}}

\newcommand{\arccosh}{\text{arccosh}}

\subheader{\today}
\title{\boldmath A Half de Sitter Holography}

\author[a]{Taishi Kawamoto,}
\author[a]{Shan-Ming Ruan,}
\author[a]{Yu-ki Suzuki,}
\author[a,b,c]{Tadashi Takayanagi}

\affiliation[a]{Center for Gravitational Physics and Quantum Information, Yukawa Institute for Theoretical Physics, Kyoto University,\\
	Kitashirakawa Oiwakecho, Sakyo-ku, Kyoto 606-8502, Japan}
\affiliation[b]{Inamori Research Institute for Science,\\
	620 Suiginya-cho, Shimogyo-ku,Kyoto 600-8411 Japan}

\affiliation[c]{Kavli Institute for the Physics and Mathematics
	of the Universe (WPI),\\
	University of Tokyo, Kashiwa, Chiba 277-8582, Japan}

\emailAdd{taishi.kawamoto@yukawa.kyoto-u.ac.jp}
\emailAdd{ruan.shanming@yukawa.kyoto-u.ac.jp}
\emailAdd{yu-ki.suzuki@yukawa.kyoto-u.ac.jp}
\emailAdd{takayana@yukawa.kyoto-u.ac.jp}

\abstract{A long-standing and intriguing question is: does the holographic principle apply to cosmologies like de Sitter spacetime? In this work, we consider a half dS spacetime wherein a timelike boundary encloses the bulk spacetime, presenting a version of de Sitter holography. By analyzing the holographic entanglement entropy in this space and comparing it with that in AdS/CFT, we argue that gravity on a half dS$_{d+1}$ is dual to a highly non-local field theory residing on dS$_d$ boundary. This non-locality induces a breach in the subadditivity of holographic entanglement entropy. Remarkably, this observation can be linked to another argument that time slices in global de Sitter space overestimate the degrees of freedom by redundantly counting the same Hilbert space multiple times.}

\begin{document} 
	
	%%%%%%%%%%%%%%%%%%%%%%%%%%%%%%%%%%%%%%%%%%%%%%%%%%%%
	\begin{flushright}
		YITP-23-73
		\\
	\end{flushright}
	%%%%%%%%%%%%%%%%%%%%%%%%%%%%%%%%%%%%%%%%%%%%%%%%%%%%
	\maketitle
	\flushbottom

	%%%%%%%%%%%%%%%%%%%%%%%%%%%%%%%%%%%%%%%%%%%%%%%%%%%%%%%%
	%%%%%%%%%%%%%%%%%%%%%%%%%%%%%%%%%%%%%%%%%%%%%%%%%%%%%%%%
	\section{Introduction}
	\label{sec:intro}
	%%%%%%%%%%%%%%%%%%%%%%%%%%%%%%%%%%%%%%%%%%%%%%%%%%%%%%%%
	%%%%%%%%%%%%%%%%%%%%%%%%%%%%%%%%%%%%%%%%%%%%%%%%%%%%%%%%
	The holographic duality has bestowed upon us a promising framework, enabling the exploration of quantum gravity \cite{tHooft:1993dmi,Susskind:1994vu}. The most outstanding example will be the AdS/CFT correspondence where quantum gravity on $d+1$-dimensional Anti-de Sitter space (AdS$_{d+1}$) becomes equivalent to $d$-dimensional conformal field theory (CFT$_{d}$) \cite{Maldacena:1997re,Gubser:1998bc,Witten:1998qj}. Despite the resounding success attained in AdS/CFT, we are still at a nascent stage in the development of the holographic duality pertaining to gravity in de Sitter space. The potential dS holography promises paramount application to realistic cosmological spacetime.

	\begin{figure}[t]
		\centering
		\includegraphics[width=3in]{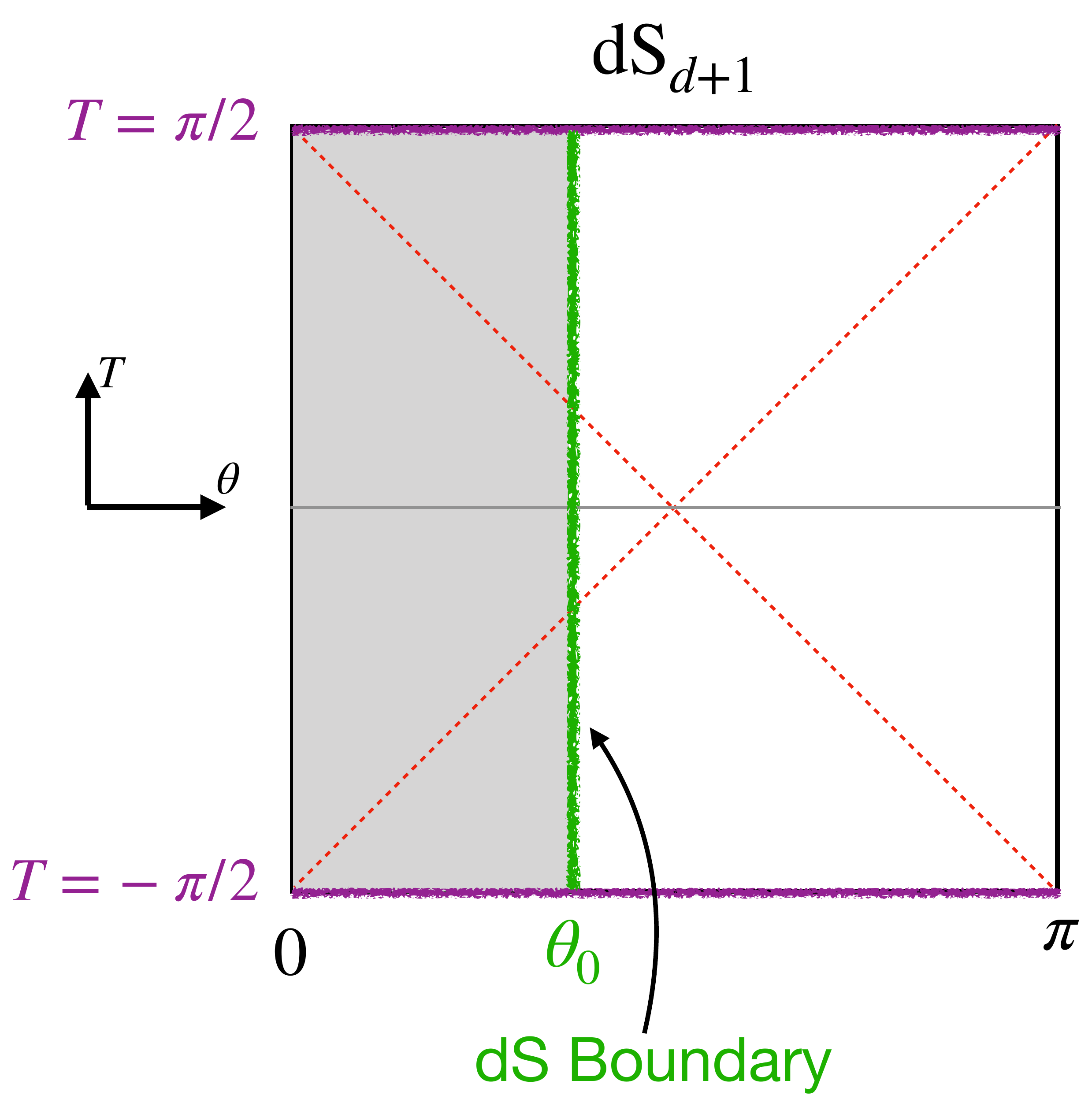}
		\caption{The Penrose diagram of dS$_{d+1}$ bulk spacetime. The conformal time $T\in [-\frac{\pi}{2},+\frac{\pi}{2}]$ is associated with the global time $t$ by $\cosh t= \frac{1}{\cos T}$. We introduce a timelike boundary at $\theta=\theta_0$ which is described by a $d-$dimensional dS spacetime. The dual bulk dS$_{d+1}$ spacetime is given by the gray shaded region.} 
		\label{fig:dSbulk}
	\end{figure}
	
	\begin{figure}[t]
		\centering
		\includegraphics[width=5in]{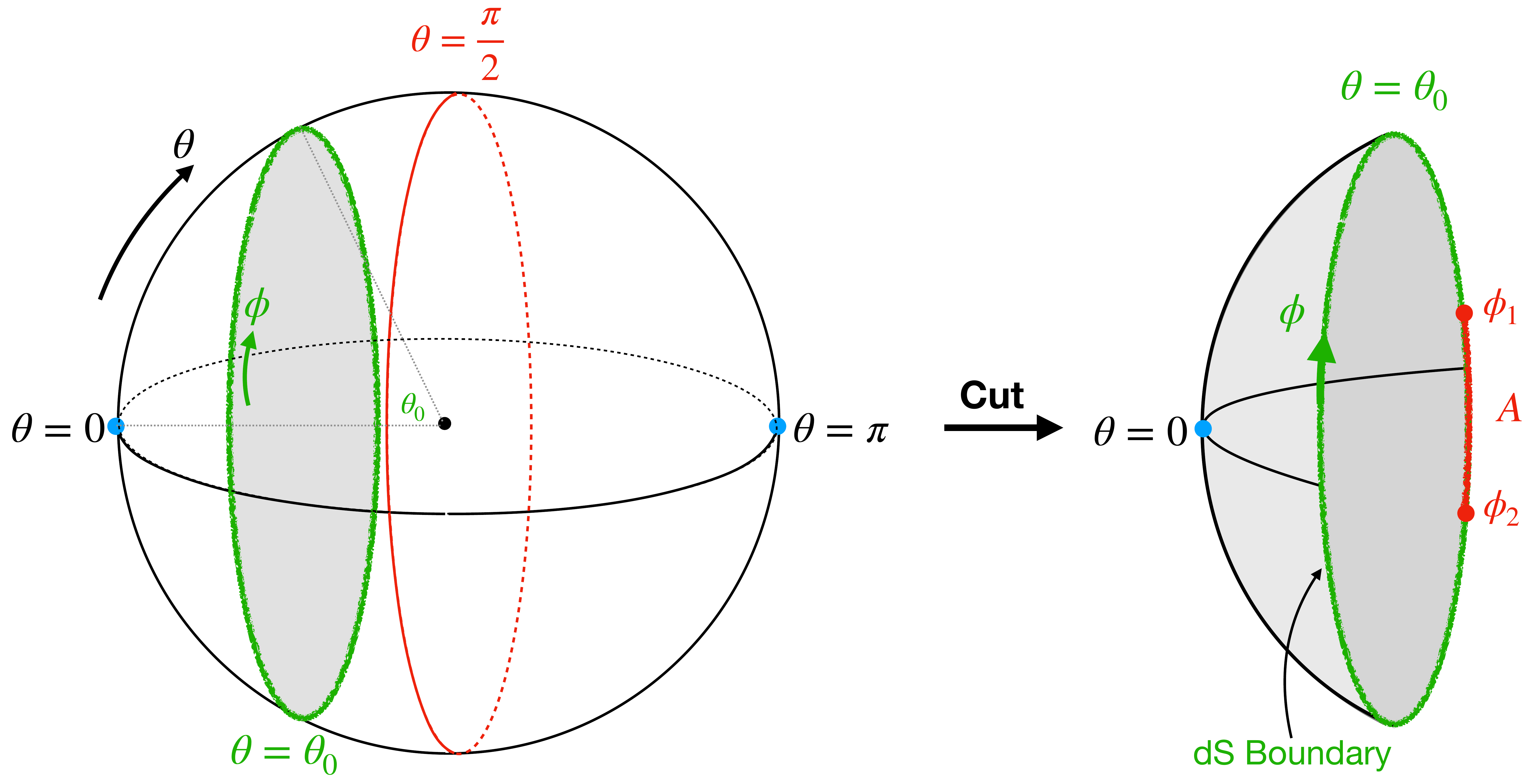}
		\caption{The left panel shows the geometry of time slices for a global dS$_3$. In the right panel we keep a half of dS$_3$ space with a dS$_2$ boundary (denoted by green circle) at $\theta=\theta_0$.} 
		\label{fig:dSHolT}
	\end{figure}

\begin{figure}[t]
		\centering
		\includegraphics[width=4.5in]{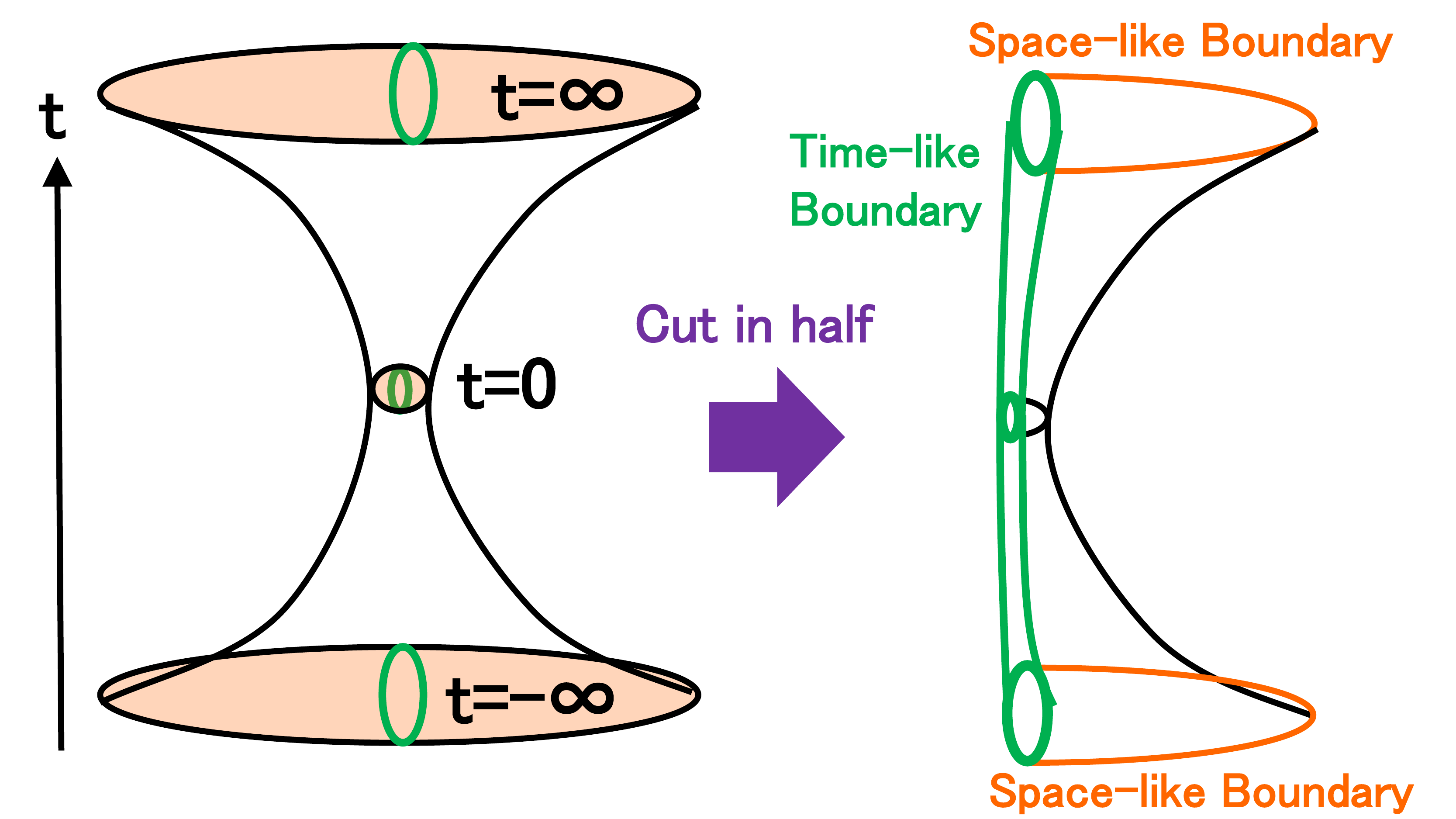}
		\caption{The left panel shows a sketch of global de Sitter space, which has only space-like boundaries at $t=\pm\infty$. The right one is a sketch of our holographic setup which is obtained by cutting in half a de Sitter space. Notice that this geometry has both the time-like and space-like boundary.} 
		\label{fig:dSHol}
	\end{figure}

	Let us commence by exploring the explicit differences between dS holography and AdS/CFT. The $d+1$-dimensional de Sitter space (dS$_{d+1}$) can be described in global coordinates as follows (see \eg \cite{Spradlin:2001pw,Galante:2023uyf} for comprehensive reviews):
	\begin{equation} \label{dsg}
		ds^2=-dt^2+\cosh^2 t d\Omega_{d}^2,.
	\end{equation}
	whose Penrose diagram is shown in figure \ref{fig:dSbulk}.
	Here, we have set the dS radius to unity, and $d\Omega_{d}^2$ represents the metric of the unit $d$-dimensional sphere:
	\begin{equation}\label{sth}
		d\Omega_{d}=d\theta^2+\sin^2\theta d\Omega_{d-1}^2\,,
	\end{equation}
	The constant time slice in global dS$_{d+1}$ coordinate is depicted in the left diagram of figure \ref{fig:dSHol}. 
	In the case of $d=2$, we simply write $ d\Omega_{d-1}^2=d\phi^2$ in this paper.  As shown in the left panel of figure \ref{fig:dSbulk}, the conformal boundaries of global dS spacetime are spacelike surfaces \ie $S^d$, located at the future and past infinity $t=\pm \infty$. The original dS/CFT correspondence postulates that the gravitational dynamics in the dS$_{d+1}$ bulk is dual to a Euclidean conformal field theory (CFT) on the sphere S$^d$ at future infinity $t\to \infty$ \cite{Spradlin:2001pw,Witten:2001kn,Maldacena:2002vr}. This proposal assumes that the quantum state is generated by the Euclidean instanton using the Hartle-Hawking prescription and is then continued to Lorentzian dS at $t=0$.  However, the dual CFT becomes non-unitary and exhibits numerous exotic characteristics. For instance, in the case of dS$_4/$CFT$_3$, the 3D CFT dual to higher-spin gravity on dS$_4$ is described by the $SP(N)$ model, which incorporates ghost fields \cite{Anninos:2011ui,Ng:2012xp}. Similarly, for dS$_3/$CFT$_2$, the 2D CFT dual to Einstein gravity on dS$_3$ is obtained through an analytical continuation of current algebra or Liouville CFT, with an imaginary central charge \cite{Hikida:2021ese,Hikida:2022ltr,Chen:2022ozy,Chen:2023prz}. Furthermore, holography for dS$_2$ has been investigated in \cite{Maldacena:2019cbz,Cotler:2019nbi,Cotler:2023eza}. Despite the peculiar holographic properties, studies from the perspective of gravity have also been developed, with a partial list of references including \cite{Castro:2011xb,Castro:2012gc,Anninos:2020hfj,Anninos:2021ihe,Castro:2023dxp}.

	One of the manifestations of the non-unitary nature inherent in the dS/CFT correspondence is the computation of holographic entanglement entropy. In the context of the AdS/CFT correspondence, the holographic entanglement entropy $S_A$ for a subsystem $A$ in the dual CFT can be determined by evaluating the area of the extremal surface $\Gamma_A$ in AdS, which ends on the the boundary of the subsystem $A$ \cite{Ryu:2006bv,Ryu:2006ef,Hubeny:2007xt}:
	\begin{equation}\label{Area}
		S_A=\frac{A(\Gamma_A)}{4\GN} \,, 
	\end{equation}
	where $\GN$ denotes the Newton constant and $A(\Gamma_A)$ represents the area of the extremal surface $\Gamma_A$.
	In principle, this geometric computation can be extended to various spacetimes, including de Sitter space. However, in de Sitter space, the absence of a spacelike extremal surface connecting two distinct points on S$^d$ at future infinity leads to the holographic entanglement entropy being complex-valued \cite{Narayan:2015vda,Sato:2015tta,Hikida:2022ltr}. In this context, the holographic entanglement entropy is computed by using the timelike extremal surface in de Sitter space. Subsequently, in \cite{Doi:2022iyj,Doi:2023zaf}, this complex-valued entropy was appropriately interpreted as the pseudo-entropy \cite{Nakata:2021ubr,Mollabashi:2020yie,Mollabashi:2021xsd}, which generalizes the notion of entanglement entropy to non-Hermitian density matrices (see \cite{Narayan:2022afv,Narayan:2023ebn} for closely related ideas). This consideration suggests a connection between the emergent time coordinate in dS/CFT and the imaginary part of the pseudo-entropy. Notably, an intriguing quantum entanglement structure of dS holography has been recently proposed in \cite{Cotler:2023xku}.
	
	Several other approaches to de Sitter holography have been explored. One notable example is the holography for de Sitter space in the static patch, which has recently garnered significant attention and discussions from multiple perspectives \cite{Susskind:2021omt,Susskind:2021esx,Shaghoulian:2022fop,Chapman:2021eyy,Jorstad:2022mls,Chandrasekaran:2022cip,Anegawa:2023wrk,Franken:2023pni}. Additionally, another approach to dS holography has been investigated, based on the TTbar deformation in AdS/CFT \cite{McGough:2016lol} and the dS/dS duality \cite{Alishahiha:2004md}, with studies conducted in \cite{Dong:2018cuv,Gorbenko:2018oov}. An interesting dS/dS duality setup can also be found in \cite{Geng:2021wcq}. The quantum information structure in de Sitter has been analyzed by applying the surface/state duality \cite{Miyaji:2015yva}. Notably, it has been observed that the state dual to the $t=0$ slice is maximally entangled.
	
	In this paper, we propose a novel approach to dS holography that adheres to the standard holographic formalism, wherein gravity in a given bulk space is dual to a non-gravitational theory on its timelike boundary. 
	However, since de Sitter space lacks timelike boundaries, we introduce a procedure to create one. Our proposal involves cutting a de Sitter space into a half at $\theta=\theta_0$ by confining the sphere $S^d$ to a semi-sphere. More generally, we can cut the bulk dS space by putting a boundary at $\theta=\theta_0$, as depicted in figure \ref{fig:dSbulk} and \ref{fig:dSHolT}. This resulting spacetime is referred to as a half de Sitter space (or simply half dS). The boundary of the $d+1$-dimensional half de Sitter space corresponds to $d$-dimensional global de Sitter spacetime dS$_d$, which is described by the metric:
	\begin{equation}\label{dSd}
		ds^2=-dt^2+\cosh t^2 \sin^2\theta_0 d\Omega^2_{d-1}\,.
	\end{equation}
	In this setup, we argue that {\it  gravity on a half dS$_{d+1}$ is dual to a field theory without gravity on dS$_d$}. We expect that such a field theory exhibits high non-locality due to the finite geometric cut-off in dS$_{d+1}$. It is worth noting that this spacetime possesses the SO$(d,1)$ symmetry, which is a subgroup of the original SO$(d+1,1)$ symmetry of dS$_{d+1}$. Although our approach shares common features with the TTbar approach \cite{Dong:2018cuv,Gorbenko:2018oov} and surface/state duality \cite{Miyaji:2015yva}, our focus lies specifically on the half dS space, as it contains a timelike boundary. Nevertheless, in principle, our analysis of half dS holography can be extended to full dS geometry by combining two copies of our holographic duality, as demonstrated in the case of gluing AdS/CFT \cite{Kawamoto:2023wzj}. To probe our holographic proposal, we would investigate the holographic entanglement entropy \eqref{Area} in dS holography as a fundamental tool. Here we would like to emphasize that we stay with the standard calculation of holographic entanglement entropy \cite{Ryu:2006bv,Ryu:2006ef,Hubeny:2007xt}, where we minimize the area, as opposed to the different prescription of maximizing the area in the version of 
	dS holography discussed in \cite{Susskind:2021esx}.
	
	It is noteworthy that we can also apply the standard AdS/CFT correspondence to study the field theory on dS$_d$ \cite{Hawking:2000da,Maldacena:2012xp}. In this case, a CFT on dS$_d$ is dual to gravity in AdS$_{d+1}$ whose conformal boundary is given by dS$_d$. Given the well-established nature of this holographic duality, we will initially focus on this situation. Subsequently, we will delve into the main subject of this paper, 
	\ie holography for the half dS. Throughout this paper, we will consider two different prescriptions for both holography in AdS and half dS, which are described below and depend on how we handle the future infinity at $t\to\infty$.
	
	{\bf Case 1: Schwinger-Keldysh prescription without EOW}
	
\begin{figure}[t]
		\centering
		\includegraphics[width=6in]{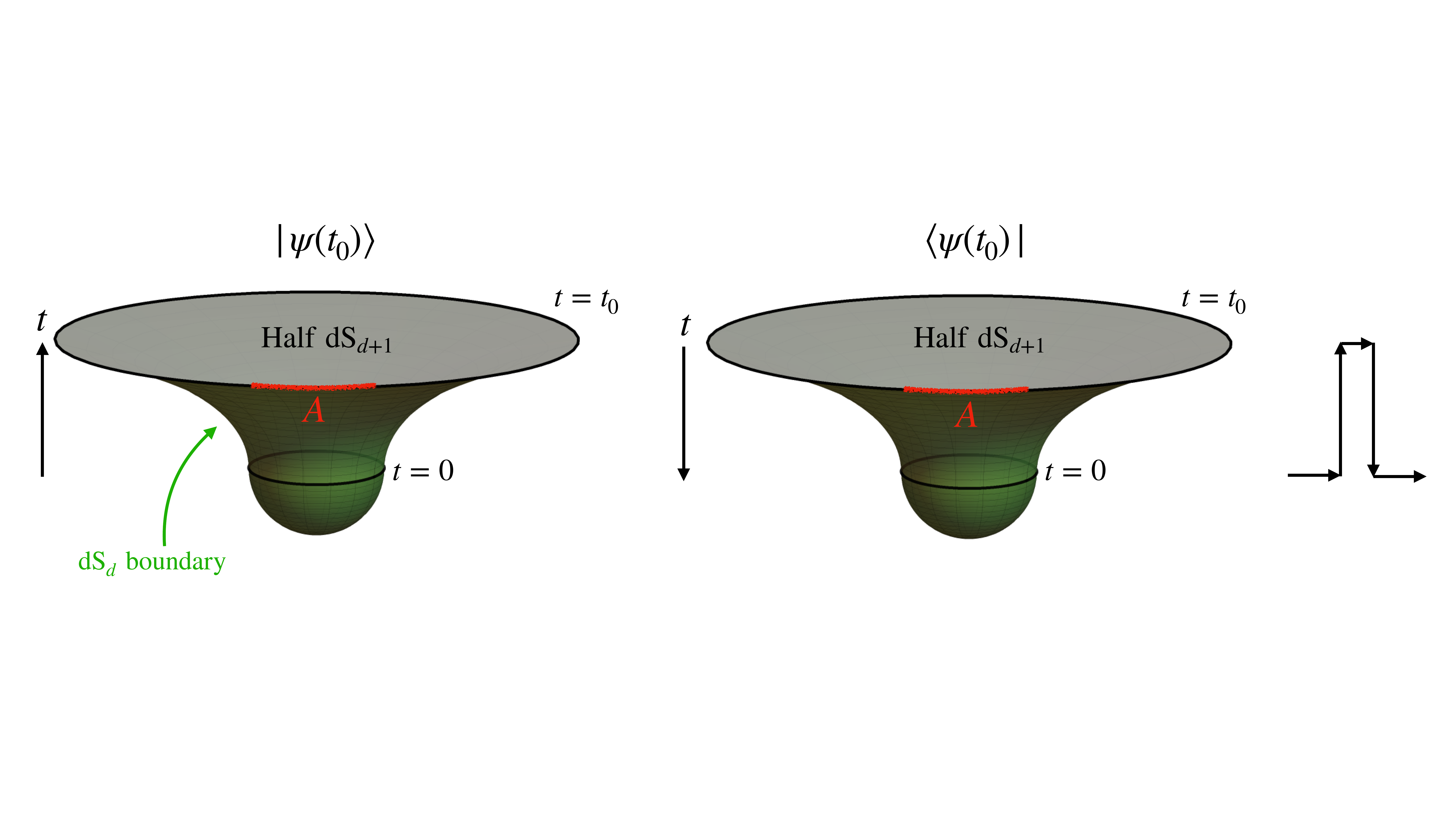}
		\caption{The geometry which describes the Schwinger-Keldysh contour of a half dS$_{d+1}$. The top and bottom region presents the Lorentzian and Euclidean evolution, respectively.  This is dual to a field living on dS$_d$ boundary which is parametrized by the green surface.} 
		\label{fig:SCK}
	\end{figure}

	We apply the Hartle-Hawking prescription to construct the initial state at $t=0$ using the Euclidean instanton geometry, which takes the form of a $d+1$ dimensional semi-sphere. Subsequently, we examine its Lorentzian time evolution. The density matrix for this state at time $t$ is determined by the corresponding geometry dictated by the Schwinger-Keldysh prescription, as depicted in figure \ref{fig:SCK}. For a detailed understanding of the Schwinger-Keldysh prescription in holography and holographic entanglement entropy, we refer readers to \cite{Skenderis:2008dg,Dong:2016hjy}. It is important to note that the asymptotic infinity as $t\to \infty$ is absent in this particular setup. In line with conventional Lorentzian holography in AdS, we anticipate that the gravitational dynamics in half dS$_{d+1}$ geometry is dual to a field theory residing on its boundary, namely, dS$_d$. We expect this field theory to exhibit strong non-locality, primarily due to the fact that the boundary at $\theta=\theta_0$ is not an asymptotic boundary. Unlike in the AdS/CFT correspondence, the metric, in this case, does not exhibit divergent behaviour, indicating that the dS boundary plays the role of a finite cutoff.
	
	{\bf Case 2: Final state projection with EOW}
	
	On the other hand, if we consider the full geometry of a half dS$_{d+1}$,  we need to impose a boundary condition at the 
	infinity $t\to \infty$. If we impose the Dirichlet b.c., then we expect a dual CFT lives there in addition to the dS$_d$ boundary. To avoid this complicated situation, we focus on the other case: Neumann boundary condition. Namely this means that the asymptotic boundary $t=\infty$ is an end-of-the-world (EOW) brane. In the context of AdS/CFT, the EOW brane is dual to the boundary conformal field theory (BCFT), whose holographic duality is called the AdS/BCFT \cite{Takayanagi:2011zk,Fujita:2011fp,Karch:2000gx}.  In our dS case, we expect that the dual theory is a non-local field theory on dS$_d$ with a final state projection at future infinity $t=\infty$.
	
	In the presence of post-selection, a useful quantity is pseudo entropy \cite{Nakata:2021ubr}, which is a natural generalization of entanglement entropy such that it depends on two different quantum states $|\psi_1\lb$ and $|\psi_2\lb$. This quantity is defined as follows. We introduce the reduced transition matrix 
	\ba
	\tau_A=\mbox{Tr}_B\left[\frac{|\psi_1\lb\la\psi_2|}{\la\psi_2|\psi_1\lb}\right].\label{transm}
	\ea
	The pseudo entropy is defined by 
	\ba
	S_A=\mbox{Tr}\left[-\tau_A\log\tau_A\right]. \label{PE}
	\ea
	Note that this quantity in general takes complex values 
	as $\tau_A$ is not hermitian.
	
	Interestingly, the gravity dual of this quantity is given by (\ref{Area}) when $\Gamma_A$ is the minimal surface in an Euclidean time-dependent asymptotically AdS \cite{Nakata:2021ubr}. We can also regard the area of extremal surface as a holographic pseudo entropy in Lorentzian AdS in the presence of final state projection \cite{Akal:2021dqt}. We will assume an extension of this correspondence to de Sitter spaces in this paper.
	As studies of quantum many-body systems and CFTs suggest \cite{Mollabashi:2020yie,Mollabashi:2021xsd,Akal:2021dqt}
	the real part of pseudo entropy typically measures the amount of quantum entanglement in the intermediate states between $|\psi_1\lb$ and $|\psi_2\lb$. 
	
	\begin{figure}[t]
		\centering
		\includegraphics[width=5.5in]{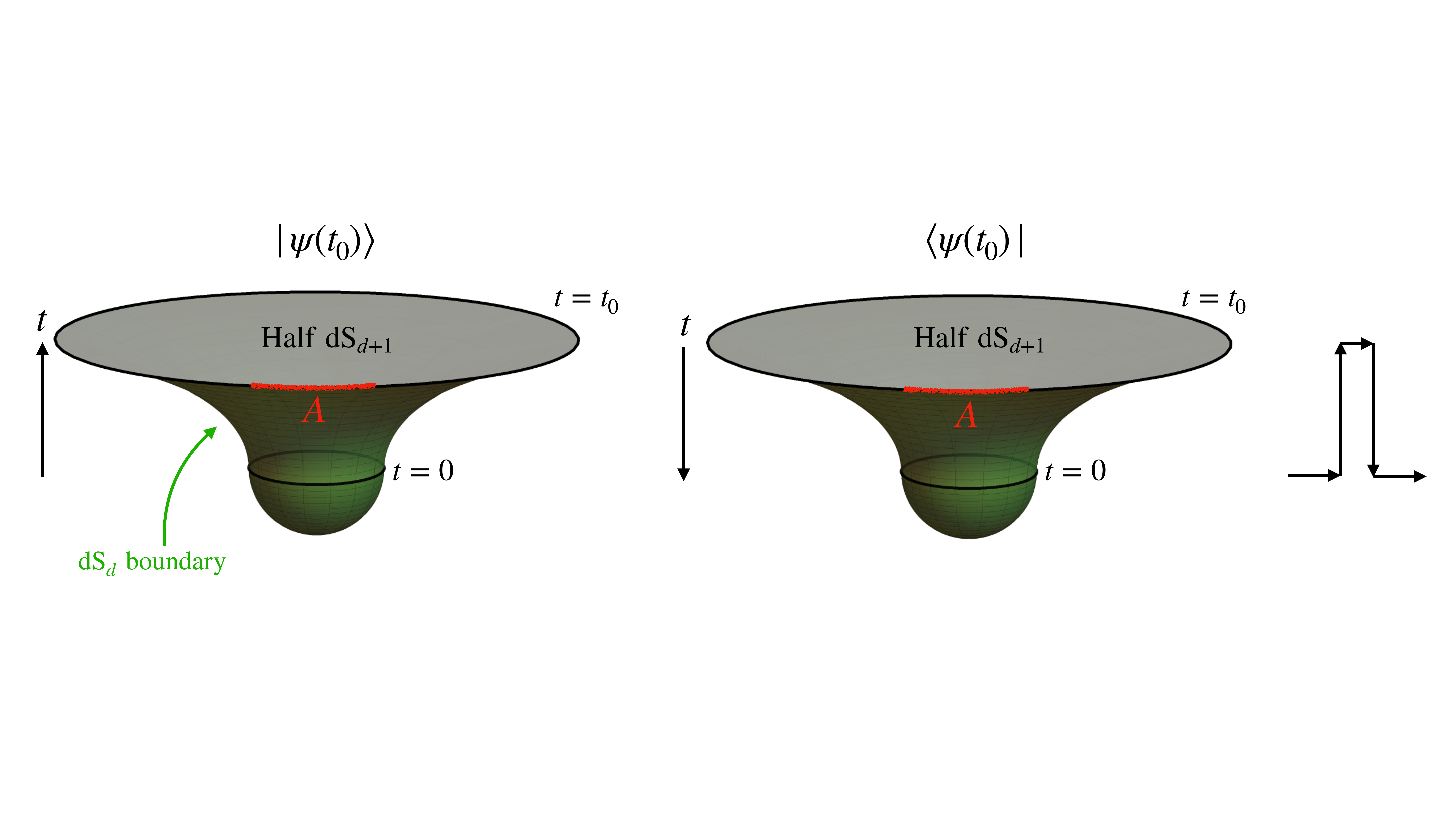}
		\caption{The geometry of a half dS$_{d+1}$. Notice that there are both timelike (green surface) and spacelike (purple surface) boundaries.} 
		\label{fig:FSP}
	\end{figure}
	
	This paper is organized as follows. In section \ref{sec:AdSCFT}, we study the AdS/CFT for CFTs on de Sitter space. We calculate the entanglement entropy in both case 1 and case 2 and discuss physical interpretations.
	In section \ref{sec:HEE} we consider the holography for a half de Sitter space in case 1. We calculate the holographic entanglement entropy and study its property including the violation of subadditivity. We will discuss its implication on the Hilbert structure of its dual field theory. In section \ref{sec:HEEW} we study the holographic duality for a half de Sitter space in case 2, taking into account the presence of EOW brane. In section \ref{sec:discussion} we summarize our conclusions and discuss future problems. In appendix 
	\ref{sec:APgeo}, we show explicit calculations of geodesics in de Sitter space.

	\section{CFT on de Sitter Space from AdS/CFT}\label{sec:AdSCFT}
	Before we work on the holography of de Sitter spaces, we would like to examine the AdS/CFT correspondence with CFT living on the de Sitter space as a warm up exercise. In this section, we focus on the holographic CFT$_2$ living on dS$_2$ spacetime whose metric is defined by 
	\begin{equation}\label{eq:dS2}
		\begin{split}ds^2_{\mathrm{dS}_2} = -dt^2 +\cosh^2{t} d\phi^2= \frac{1}{\cos^2{T}}\left(-dT^2+d{\phi}^2\right)\,.
		\end{split}
	\end{equation}
	In the following dS metrics, we have chosen the de Sitter radius to be a unit, 
	and use the transformation between the global time $t$ and conformal time $T$ given by $\cosh{t}=\frac{1}{\cos{T}}$. We note that the conformal time is compactified due to $T \in [-\frac{\pi}{2}, \frac{\pi}{2}]$, while the global time is not, i.e. $t\in (-\infty, +\infty)$.
	
	For simplicity, we mainly consider AdS$_3/$CFT$_2$ where the CFT lives on dS$_2$. The holographic dual is described by the global AdS$_3$,
	\begin{equation}\label{eq:globalAdS}
		ds^2_{(g)}=-\cosh^2\rho d\tau^2+d\rho^2+\sinh^2\rho d\phi^2\,.
	\end{equation}
	where $\rho$ is the radial coordinate and $\tau$ is the global time coordinate of AdS bulk spacetime. Since the extremal surface in AdS$_3$ is nothing but a geodesic, we can evaluate the area of the extremal surface using the geodesic distance $D^{(g)}_{12}$ in the global AdS metric between two points $(\rho_1,\tau_1,\phi_1)$ and $(\rho_2,\tau_2,\phi_2)$, as given by 
	\begin{equation}\label{GAdSgeo}
		\cosh D^{(g)}_{12}=\cos(\tau_1-\tau_2)\cosh\rho_1\cosh\rho_2-\cos(\phi_1-\phi_2)\sinh\rho_1\sinh\rho_2\,.
	\end{equation}

	\subsection{de Sitter and Hyperbolic Slicing of AdS
		\texorpdfstring{$_3$}{Lg}
	}
	
	\begin{figure}[h!]
		\centering
		\includegraphics[width=3in]{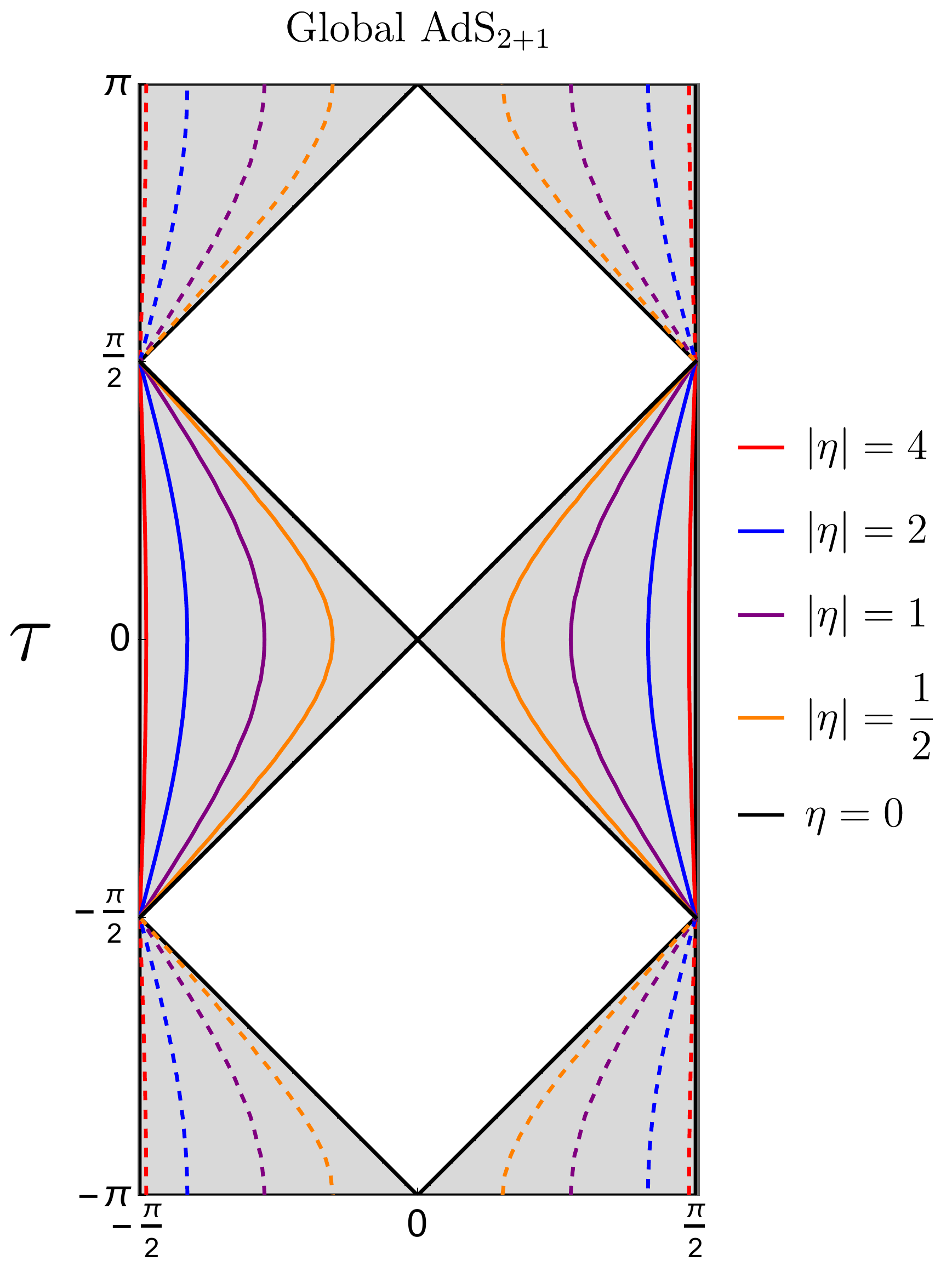}
		\includegraphics[width=3in]{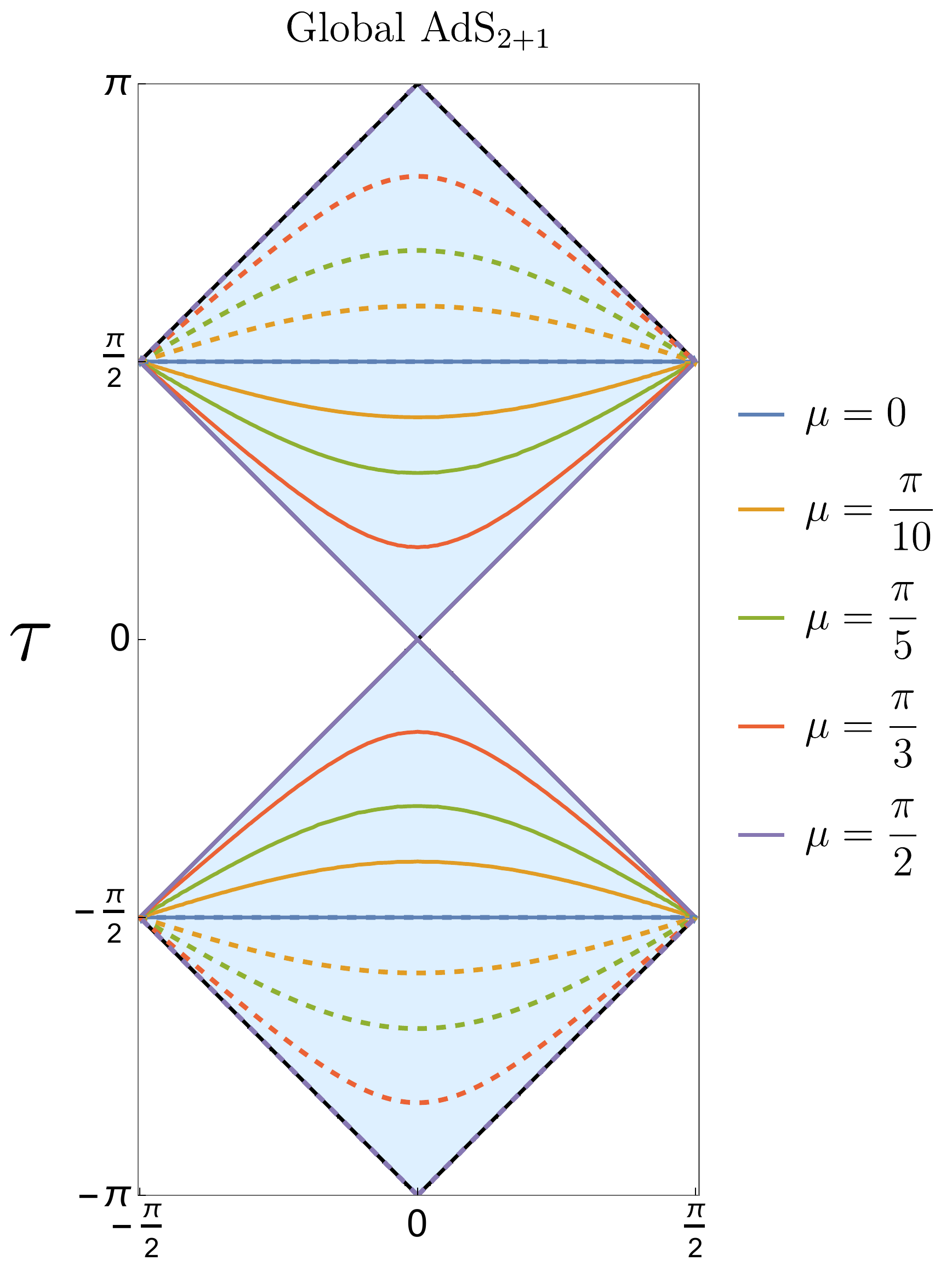}
		\caption{Left: Penrose diagram of AdS$_3$ with de Sitter slicing which is defined in eq.~\eqref{eq:dSSlicing}. The solid curves denote the constant $\eta$ surfaces with $\eta>0$ and the negative ones are described by the dashed curves. Right: Penrose diagram of AdS$_3$ with hyperbolic slicing as described by the metric \eqref{eq:H2Slicing}. The solid and dashed curves present constant $\mu$ surfaces with positive and negative values, respectively.}
		\label{fig:AdS3}
	\end{figure}

	To describe a CFT on dS$_2$, we employ the de Sitter sliced AdS$_3$ (refer to the left plot in figure \ref{fig:AdS3}):
	\begin{equation}\label{eq:dSSlicing}
		ds^2_{(d)}=d\eta^2+\sinh^2\eta(-dt^2+\cosh^2 t d\phi^2),
	\end{equation}
	where $(t,\eta)$ is related to the global coordinates $(\tau,\rho)$ via
	\begin{equation}\label{trasa}
		\sinh\rho =\cosh t\sinh |\eta|,\ \ \ \  \tan\tau =\tanh\eta\,\sinh t \,
	\end{equation}
	The coordinate transformation (\ref{trasa}) leads to the geodesic length between two points $(t_1,\eta_1,\phi_1)$ and 
	$(t_2,\eta_2,\phi_2)$ in the dS sliced metric:
	\begin{equation}\label{dSSgeo}
		\cosh D^{(d)}_{12}=\cosh\eta_1 \cosh\eta_2 - \sinh \eta_1 \sinh\eta_2\(   \cosh t_1 \cosh t_2 \cos \( \phi_2 -\phi_1\)  - \sinh t_1 \sinh t_2\)\,. 
	\end{equation}
	
	The holographic CFT$_2$ on dS$_2$ with a physical metric \eqref{eq:dS2} is living on the conformal boundary of AdS$_3$ which is defined by $\eta \to \infty$. We can fix the UV cut-off $\ep$ in the CFT$_2$ by taking the cut-off surface at $\eta = \eta_{\infty}$ with 
	\begin{equation}\label{eq:cutoff}
		e^{\eta_\infty} = \frac{1}{\tanh \frac{\epsilon}{2}} \approx \frac{2}{\epsilon} \,. 
	\end{equation} 
	
	For a later purpose it is also useful to consider a hyperbolic sliced AdS$_3$  (refer to the right plot in figure \ref{fig:AdS3}), whose metric reads 
	\begin{equation}\label{eq:H2Slicing}
		ds^2_{(h)}=-d\mu^2+\cos^2\mu(d\xi^2+\sinh^2\xi d\phi^2)\,.
	\end{equation}
	The coordinate $(t,\eta)$ is related to that of the global AdS$_3$ $(\tau,\rho)$ via
	\begin{equation}\label{trasaq}
		\sinh\rho =\sinh \xi\cos\mu , \ \ \ \ 
		\tan\tau=\cot\mu \, \cosh \xi ,
	\end{equation}
	The geodesic length $D^{(h)}_{12}$ between two points at $(\mu_1,\xi_1,\phi_1)$
	and $(\mu_2,\xi_2,\phi_2)$ can be derived from 
	\begin{equation}
		\cosh D^{(h)}_{12}=\cos\mu_1\cos\mu_2\left(\cosh\xi_1\cosh\xi_2-\cos(\phi_1-\phi_2)\sinh\xi_1\sinh\xi_2\right)+\sin\mu_1\sin\mu_2 \,.
	\end{equation}
	It is also useful to note that the hyperbolic sliced metric and the de Sitter sliced one are related by the following analytical continuation:
	\begin{equation}
		\eta=i\left(\mu-\frac{\pi}{2}\right)\,, \qquad \xi=i\frac{\pi}{2}-t\,.
	\end{equation}

	\begin{figure}[t]
		\centering
		\includegraphics[width=2.5in]{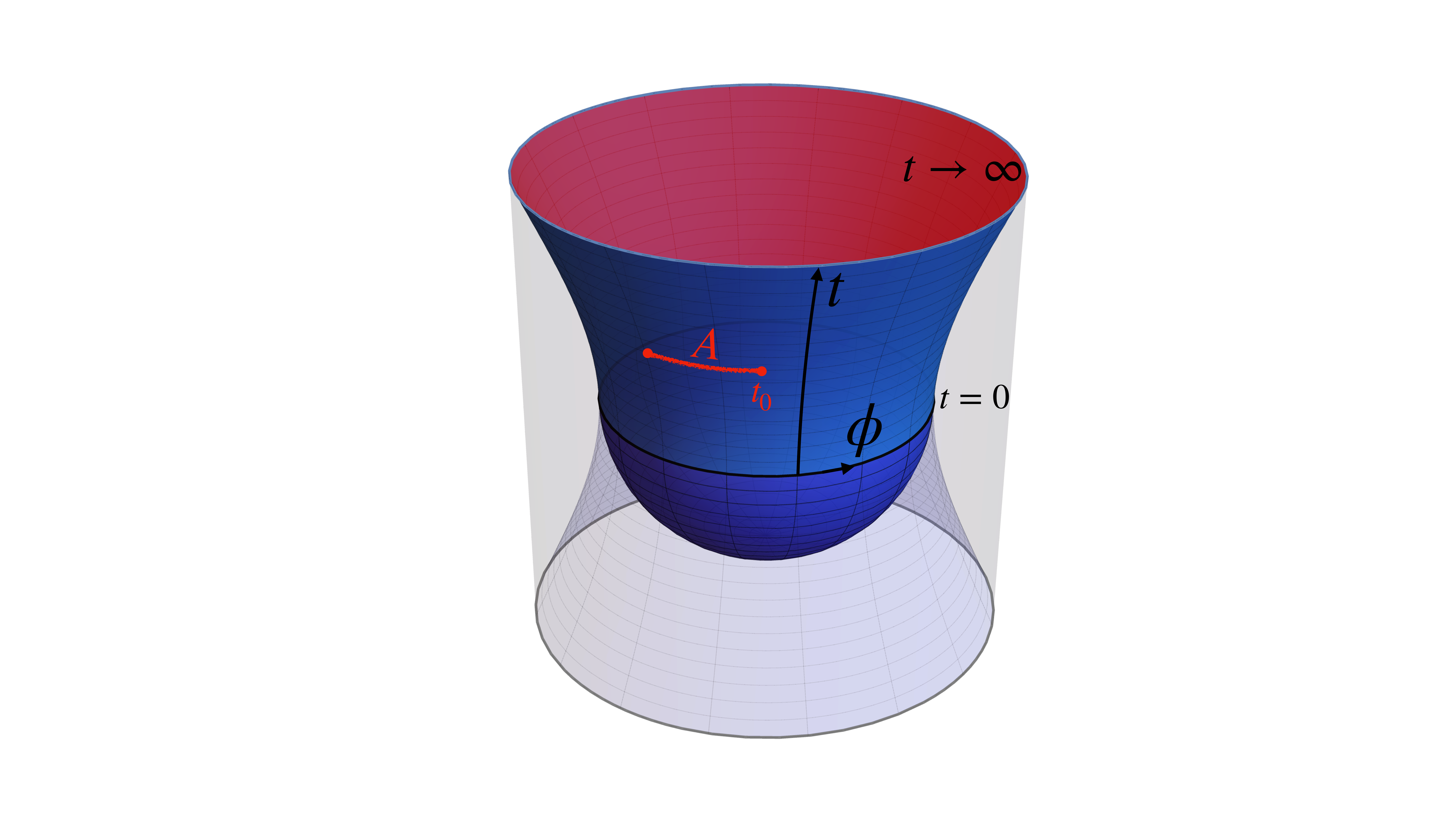}
		\caption{The setup of computing entanglement entropy of CFT$_2$ on dS$_2$ via AdS$_3$/CFT$_2$. We replace the region $t<0$ by the Euclidean instanton namely the half sphere, following Hartle-Hawking prescription (blue shaded region). In the case of Schwinger-Keldysh setup, we restrict the Lorentzian path-integral to the region $0\leq t\leq t_0$ and glue the same path-integral at $t=t_0$.}
		\label{fig:AdS3D}
	\end{figure}
	
	\subsection{Case 1:Entanglement entropy in Schwinger-Keldysh prescription}
	
	In this section, we aim to compute the holographic entanglement entropy \cite{Ryu:2006bv,Ryu:2006ef,Hubeny:2007xt} for a CFT living on dS$_2$ spacetime, following the Schwinger-Keldysh prescription (case 1) of the CFT on a time-dependent background. Refer to \cite{Maldacena:2012xp} for calculations of holographic entanglement entropy for CFTs on de Sitter spaces in higher dimensions.
	
	Specifically, we assume that the quantum state at $t=0$ is generated by an Euclidean path integral on a semi-sphere using the Hartle-Hawking prescription, as depicted in figure \ref{fig:AdS3D}. To define the entanglement entropy $S_A$, we take the subsystem $A$ as an interval with endpoints $(t_0,\phi_1)$ and $(t_0,\phi_2)$, as shown in figure \ref{fig:AdS3D}.
	
	To derive the holographic entanglement entropy, we construct the gravity dual by gluing two copies of AdS$_3$, each truncated at a time $t=t_0$, following the holographic Schwinger-Keldysh prescription \cite{Skenderis:2008dg,Dong:2016hjy}. Hence, there is no need to consider the treatment of the conformal boundary of the dS$_2$ at $t=\infty$. Utilizing the geodesic length given in eq.~\eqref{dSSgeo}, we can obtain the holographic entanglement entropy, \viz 
	\begin{equation}\label{eq:holcon}
		S^{\rm con}_A=\frac{c}{3}\log\left[\frac{2\cosh t_0\sin \frac{|\phi_1-\phi_2|}{2}}{\epsilon}\right]\,,
	\end{equation}
	where we adopt the cutoff surface specified by eq.~\eqref{eq:cutoff}. Notably, this entanglement entropy exhibits a linear growth as $S_A \simeq \frac{c}{3}t_0$ at late times. We attribute this contribution to a connected geodesic that connects two boundary points. Thus, in the Schwinger-Keldysh setup, this result \eqref{eq:holcon} provides the final expression for the holographic entanglement entropy.

	\subsection{Case 2: Holographic pseudo entropy with a final state projection}\label{catwadS}
	
	\begin{figure}[t]
		\centering
		\includegraphics[width=4.5in]{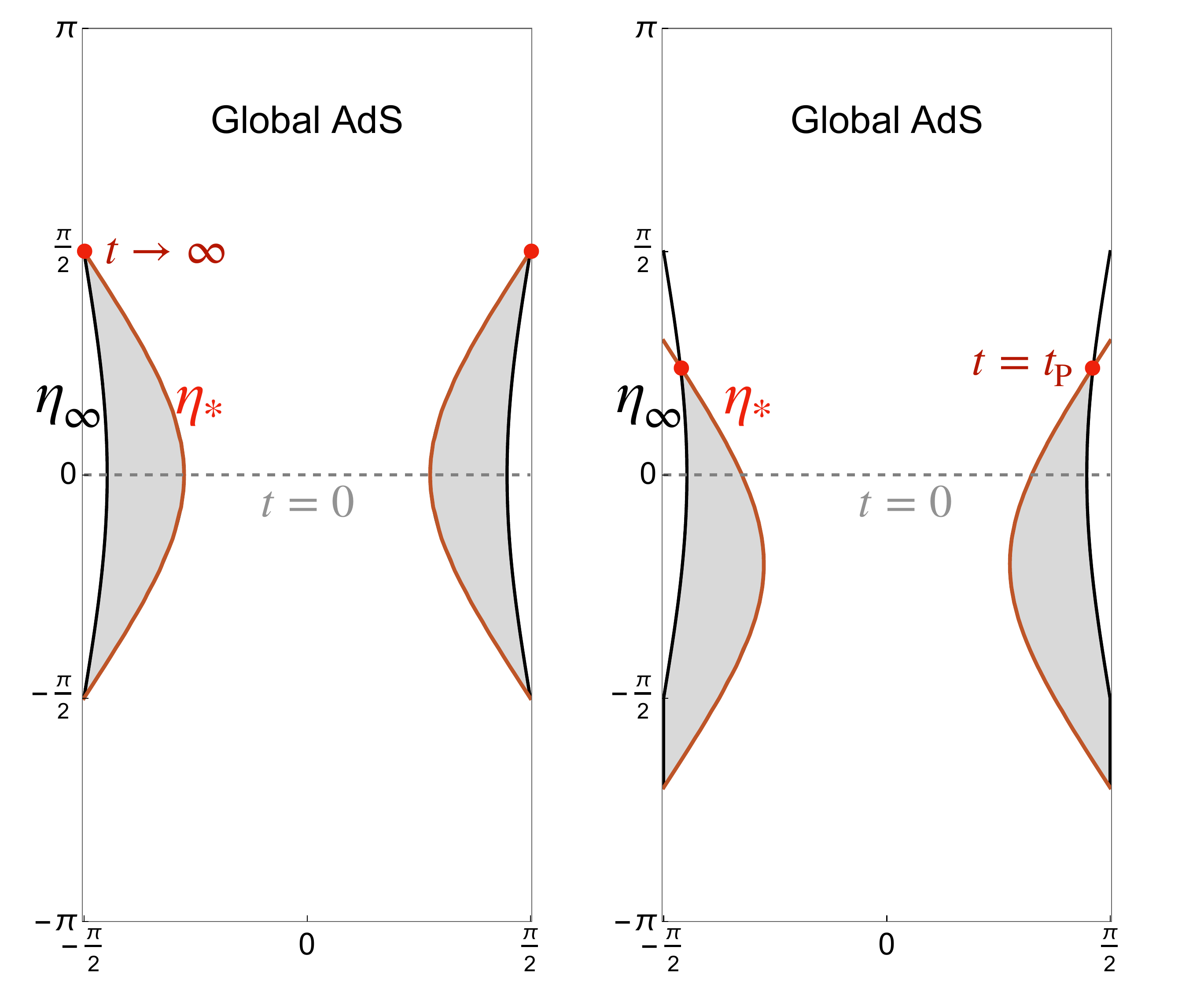}
		\caption{The holographic dual of CFT$_2$ living on dS$_2$ with a projection is shown as the gray shaded region. The angular direction of AdS$_3$ is not shown in this figure. The solid black curve represents the two dimensional dS spacetime where CFT$_2$ lives. Left: The projection is performed at the infinity time $t_{\mt{P}} \to \infty$ with a dS$_2$ brane (red curve) parametrized by a constant $\eta_\ast$. Right: The projection at a finite time $t_{\mt{P}}$.
		}\label{fig:projection}
	\end{figure}
	
	\begin{figure}[t]
		\centering
		\includegraphics[width=5in]{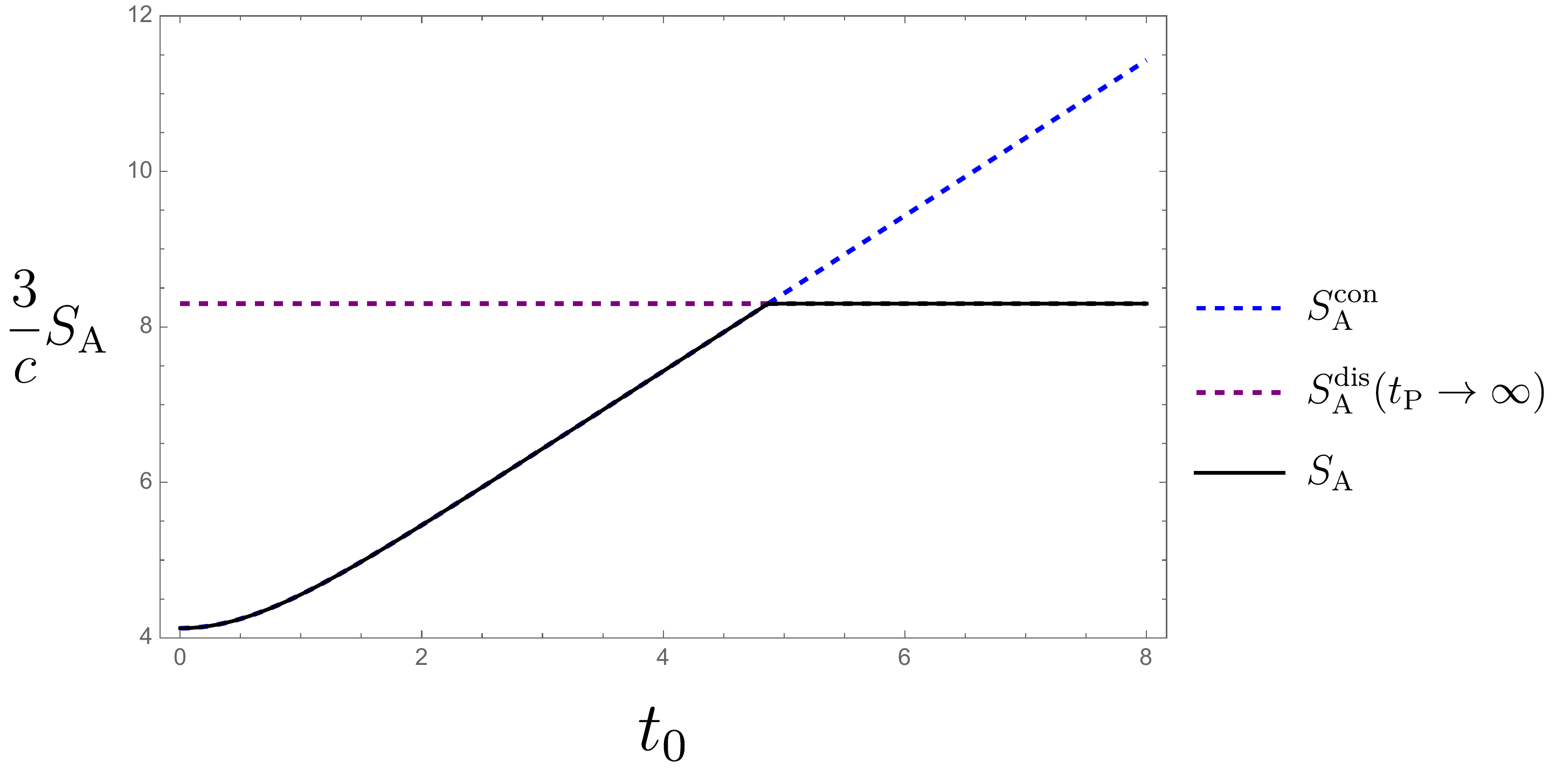}
		\caption{The plot of $S^{\rm con}_A$ (blue) and $S^{\rm dis}_A$ (purple) as a function of time $t$. We set $\eta_\ast=3, \phi_2-\phi_1= \frac{\pi}{5},\epsilon=\frac{1}{100}$ for this numerical plot.}
		\label{fig:SEEinfinity}
	\end{figure}
	
	%%%%%%%%%%%%%%%%%%%%%%%%%%%%%%%%%%%%%%%%%%%%%%%%%%%%%%%%%%%%%%%
	%%%%%%%%%%%%%%%%%%%%%%%%%%%%%%%%%%%%%%%%%%%%%%%%%%%%%%%%%%%%%%%
	We now consider the setup in which there is a final state projection with a boundary state $|\rm{B}\rangle$ at future infinity $t=\infty$, and interpret the area of the extremal surface, \ie $S_A$ as the holographic pseudo entropy \cite{Nakata:2021ubr}. To construct the gravity dual of the holographic CFT$_2$ with a projection, we insert an end-of-the-world brane (EOW) brane on which we impose Neumann boundary condition, \ie 
	\begin{equation}\label{eq:Neumann}
		K_{ij} - K h_{ij} + T h_{ij} =0 \,, \qquad \text{with} \qquad K= 2T\,,
	\end{equation}
	where $K_{ij}$ is the extrinsic curvature of the brane and the constant $T$ is the brane tension. Since the bulk is AdS$_3$ spacetime, one can show that the solutions of eq.~\eqref{eq:Neumann} have to be maximally symmetric spacetime as well. More explicitly, one can obtain (see \eg \cite{Kawamoto:2023wzj})
	\begin{equation}
		R_{ij}[h] = h_{ij}\( \varepsilon\,  T^2 -1  \) \,, \qquad \text{with} \quad R[h] =2 ( \varepsilon \, T^2 -1)\,,
	\end{equation}
	for a timelike brane with $\varepsilon=+1$ and a spacelike brane with $\varepsilon=-1$, respectively. In order words, the intrinsic geometry of the brane is nothing but two dimensional Minkowski ($|T|=1$), de Sitter ($|T|>1$), AdS ($|T|<1$) spacetime or hyperbolic space ($\varepsilon =-1$).
	The de Sitter EOW brane in AdS/BCFT was introduced in \cite{Akal:2020wfl} and was applied to the calculation of holographic pseudo entropy in \cite{Akal:2021dqt} in the context of black hole final state proposal \cite{Horowitz:2003he}.
	
	With taking the intersection of the brane and the dS$_2$ boundary at $t \to \infty$, we can find that there are two types of solutions of the brane: a timelike dS brane and a spacelike hyperbolic brane. From the dS slicing and hyperbolic slicing of AdS$_3$, one can easily read the corresponding brane profiles in global AdS$_3$ (as shown in figure \ref{fig:AdS3}), \ie 
	\begin{equation}
		\begin{split}
			\text{dS$_2$ brane:} \qquad  \cosh \rho \cos \tau &= \pm \cosh \eta_\ast \,, \\ 
			\text{H$_2$ brane:} \qquad  \cosh \rho \cos \tau &= \sin \mu_\ast \,, \\ 
		\end{split}
	\end{equation}
	where $\eta_\ast, \mu_\ast$ is a constant along the brane and is determined by the tension which is given by the boundary entropy of the boundary state. Furthermore, the parameter $\eta_\ast$ and $\mu_\ast$ should be determined by the boundary condition at $t=\infty$.

	We begin with the case with a dS$_2$ EOW brane. Let us first note that the trace of the extrinsic curvature of EOW brane can be expressed as follows:
	\begin{equation}
		K \big|_{\eta=\eta_\ast} = -\frac{\partial_\eta \sqrt{-\gamma}}{\sqrt{-\gamma}} \Big|_{\eta=\eta_\ast} =  -\frac{2 \cosh \eta_\ast}{\sinh \eta_\ast}\,,
	\end{equation}
	which indicates that the tension and curvature of the EOW brane are given by $T = -\coth \eta_\ast< 0$ and $R=\frac{2}{\sinh^2 \eta_\ast}$, respectively. The corresponding boundary entropy associated with the boundary state $\ket{B}$ projected at $t \to \infty$ can be calculated as \cite{Akal:2021dqt}
	\begin{equation}\label{eq:bdyent}
		S_{\rm bdy}=\frac{c}{6}\log \sqrt{\frac{ 1-\mathcal{T}}{1+\mathcal{T}}} =\frac{c}{6}\log \sqrt{\frac{ |\mathcal{T}|-1}{|\mathcal{T}|+1}}-i\,\frac{\pi c}{12} = -\frac{c}{6}\eta_\ast -i\,\frac{\pi c}{12} \,. 
	\end{equation} 
	which decodes the information of the boundary condition at $t \to \infty$. 
	
	Due to the appearance of the EOW brane in the bulk, we must account for the disconnected geodesic that connects a endpoint $(t_0,\eta_\infty, \phi_i)$ of boundary interval $A$ to a point $(t_\ast,\eta_\ast, \phi_\ast)$ on the de Sitter EOW brane. Thanks to the rotation invariance, it is straightforward to get the extremal surface is given by $\phi_\ast=\phi_i$. Furthermore, we can fix the coordinate values of $t_\ast$ on the brane by ensuring that the geodesic length remains stationary. For the de Sitter EOW brane, the disconnected contribution for pseudo entropy can be derived from eq.~\eqref{dSSgeo} as follows:
	\begin{equation}
		S^{\rm dis}_A=\frac{c}{3}(\eta_\infty-\eta_\ast)=\frac{c}{3}\log\frac{2}{\ep}-\frac{c}{3}\eta_\ast\, \label{HPED1}
	\end{equation}
	where the constant part is the real part of the boundary entropy defined in eq.~\eqref{eq:bdyent}.

	If we compare the connected contribution $S^{\rm con}_A$ and the disconnected part $S^{\rm dis}_A$, we find that in the early time $S^{\rm con}_A$ is favored and holographic pseudo entropy of the interval $A$ grows linearly in time. However, $S^{d\rm dis}_A$ is always favored in the late time and holographic pseudo entropy becomes a constant. This phase transition is plotted in figure \ref{fig:SEEinfinity}. At $\eta_\ast=0$, this phase transition happens when
	\begin{equation}\label{tmaxt}
		\cosh t_0\cdot \sin \frac{|\phi_1-\phi_2|}{2}=1.
	\end{equation}
	This describes the null geodesic in dS$_2$ as sketched in figure \ref{fig:horizon}, which is the same as \eqref{boundnu}.

	For the hyperbolic EOW brane, the disconnected PE is estimated from (\ref{dSSgeo}) as follows:
	\begin{equation}
		S^{\rm dis}_A=\frac{c}{3}\log\frac{2}{\ep}+i\frac{c}{3}\left(\frac{\pi}{2}-\mu_*\right),
	\end{equation}
	where $\mu_*$ takes the range $-\frac{\pi}{2}\leq \mu_*\leq \frac{\pi}{2}$.
	
	Before we go on, we would like to mention a possibility that this holographic calculation of pseudo entropy may actually be interpreted as genuine entanglement entropy. This is because the global AdS$_3$ has the periodicity in the time direction $\tau$. This may imply the periodicity for the time evolution by $\pi$ of the boundary state $e^{-i\pi H}|\text{B}\lb\propto |\text{B}\lb$. Even though this is suggested by the classical geometry, we are not completely sure if this is true at the quantum level which is dual to the full dynamics of the CFT. We leave this issue for a future problem.

	\subsection{Holographic pseudo entropy under final state projection at \texorpdfstring{$t=\tP$}{Lg}}\label{24}
	
\begin{figure}[t]
		\centering
		\includegraphics[width=5in]{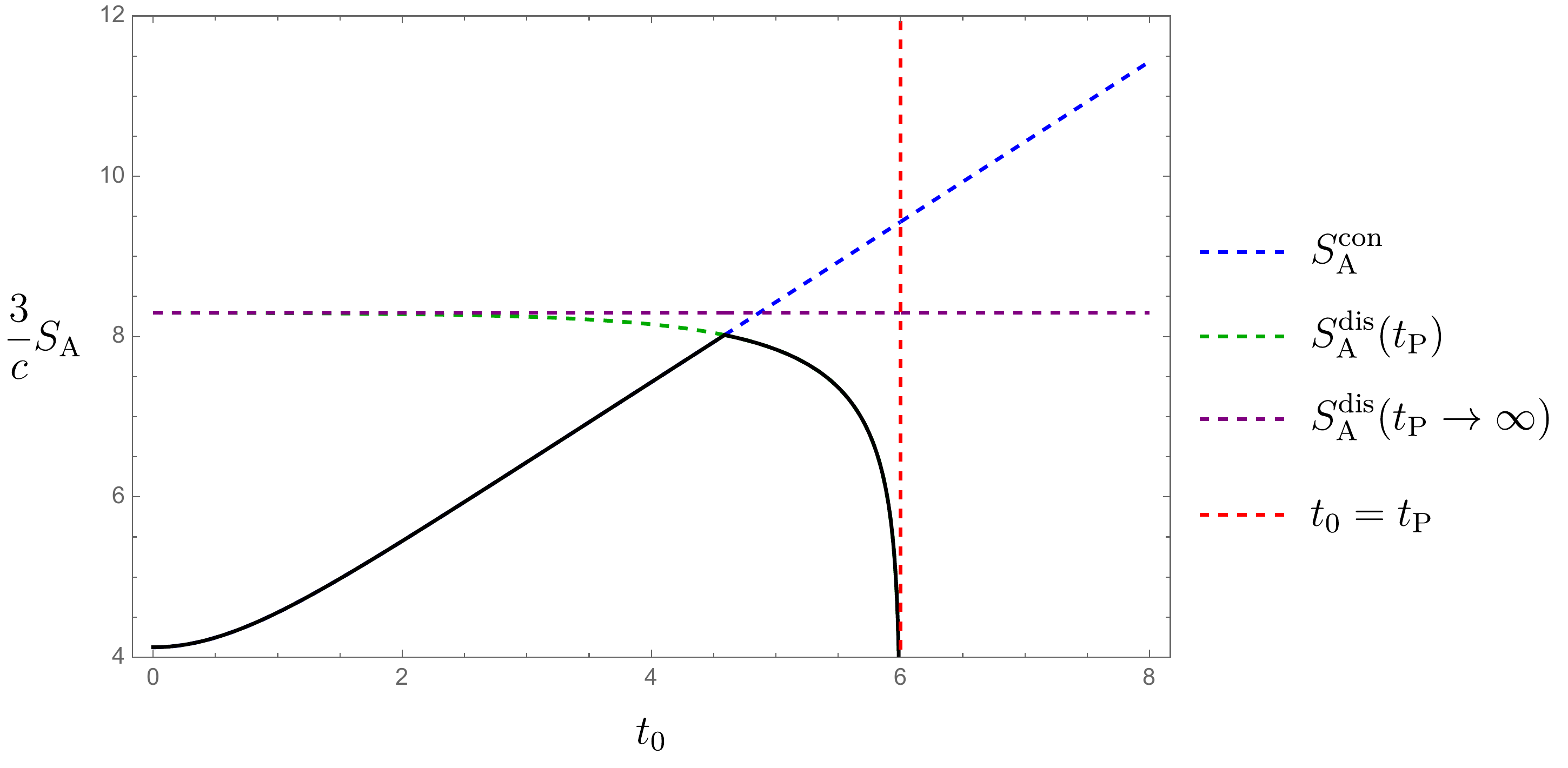}
		\caption{Time evolution of holographic entanglement entropy (denoted by the black curve) with respect to the global time $t_0$. The projection time is chosen as $t_\mt{P}=6$ with $\eta_\ast =3$. We set $\phi_2-\phi_1= \frac{\pi}{5},\epsilon=\frac{1}{100}$ for this numerical plot.}
		\label{fig:entropyprojection}
	\end{figure}

	For later purpose, it is useful to consider a CFT on dS$_2$ with a final state projection at a finite time $t=t_{\mt P}$.
	In the AdS$_3/$BCFT$_2$, this is dual to inserting the EOW brane earlier as depicted in the right panel of figure \ref{fig:projection}. It is obvious that $S_A$ vanishes at $t=\tP$. Therefore the time evolution of $S_A$ looks like a Page curve as sketched in the right panel of figure \ref{fig:entropyprojection}.
	
	Since the brane profile is not completely covered by the dS slicing coordinates, it is more convenient to work on the global coordinates (\ref{eq:globalAdS}). We denote the points on the dS boundary as $(\tau_0, \rho_\infty, \phi_0)$. The dS boundary is thus given by
	\begin{equation}
		\cosh \rho_\infty \cos \tau_0 = \text{constant}= \cosh \eta_\infty\,, 
	\end{equation}
	where $\eta_\infty$ determines the position of the dS boundary. Corresponding, the point on the brane is referred to as  $(\tau_*, \rho_*, \phi_*)$. Since the brane is still represents a two-dimensional dS spacetime, we can parametrize the brane profile as 
	\begin{equation}
		\cosh \rho_* \cos \(  \tau_* + \tau_{\rm shift}  \) = \text{constant}= \cosh \eta_\ast\,, 
	\end{equation}
	where $\eta_\ast$ is related to the tension of the brane. Since we parametrize the projection time is $t=\tP$ or $T=T_{\mt P}$ in terms of the time coordinate on the dS boundary, we can find that this fix the brane parameter $\tau_{\rm shift}$ as 
	\begin{equation}
		\tau_{\rm shift} = \arccos \( \frac{\cosh \eta_\ast \cos T_0 }{ \cosh \eta_\infty}  \)  - \tau{\mt{P}} =\arccos \( \frac{\cosh \eta_\ast \cos T_0 }{ \cosh \eta_\infty}  \)  - \arctan \( \tanh \eta_\infty \tan \TP  \)  \,, 
	\end{equation}
	with using the coordinate transformations for the points on dS boundary as  
	\begin{equation}
		\begin{cases}
			\cosh \rho =  \sqrt{ \cosh^2 \eta  +\sinh^2 \eta \tan^2 T }   \,,\\
			\tan \tau = \tanh \eta \tan T  \,. \\
		\end{cases}
	\end{equation}

	On the other hand, the geodesic distance $D_{0b}$ from the dS boundary to dS brane is derived as 
	\begin{equation}
		\cosh D_{0b}=\cos(\tau_0-\tau_*)\cosh\rho_\infty\cosh\rho_*-\cos(\phi_0-\phi_*)\sinh\rho_\infty\sinh\rho_*.
	\end{equation}
	Extremization over the spatial direction simply results in $\phi_0=\phi_*$, which is expected by symmetries. Furthermore, we need to find that maximal value over the timelike direction by solving 
	\begin{equation}
		\frac{\partial D_{0b}}{\partial \tau_*} =0 \,, \quad \text{with} \quad \rho_* = \rho_* (\tau_*)\,,
	\end{equation}
	which reduces to 
	\begin{equation}
		\cosh \eta_\ast \sinh \rho_\infty \tan\( \tau_* + \tau_{\rm shift} \) -  \frac{\sin\( \tau_* + \tau_{\rm shift} \)}{\cos\( \tau_0 + \tau_{\rm shift} \)}\cosh \rho_\infty \sqrt{ \cosh^2 \eta_\ast -  \cos^2 \( \tau_0 + \tau_{\rm shift} \)}=0 \,. 
	\end{equation}
	The advantage of working on global coordinates is obvious in the above extremization equation. After straightforward algebras, one can derive the solutions as 
	\begin{equation}
		\begin{split}
			\cos\( \tau_* + \tau_{\rm shift} \) &= \cosh \eta_\ast  \sqrt{\frac{ \cosh^2 \rho_\infty \sin^2 (\tau_0 +\tau_{\rm shift}  )  - \sinh^2 \rho_\infty  }{ \cosh^2 \rho_\infty \sin^2 (\tau_0 +\tau_{\rm shift}  )  - \sinh^2 \rho_\infty  \cosh^2 \eta_\ast }  }\,,\\
			\cosh \rho_* &= \sqrt{\frac{ \cosh^2 \rho_\infty \sin^2 (\tau_0 +\tau_{\rm shift}  )  - \sinh^2 \rho_\infty  \cosh^2 \eta_\ast }{ \cosh^2 \rho_\infty \sin^2 (\tau_0 +\tau_{\rm shift}  )  - \sinh^2 \rho_\infty  }  }\,.\\
		\end{split}
	\end{equation}
	Substituting the above solutions to the expression for the distance of geodesics, we can obtain the distance of the extremal geodesic between the dS boundary and brane, \ie the area of the disconnected HRT surface.
	
	We are interested in the case where the dS boundary is taken as the cut-off surface located at $\eta_\infty \sim \frac{2}{\epsilon}$. In this limit, we can rewrite the global coordinates on the boundary as 
	\begin{equation}
		\begin{split}
			\tau_0 &= T_0 \,, \qquad  \rho_\infty \approx \log \( \frac{2}{\epsilon \cos T_0} \)\,, \qquad \tau_{\rm shift} = \frac{\pi}{2} - \TP \,.\\
		\end{split}
	\end{equation}
	The extremal point on the brane reduces to 
	\begin{equation}
		\begin{split}
			\sin\( \TP - \tau_* \) &= \cosh \eta_\ast \sqrt{\frac{\sin^2(\TP - T_0) }{ \cosh^2 \eta_\ast - \cos^2 (\TP -T_0)}}\,,\\
			\cosh \rho_* &= \sqrt{1 + \frac{\sinh^2 \eta_\ast}{\sin^2(\TP-T_0)} } \,,\\
		\end{split}
	\end{equation}
	and the corresponding geodesic distance is given by  
	\begin{equation}
		D_{0b} \approx \log \( \frac{2}{\epsilon} \frac{\cos (T_0 + \tau_{\rm shift})  }{\cos T_0} e^{-\eta_\ast} \) \,.
	\end{equation}
	Thus, the holographic entanglement entropy from the disconnected HRT surface is recast as 
	\begin{equation}
		S^{\rm dis}_{A} = \frac{2 D_{0b}}{4 \GN} = \frac{c}{3} \log \( \frac{2}{\epsilon} \frac{\sin \( \TP - T_0 \)  }{\cos T_0}     \)  - \frac{c}{3}\eta_\ast  \,.
	\end{equation}
	In terms of the de Sitter time $t$, we finally find the following expression:
	\begin{equation}
		S^{\rm dis}_A=\frac{c}{3}\log\left[\frac{2}{\ep}\cdot \frac{\sinh \tP-\sinh t_0}{\cosh \tP}\right]-\frac{c}{3}\eta_\ast,  \label{HPED2}
	\end{equation}
	where we have used $\sinh \tP=\tan \TP$. The holographic pseudo entropy in the presence of final state projection is given by the smaller one among the two: $S_A=\mbox{min}[S^{\mathrm{con}}_A, S^{\mathrm{dis}}_A]$.

	\subsection{Entanglement/Pseudo entropy of CFT\texorpdfstring{$_2$}{Lg} on dS\texorpdfstring{$_2$}{Lg}}
	
	It is intriguing to compare previous holographic results of entanglement entropy and pseudo entropy with that derived in 2d CFT. 
	Indeed, as in the general replica method for 2d CFTs \cite{Holzhey:1994we,Calabrese:2004eu},
	we can easily compute the two-point function of twist operators of local CFT on the $\mathrm{dS}_2$ by considering the Weyl scaling:
	\begin{equation}
		\la\sigma(t_0,\phi_1)\overline{\sigma}(t_0,\phi_2)\ra_{\mathrm{dS}_2}= (\cos{T_0})^{2\Delta_n}\la\sigma(T_0,\phi_1)\overline{\sigma}(T_0,\phi_2)\ra_{\mathrm{cylinder}} \,.
	\end{equation}
	Here we have define the conformal coordinates $(T,\phi)$ by
	\begin{equation}
		\begin{split}
			ds^2_{\mathrm{dS}_2} &= -dt^2 +\cosh^2{t}d\phi^2= \frac{1}{\cos^2{T}}\left(-dT^2+d{\phi}^2\right)\,\\
		\end{split}
	\end{equation}
	with the transformation between the global time and conformal time given by $\cosh{t}=\frac{1}{\cos{T}}$. 
	The Euclidean part of the Hartle-Hawking state is described by a half sphere whose metric reads 
	\begin{equation}
		ds^2_{\mathrm{HH}}= dt_{\mt {E}}^2 +\cos^2{t_{\mt{E}}}d\phi^2
		= \frac{1}{\cosh^2 T_{\mt{E}}} \left(dT_{\mt{E}}^2+d{\phi}^2\right)\,,
	\end{equation}
	where the Euclidean conformal time $T_{\mt{E}}$ is defined by 
	$\cos{t_{\mt{E}}}=\frac{1}{\cosh{T_{\mt{E}}}} $ with $-\frac{\pi}{2} \leq t_{\mt{E}}\leq 0,-\infty\leq T_{\mt{E}}\leq 0$.

	\begin{figure}
		\centering
		\includegraphics[width=6in]{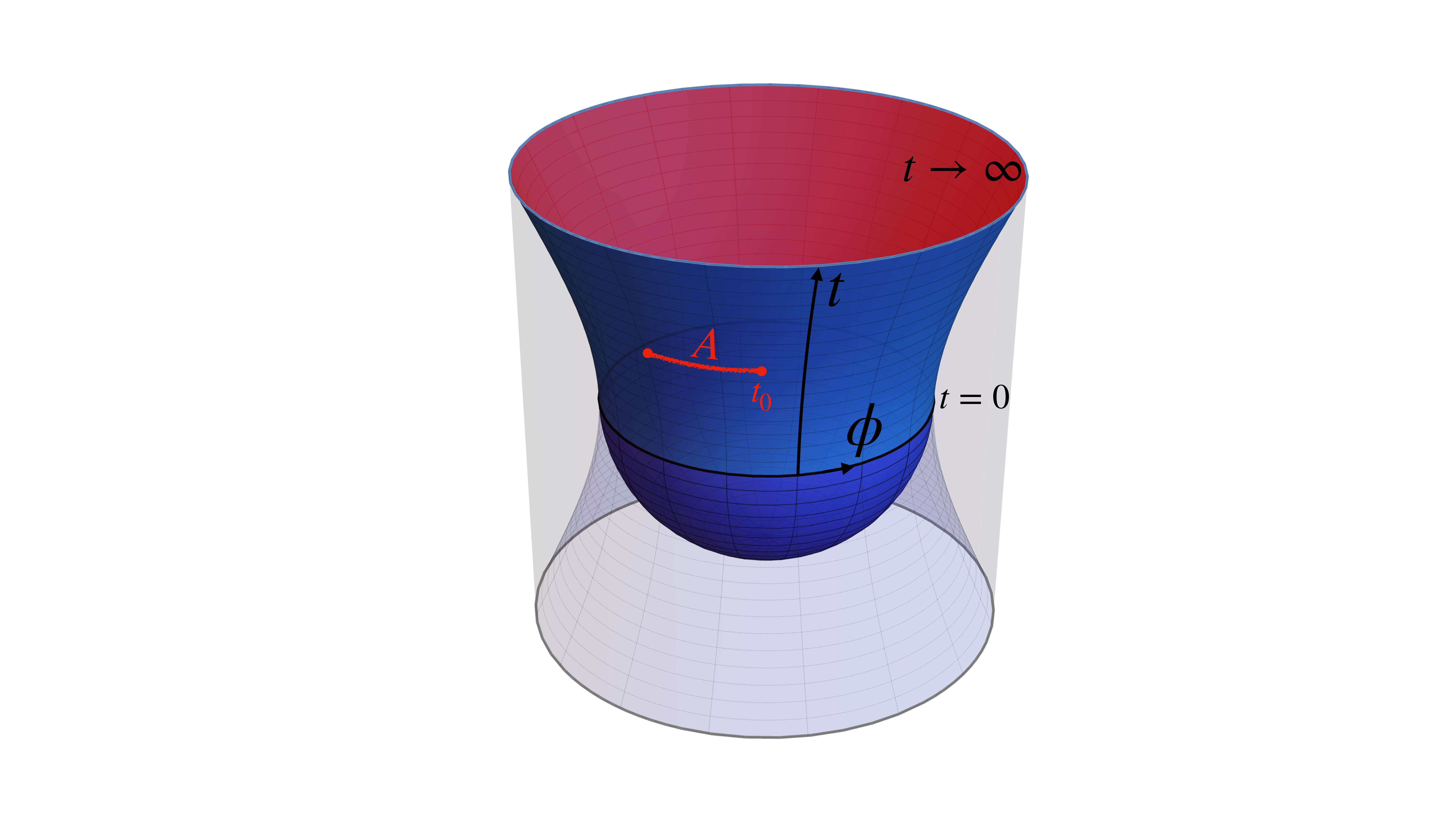}
		\caption{The conformal maps used for computing the pseudo entropy. The projection is performed at a fixed time $t_{\mt{P}}$ which is indicated by the red circle.}
		\label{fig:map}
	\end{figure}

	\subsubsection*{Case 1: Schwinger-Keldysh description without EOW}
	First of all, it is known that the entanglement entropy for an interval in the cylinder is shown as  
	\begin{equation}
		S_A^{\mathrm{cylinder}}=\frac{c}{3}\log{\left(\frac{2}{\epsilon}\sin{\frac{|\phi_1-\phi_2|}{2}}\right)}\,.
	\end{equation}
	where we have used the fact that the periodicity of the cylinder is $2\pi$. Noting the contribution from Weyl scaling, we can obtain 
	\begin{equation}\label{eq:EE case 1}
		S_A^{\mathrm{dS}_2}=\frac{c}{3}\log{\cosh{t_0}}+\frac{c}{3}\log{\left(\frac{2}{\epsilon}\sin{\frac{|\phi_1-\phi_2|}{2}}\right)}\,,
	\end{equation}
	which reproduces the holographic result shown in eq.~\eqref{eq:holcon}.
	
	\subsubsection*{Case 2: Final projection at $t=\tP$}
	To compute the pseudo entropy of CFT$_2$ living on $\mathrm{dS}_2$, we follow the recipe illustrated in figure \ref{fig:map}. Let us  begin with the following two-point function and its Weyl scaling: 
	\begin{equation}
		\begin{split}
			&\bra{\rm B}e^{iH(T_p-T)} \sigma(T_0,\phi_1)\overline{\sigma}(T_0,\phi_2)\ket{\rm HH}_{\mathrm{dS}_2}\\= &(\cos{T_0})^{2\Delta_n} \bra{\rm B}e^{-iH(\TP-T)}\sigma(T_0,\phi_1)\overline{\sigma}(T_0,\phi_2)\ket{0}_{\mathrm{cylinder}}\,,
		\end{split}
	\end{equation}
	where $|\rm B\lb$ is the boundary state (Cardy state) in the boundary conformal field theory (BCFT) \cite{Cardy:2004hm} and $|\rm HH\lb$
	denotes the Hartle-Hawking state of dS$_2$. We can first focus on the transition matrix, \ie 
	\begin{equation}\label{eq:transition}
		\bra{\rm B}e^{-iH(\TP-T)}\sigma(T_0,\phi_1)\overline{\sigma}(T_0,\phi_2)\ket{0}_{\mathrm{cylinder}}\,,
	\end{equation}
	with ignoring the Weyl factor. 
	The corresponding cylinder for evaluating the transition matrix is composed of the infinitely long Euclidean cylinder $-\infty\leq \TE\leq 0$ and also a finite Lorentzian cylinder with $0\leq T \leq \TP$. By Wick rotation, we consider the Euclidean cylinder $-\infty\leq \TE \leq \TP^E=i \TP$. With denoting the complex coordinate for the cylinder as 
	\begin{equation}
		w=\TE+i\phi\,,
	\end{equation}
	we can map the Euclidean cylinder to the upper half plane. The explicit conformal map is given by 
	\begin{equation}
		\frac{e^w}{e^{\TP^E}}= \frac{z-i}{z+i} \Leftrightarrow z=f(w)= -i\frac{e^{w-\TP^E}+1}{e^{w-\TP^E}-1}\,.
	\end{equation}
	See the figure \ref{fig:map} for the illustration.

	By assuming the mirror trick and the factorization, it is straightforward to evaluate the transition \eqref{eq:transition}. Similar to the connected and disconnected geodesics, one can also obtain two distinct contributions, \ie 
	\begin{enumerate}
		\item Connected part:
		\begin{equation}
			S^{\mathrm{con,cyl}}_{A}=\frac{c}{6}\log\frac{|f(w_1)-f(w_2)|^2}{\ep^2 |f'(w_1)||f'(w_2)|} .
		\end{equation}
		After substituting 
		\begin{equation}\label{eq:coordinates}
			w_1=i(T_0+\phi_1),w_1^*=i(T_0-\phi_1),w_2=i(T_0+\phi_2),w_2^*=i(T_0-\phi_2)\,, 
		\end{equation}
		we obtain the same formula \eqref{eq:EE case 1} which it was derived in holographic spacetime for Case 1. 
		\item  Disconnected part: 
		\begin{equation}S^{\mathrm{dis,cyl}}_A=\frac{c}{6}\log\frac{|f(w_1)-\bar{f}(\bar{w}_1)||f(w_2)-\bar{f}(\bar{w}_2)|}{\ep^2 |f'(w_1)||f'(w_2)|}
			+2S_{\mathrm{bdy}}.
		\end{equation}
		Again, by substituting eq.~\eqref{eq:coordinates}, we can derive the contribution from the disconnected part, \viz 
		\begin{equation}
			S^{\mathrm{dis,cyl}}_A = \frac{c}{6}\log{\frac{-4}{\ep^2}\sin^2{(T_p-T_0)}} +2S_{\mathrm{bdy}} = \frac{c}{3}\log\left(\frac{2}{\ep}\sin{(T_p-T_0)}\right) + \frac{\pi i}{12}c + 2S_{\mathrm{bdy}}\,.
		\end{equation}
		Finally, we need to consider the extra contribution from Weyl scaling and rewrite the final answer as 
		\begin{equation}
			S_A^{\mathrm{dis,dS}}= \frac{c}{3}\log\left(\frac{2}{\ep}\frac{\sin{(T_p-T_0)}}{\cos{T_0}}\right) + \frac{\pi i}{12}c + 2S_{\mathrm{bdy}}\,.
		\end{equation}
		Combining the two distinct results, we can also find that the pseudo entropy present the same behavior as shown in figure \ref{fig:entropyprojection}. Especially, when the post-selection is fixed at the future infinity, \ie $T_p=\frac{\pi}{2}$, the disconnected part reduces to a constant
		\begin{equation}
			S_A^{\mathrm{dis,dS}}= \frac{c}{3}\log\left(\frac{2}{\ep}\right) + \frac{\pi i}{12}c + 2S_{\mathrm{bdy}}\,,
		\end{equation}
	\end{enumerate}
	As a summary, we find that pseudo entropy derived from boundary 2d CFT agree with the holographic results for $S^{dis}_A$  \eqref{HPED1} and \eqref{HPED2} via the prescription \eqref{eq:bdyent}.

	\subsection{Holographic entanglement in higher dimensions}
	In this section, we study the holographic entanglement (pseudo) entropy in higher dimensional setups, which can be found from the area of extremal surfaces in the $(d+1)$-dim AdS space.
	The de Sitter sliced metric is given by
	\be
	ds^2=d\eta^2+\sinh^2\eta(-dt^2+\cosh^2t(d\phi^2+\sin^2\phi d\Omega_{d-2})).
	\ee
	
	\subsubsection*{Case 1: Schwinger-Keldysh without EOW}
	\begin{figure}[t]
		\centering
		\includegraphics[height=1.6in]{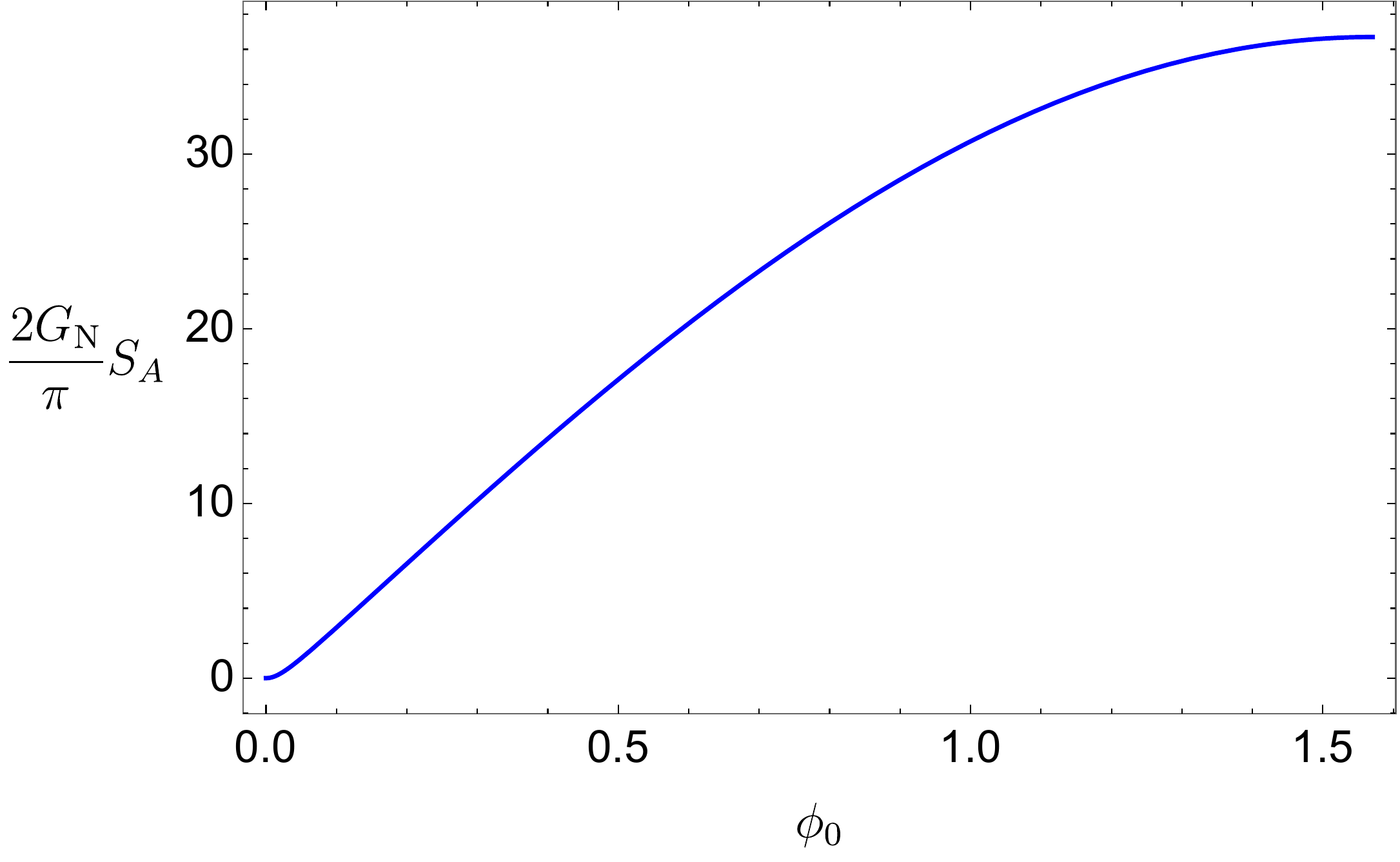}\quad 
		\includegraphics[height=1.6in]{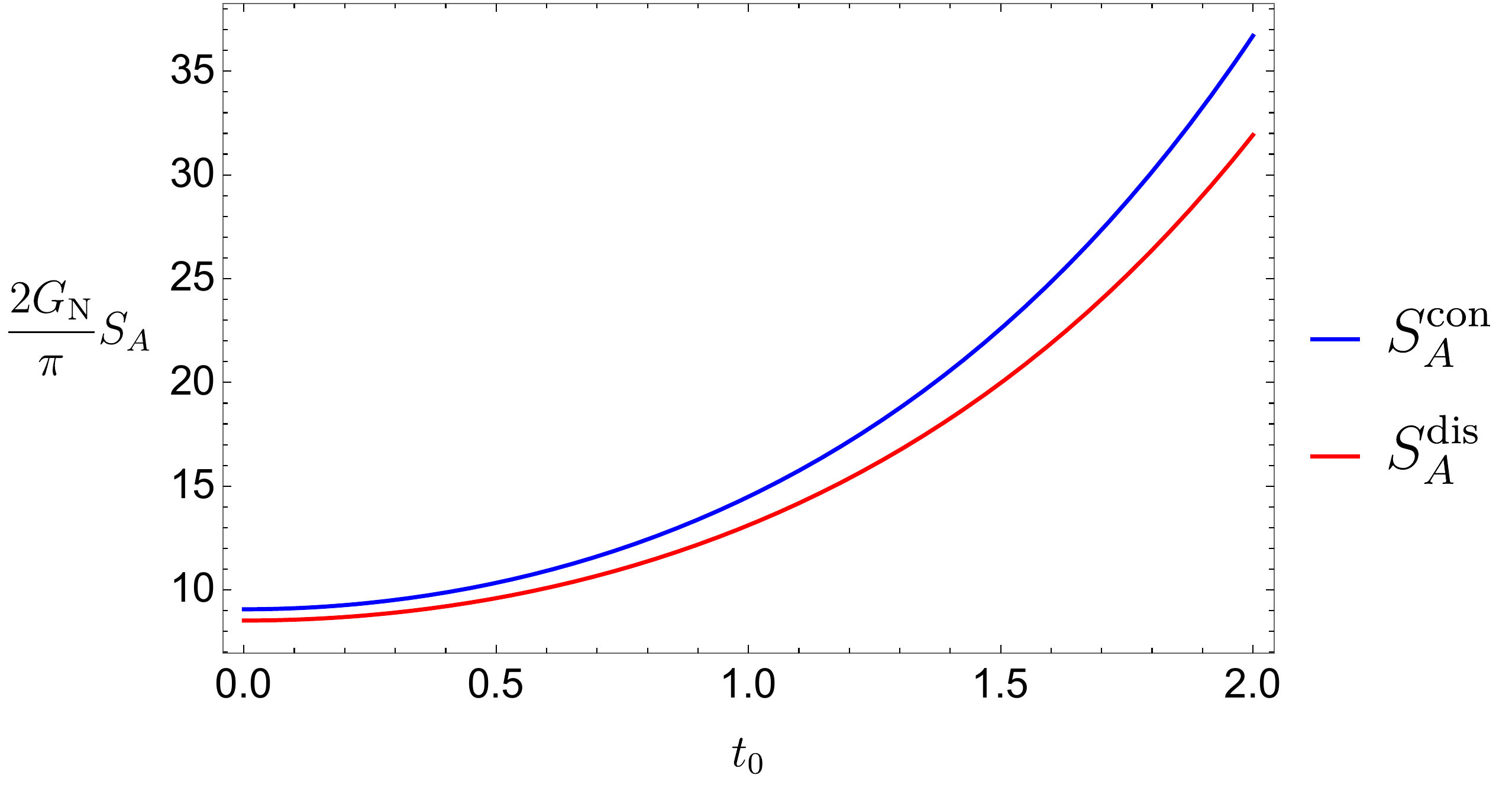}
		\caption{Holographic entanglement entropy $S^{\rm con}_A$ \eqref{eq:Scond3} and  $S^{\rm dis}_A$ \eqref{eq:Sdisd3} from Schwinger-Keldysh prescription without EOW brane. We choose $d=3, \phi_0=\frac{\pi}{2}, \eta_\infty=3$ and $\eta_*=1$for both plots.}
		\label{fig:Scon}
	\end{figure}

	First let us evaluate the connected entropy $S^{con}_A$. We consider the global AdS coordinates
	\be
	ds^2=d\rho^2-\cosh^2\rho d\tau^2+\sinh^2\rho (d\phi^2+\sin^2\phi d\Omega_{d-2}^2)
	\ee
	and take the subsystem $A$ as
	\be
	0\leq \phi\leq \phi_0,\quad
	\tau=\tau_0,\quad
	\rho=\rho_\infty.\label{bc}
	\ee
	In this coordinate, the extremal surface is labeled by $\phi=\phi(\rho)$. The holographic entanglement entropy is the area of this surface divided by $4G_N$
	\be
	S^{\rm con}_A=\frac{\mathrm{Vol}(\mathrm{S}^{d-2})}{4G_N}\int d\rho\sqrt{1+\sinh^2\rho \phi^{'2}}(\sinh\rho\sin\phi)^{d-2},
	\ee
	where we set $\phi'=\frac{\partial\phi}{\partial \rho}$. By taking the variation, we obtain the Euler-Lagrange equation
	\be
	\phi''-\frac{d-2}{\tan\phi\sinh^2\rho}+\frac{d\phi'}{\tanh\rho}-\frac{(d-2)\phi^{'2}}{\tan\phi}+(d-1)\sinh\rho\cosh\rho\phi^{'3}=0.
	\ee

	We find a suitable solution $\tanh\rho\cos\phi=C,$ where the $C$ is integration constant.
	We impose the boundary condition (\ref{bc}) and obtain the solution
	\be
	\tanh\rho\cos\phi=\tanh\rho_\infty\cos\phi_0.
	\ee
	Then, the holographic entanglement entropy reads
	\begin{equation}
		S^{\rm con}_A
		=\frac{\mathrm{Vol}(\mathrm{S}^{d-2})}{4G_N}\int_1^{\frac{\cosh\rho_\infty}{\cosh\rho_{\rm min}}}du(u^2-1)^{\frac{d-3}{2}}\,,
	\end{equation}
	where $\rho_{\rm min}$ is defined by $\tanh\rho_{\rm min}=\tanh\rho_\inf\cos\phi_0$. In particular, $S^{\rm con}_A$ for $d=3$ reduces to 
	\begin{equation}\label{eq:Scond3}
		\begin{split}
			S^{\rm con}_A=\frac{\pi}{2G_N}\(\frac{\cosh\rho_\infty}{\cosh \rho_{\rm min}}-1\)&= \frac{\pi}{2G_N} \( \sqrt{ \cosh^2 t_0 \sin^2 \phi_0 \sinh^2 \eta_\inf +1 } -1   \)  \\
			&\approx \frac{\pi}{2G_N} \( \frac{\cosh t_0 \sin \phi_0}{\epsilon} -1  \)   \,,\\
		\end{split}
	\end{equation}
	We plot this as a function of $t_0$ and $\phi_0$ in figure \ref{fig:Scon}.
	
	\subsubsection*{Case 2: Final projection at $t=\infty$}
	
	Next, let us evaluate the disconnected extremal surfaces stretching between the dS boundary and the dS EOW brane, which contributes to the holographic pseudo entropy. Particularly, we consider the extremal surface connecting the two points between $(\eta_\infty,t_0,\phi_0,\Omega_0)$ and $(\eta_*,t_*,\phi_*,\Omega_0)$ and then maximize the area by taking variation with respect to $t_*$. In the AdS$_3$ case, the extremal surface satisfies $t=$const and $\phi$=const. However, in higher-dimensional cases this is no more true. Suppose that $\phi_*=\frac{\pi}{2}$ along the extremal surface\footnote{This can be justified only in the $\phi=\frac{\pi}{2}$ case since we have a $\mathbf{Z}_2$ symmetry $\theta\rightarrow\frac{\pi}{2}-\phi$. In general the extremal surface does depend on $(\phi,t,\eta)$.}. Then, its area is given by
	\be
	S^{\rm dis}_A=\frac{\mathrm{Vol}(\mathrm{S}^{d-2})}{4G_N}\int d\eta \sqrt{1-\sinh^2\eta t^{'2}}(\sinh\eta\cosh t)^{d-2}.
	\ee
	%\begin{figure}[h]
	%    \begin{tabular}{cc}
		%  
		%      \begin{minipage}[t]{0.5\hsize}
			%        \centering
			%        \includegraphics[keepaspectratio, scale=0.3]{half dS infinite projection.pdf}
			%        \caption{dS branes in the global AdS in higher dimension. The boundary dS branes are %described by the blue curve (it is indeed a surface), in which we take a subregion A, and the %projection brane is represented by the green one. The orange curve denotes the extremal surface for %S$_{\rm dis}$. The projection time $t_p$ is given by the intersection of these curves, which is at %the infinity.}
			%       \label{fig:half dS infinite projection}
			%      \end{minipage} &
		%      \begin{minipage}[t]{0.5\hsize}
			%        \centering
			%        \includegraphics[keepaspectratio, scale=0.3]{half dS finite projection.pdf}
			%        \caption{The set up of the final projection at the finite time $t_p$. We shift the projection %brane so that it intersects with the boundary dS brane. The orange curve represents the extremal %surface, which gives the $S^{\mathrm{dis}}_A$.}
			%       \label{fig:half dS finite projection}
			%      \end{minipage} 
		%    \end{tabular}
	%  \end{figure}

The Euler-Lagrange equation reads
\be
\frac{(d-2)\tanh t}{\sinh^3\eta}+\frac{dt'}{\tanh\eta\sinh\eta}-\frac{(d-2)\tanh t t^{'2}}{\sinh\eta}-(d-1)\cosh\eta t^{'3}+\frac{t''}{\sinh\eta}=0.
\ee
A particular solution, which is the one at $\tau=$const. in the global AdS, reads
\be
\tanh\eta\sinh t=\tanh\eta_\infty\sinh t_0=\tanh\eta_*\sinh t_*.
\label{soldisf}
\ee
We expect more general solutions and we need to maximize the areas of such solutions with respect to the end points on the EOW brane. However, this is highly complicated and we will focus on the above simple solution, which gives the minimal value of S$_A^{\rm dis}$.

By substituting this into the functional we obtain
\begin{align}
	S_A^{\rm dis}&=\frac{\mathrm{Vol}(\mathrm{S}^{d-2})}{4G_N}\int d\eta \sqrt{1-\sinh^2\eta t^{'2}}(\sinh\eta\cosh t)^{d-2}\nn\\
	&=\frac{\mathrm{Vol}(\mathrm{S}^{d-2})}{4G_N}\int^{\s{1+\cosh^2 t_0\sinh^2\eta_\infty}}_{\s{1+\cosh^2 t_*\sinh^2\eta_*}}dv(v^2-1)^{\frac{d-3}{2}}.
\end{align}

Especially, in the $d=3$ case we find
\begin{equation}\label{eq:Sdisd3}
	S_A^{\rm dis}=\frac{\pi\sqrt{\tanh^2\eta_\infty\sinh^2 t_0+1}}{2G_N}(\cosh\eta_\infty-\cosh\eta_*)
	\simeq \frac{\pi}{2G_N}\cosh t_0\( \frac{1}{\epsilon} -\cosh\eta_\ast\)\,,
\end{equation}
The results are plotted in the  figure \ref{fig:Scon}. Unlike the AdS$_3$ case, both $S_A^{\rm con}$ and $S_A^{\rm dis}$ diverge exponentially as a function of $t_0$. Comparing two expressions in eqs.~\eqref{eq:Scond3} and \eqref{eq:Sdisd3}, we always have 
\begin{equation}
	S_A^{\rm dis} \le S_A^{\rm con} \,, 
\end{equation}
due to $\cosh X \ge 1$.

%In total in the  $d=3$ and $\phi=\frac{\pi}{2}$ case we obtain
%\begin{align}
%    S^{\rm con}_A&=\frac{\pi}{2G_N}(\cosh\rho_\infty-1)\nn\\
%    &\sim \frac{\pi}{4G_N}e^{\eta_\infty}\cosh t_0 ,\nn\\
%S^{\rm dis}_A&=\frac{\pi\sqrt{\tanh^2\eta_\infty\sinh^2 t_0+1}}{2G_N}(\cosh\eta_\infty-%\cosh\eta_*)\nn\\
%&\sim\frac{\pi}{2G_N}\cosh t_0(\cosh\eta_\infty-\cosh\eta_*).
%\end{align}
%The results are plotted in the  figure \ref{fill}. 

\begin{comment}
	************************************************************
	
	{\bf Tadashi's Comment:}
	
	For the connected extremal surface, Iin $d=3$ we simply have 
	\ba
	S^{con}_A=\frac{\pi}{2G_N}(u_\infty-1)=\frac{\pi}{2G_N}\left[
	\frac{1}{2}\sin\theta_0\cosh t\cdot e^{\eta_\infty}-1\right],
	\ea
	where $u_\infty$ is given by $u_\infty=\sin\theta\cosh\rho_\infty\simeq \frac{1}{2}\sin\theta e^{\rho_\infty}$.
	Here $(t,\eta_\infty)$ is the boundary de Sitter coordinate where $t$ is the time we measure the value of EE and $\eta_\infty$ is the UV cut off of the CFT on de Sitter. Note that the relation between the UV cut off $\rho_\infty$ in the global AdS is given by $e^{\rho_\infty}\simeq \cosh t\cdot e^{\eta_\infty}.$
	
	*************************************************************
\end{comment}

\subsubsection*{More on Case 2: Final projection at $t=t_{\mt P}$}
\begin{figure}[t]
	\centering
	\includegraphics[width=5in]{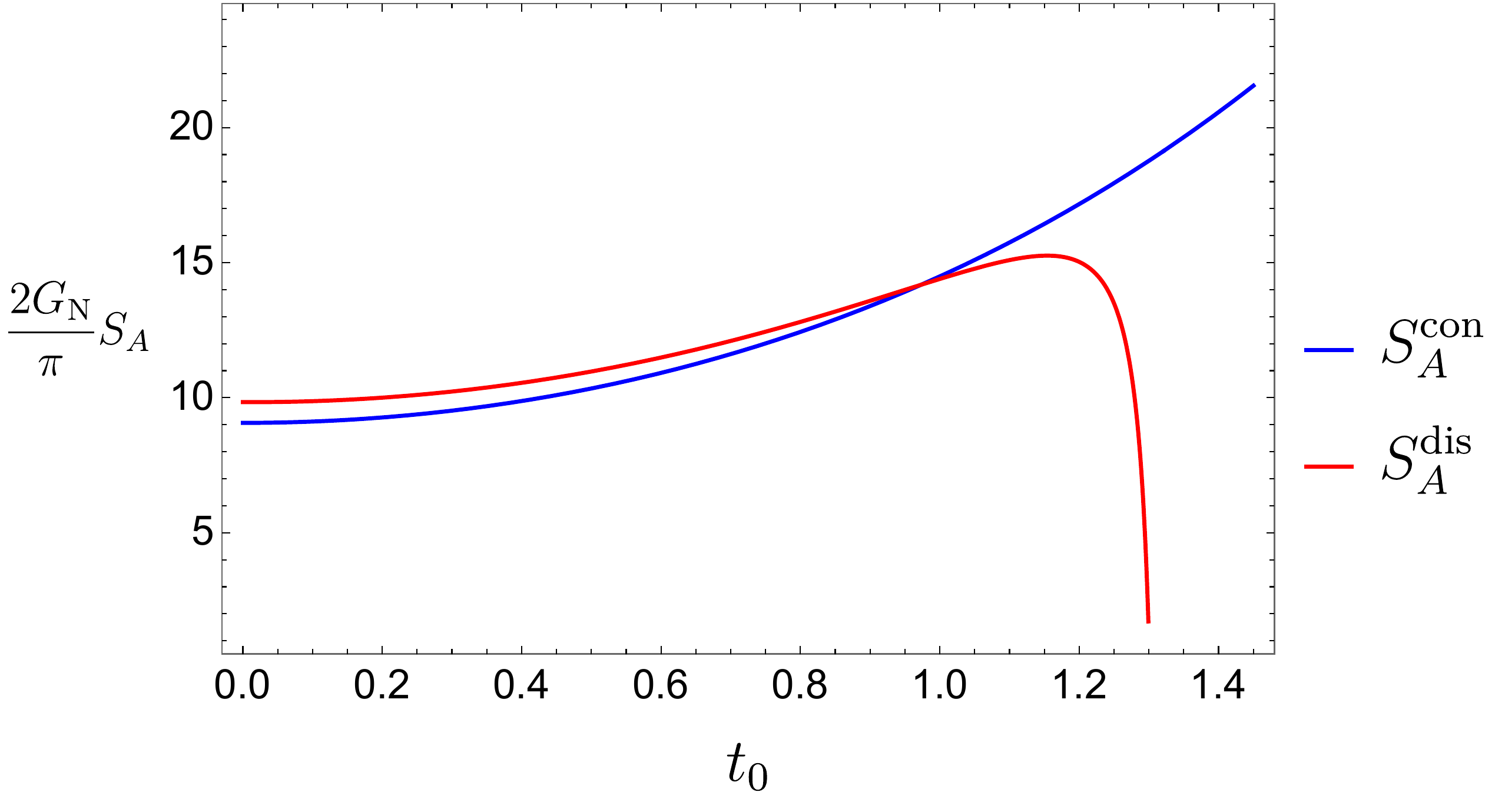}
	\caption{The time evolution of $S^{\rm con}_A$ \eqref{eq:Scond3} and $S^{\rm dis}_A$ \eqref{eq:SdistP}. We take $\eta_\infty=3$, $\phi_0=\frac{\pi}{2}$ , $\tau_{\rm shift}=\frac{\pi}{6}$ and  $\cosh\eta_c=0.2$. At early times, the connected part is dominated. After the phase transition, the disconnected part is dominated and approaches zero at the projection time $t_0=t_{\mt P}$ , where the two branes intersect.}
	\label{fig:SconSdistP}
\end{figure}

Next, we consider the final projection at finite time. For this, we shift the projection brane along the global time as in the section \ref{24}. If we shift the brane backward in $\tau_{\rm shift}$, the projection brane is defined by 
$\cosh\rho\cos(\tau+\tau_{\rm shift})=\cosh\eta_c,$
where $\eta_c$ is related with the tension of the dS brane. 
The special solution to the extremal surface condition is already obtained in the previous section as (\ref{soldisf}) for $\phi_0=\frac{\pi}{2}$. Therefore the intersection of the extremal surface and the projection brane defines the end point of the surface
$\tanh\eta_*\sinh t_*=\tanh\eta_\infty\sinh t_0$ and 
$\cosh\rho_*\cos(\tau_*+\tau_{\rm shift})=\cosh\eta_c.$

After short algebra we obtain 
\be
\cosh\eta_*=\frac{\cosh\eta_c}{\cos\tau_{\rm shift}-\tanh\eta_\infty\sinh t_0\sin\tau_{\rm shift}},
\ee
which gives for $d=3$
\begin{equation}\label{eq:SdistP}
	S^{\rm dis}_A=\frac{\pi\sqrt{\tanh^2\eta_\infty\sinh^2 t_0+1}}{2G_N}\left(\cosh\eta_\infty-\frac{\cosh\eta_c}{\cos\tau_{\rm shift}-\tanh\eta_\infty\sinh t_0\sin\tau_{\rm shift}}\right)\,.
\end{equation}
The figure \ref{fig:SconSdistP} shows the plots. As in the AdS$_3$ case, the disconnected entropy reaches zero at the projection time $t_{\mt P}$.

%%%%%%%%%%%%%%%%%%%%%%%%%%%%%%%%%%%%%%%%%%%%%%%%%%%%%%%%
%%%%%%%%%%%%%%%%%%%%%%%%%%%%%%%%%%%%%%%%%%%%%%%%%%%%%%%%
\section{Holography for a half dS without EOW brane (Case 1)}
\label{sec:HEE}
%%%%%%%%%%%%%%%%%%%%%%%%%%%%%%%%%%%%%%%%%%%%%%%%%%%%%%%%
%%%%%%%%%%%%%%%%%%%%%%%%%%%%%%%%%%%%%%%%%%%%%%%%%%%%%%%%

Now we would like to move on to the our main target: holography with a positive cosmological constant. In order to interpret the holography in de Sitter space as in the standard framework where the bulk gravity is dual to a quantum system on its time-like boundary, we focus on a half of de Sitter space defined by restricting the global dS$_{d+1}$ given by the metric (\ref{dsg}) and (\ref{sth}) to the region 
\ba
0\leq \theta\leq \theta_0,\label{rangeth}
\ea
as depicted in figure \ref{fig:dSHolT} and figure \ref{fig:dSHol}. The standard idea of holographic principle predicts that the bulk gravity on the half dS$_{d+1}$
(\ref{rangeth}) is dual to a certain quantum system on its boundary, namely, dS$_{d}$ at $\theta=\theta_0$, given by the metric (\ref{dSd}). We assume the range $0\leq \theta_0\leq \frac{\pi}{2}$. If we choose $\theta_0=\frac{\pi}{2}$ in particular, it is exactly a half of the original dS.

Below we will study how the holography for a half dS looks like by analyzing the holographic entanglement entropy. For simplicity we will mainly focus on the three dimensional space i.e. dS$_3$. In this case, the holographic entanglement entropy $S_A$, defined for an interval $A$, can be computed from the geodesic length which connects two end points of $A$. This consideration 
also suggests the range $0\leq \theta_0\leq \frac{\pi}{2}$. This is because the geodesics which compute the holographic entanglement entropy are not included in the half dS if $\theta_0\geq \frac{\pi}{2}$, even if the subsystem is very small.

\subsection{Geodesic Length in dS\texorpdfstring{$_3$}{Lg}}
\begin{figure}[t]
	\centering
	\includegraphics[width=3in]{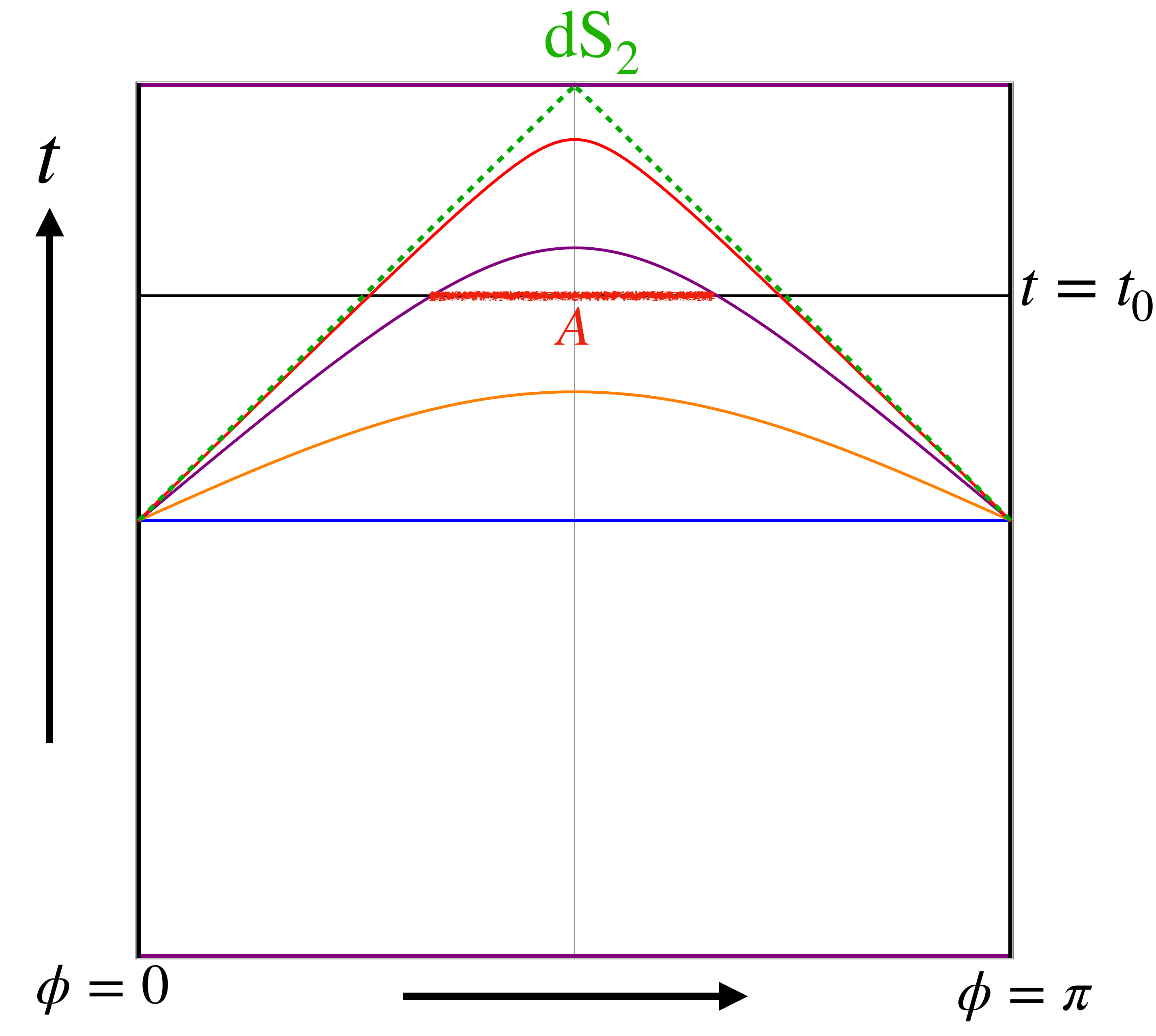}
	\caption{The Penrose diagram of dS$_2$ boundary. The colorful curves denote various spacelike geodesics on dS$_2$ where the null geodesic is presented by the green dashed curves.}
	\label{fig:horizon}
\end{figure}

To prepare for our analysis of holographic entanglement entropy $S_A$, let us see the behavior of the geodesics in a dS$_3$ whose metric reads
\ba
ds^2=-dt^2+\cosh^2 t(d\theta^2+\sin^2\theta d\phi^2).\label{dstheq}
\ea
If we choose two points $P_1$ and $P_2$ on dS$_3$: 
\begin{equation}
	P_1=(t_1,\theta_1,\phi_1)\,,\qquad   P_2=(t_2,\theta_2,\phi_2)\, ,
\end{equation}
then the geodesic distance between 
$P_1$ and $P_2$, denoted by $D_{12}$, can be found as 
\begin{equation}
	D_{12}=\left(\cos\theta_1\cos\theta_2+\sin\theta_1\sin\theta_2\cos(\phi_1-\phi_2)\right)\cosh t_1\cosh t_2-\sinh t_1\sinh t_2\,.
\end{equation}
Especially for the geodesic which connected two points $\phi=\phi_1$ and $\phi=\phi_2$ at the boundary $\theta=\theta_0$ and at the same time $t=t_0$, we have:
\ba
\cos D_{12}=\left(\cos^2\theta_0+\sin^2\theta_0\cos(\phi_1-\phi_2)\right)\cosh^2 t_0-\sinh^2 t_0.
\ea

\begin{figure}
	\centering
	\includegraphics[width=5cm]{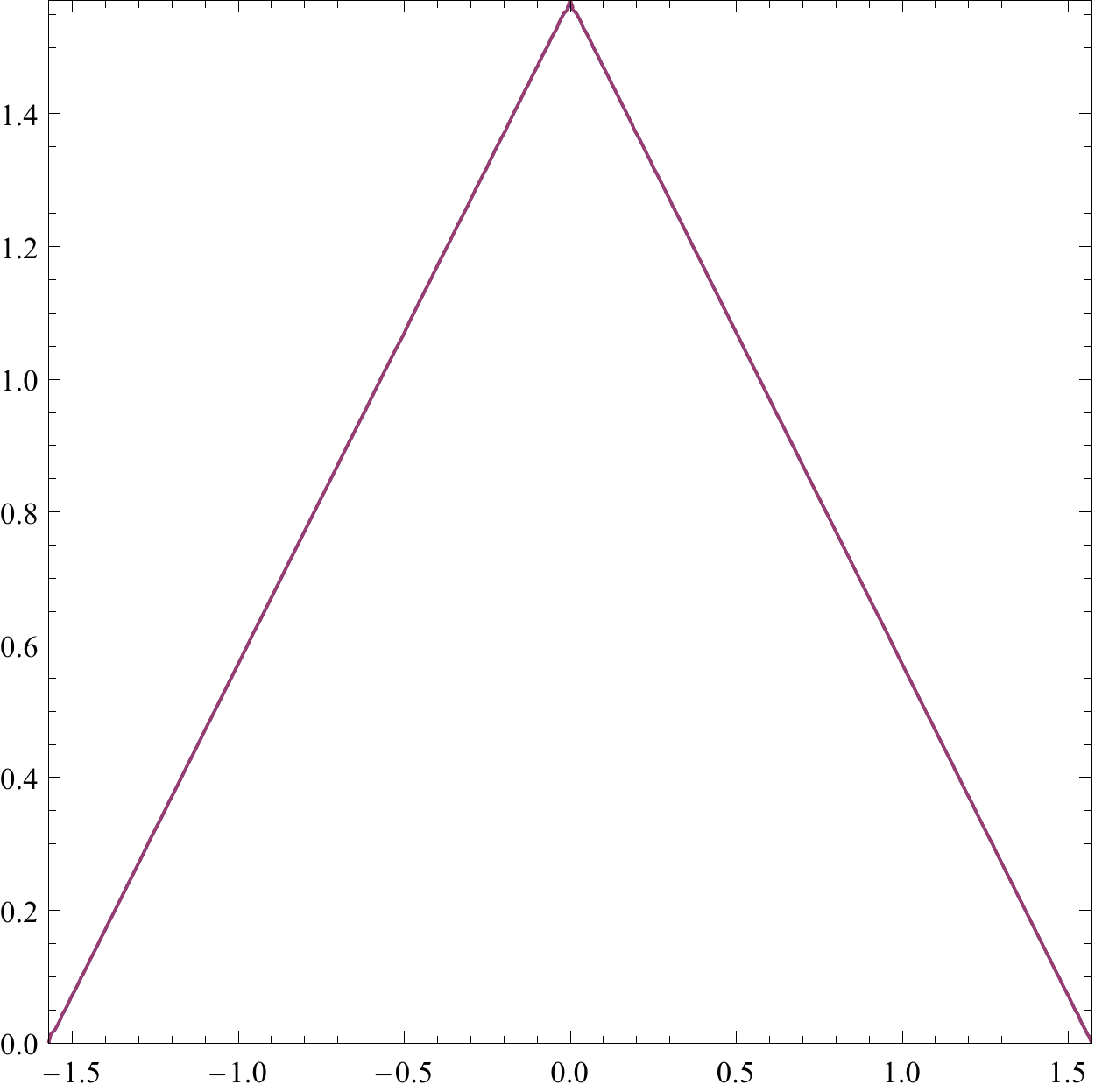}
	\includegraphics[width=5cm]{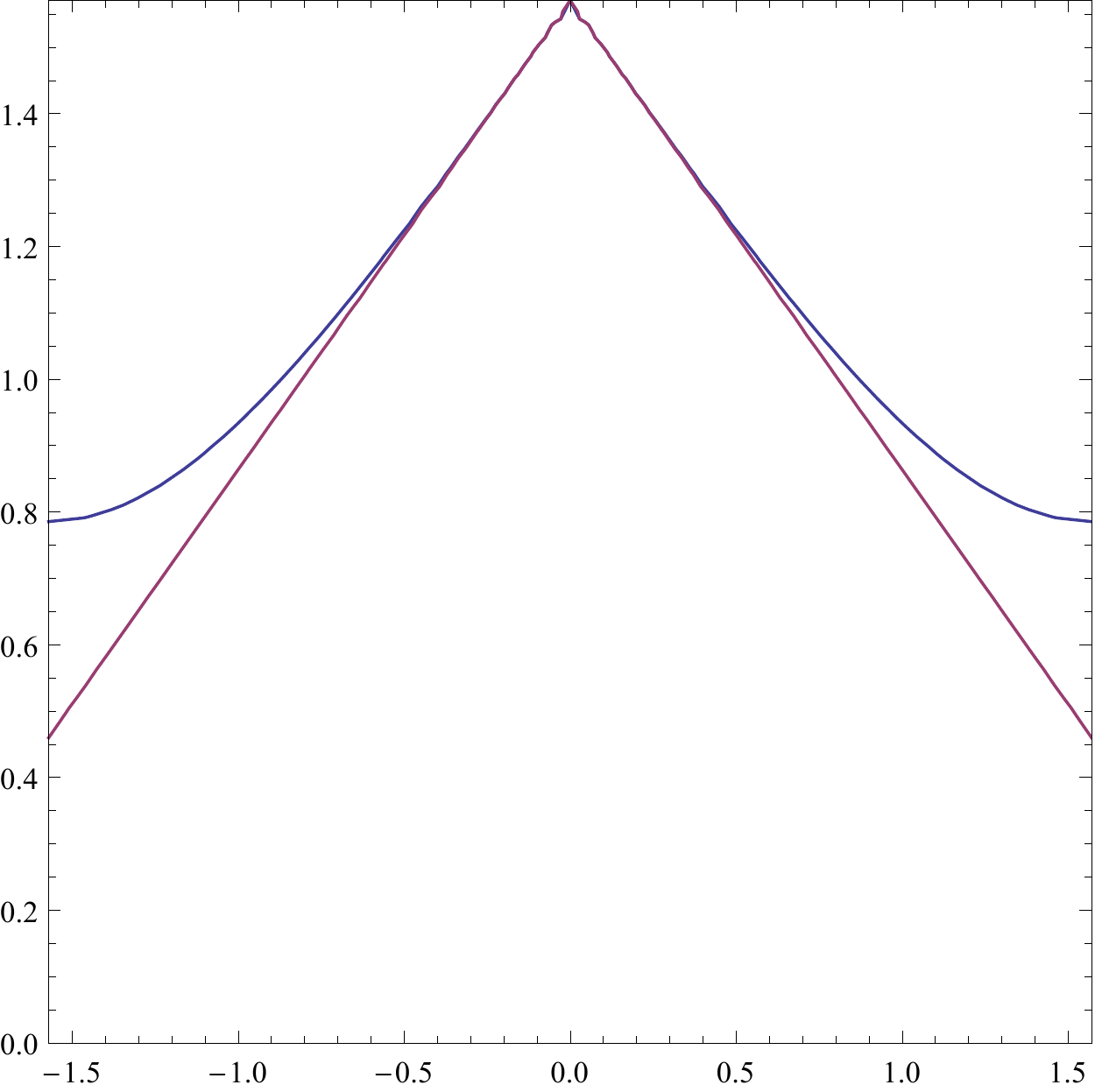}
	\includegraphics[width=5cm]{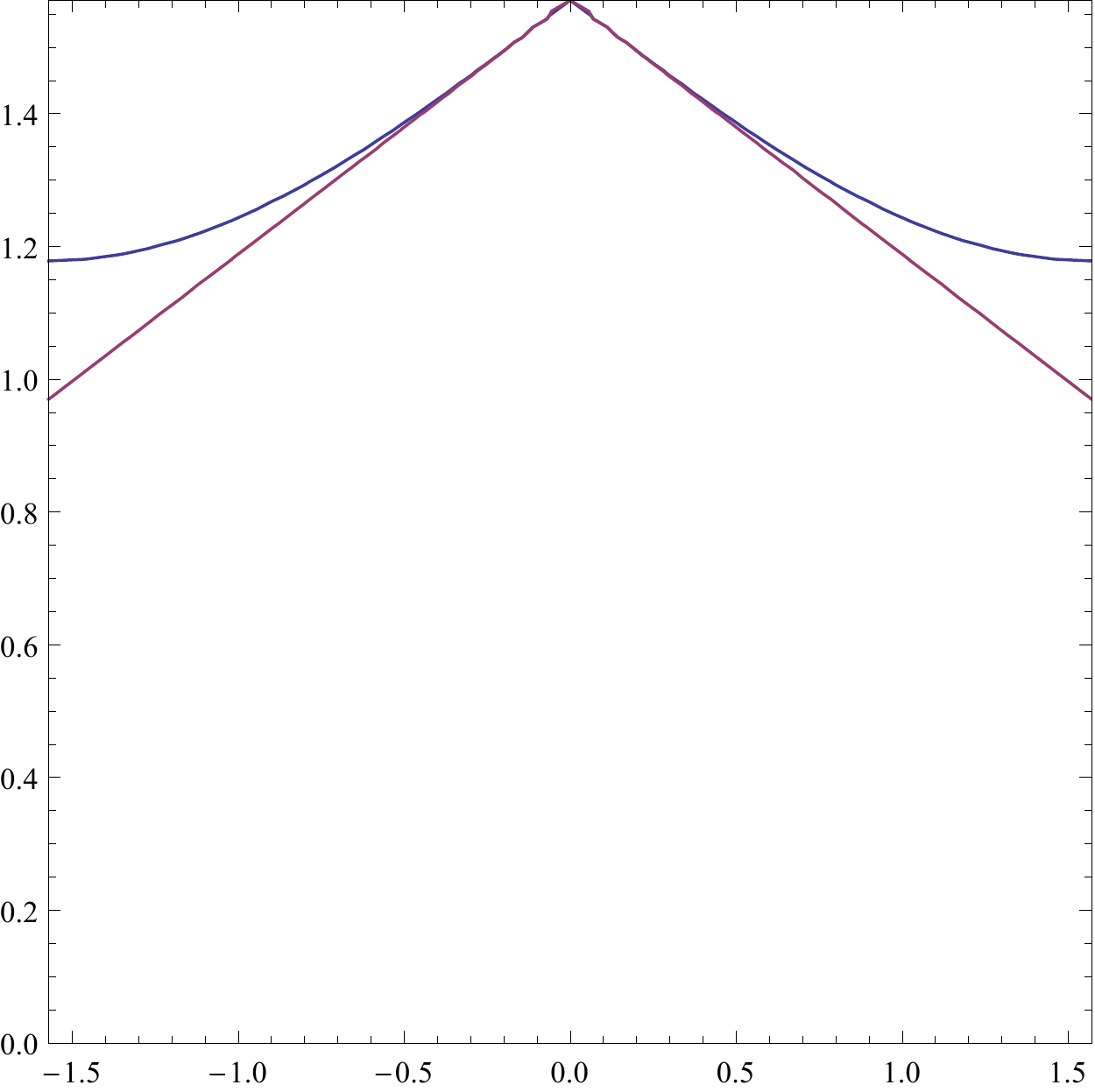}
	\caption{Plots of the boundaries beyond which the space-like geodesic does not exist for $\theta=\frac{\pi}{2}$ (left), $\theta_0=\frac{\pi}{4}$ (middle) and $\theta_0=\frac{\pi}{8}$ (right).
		The hoirzontal and vertical coordinates are $\phi$ and $T$ in the range $-\pi/2\leq \phi\leq \pi/2$ and $0\leq T\leq \pi/2$, respectively. We consider the spacelike geodesic which connects two points with $\phi_1=-\phi_2$ for each time. The blue curves are the boundaries for the dS$_3$ geodesic length $D_{12}$, while the red ones are those for the dS$_2$ metric (\ref{dSd}). Both of which coincide only when $\theta=\frac{\pi}{2}$.}
	\label{fig:lightconetd}
\end{figure}

In order for a space-like geodesic to be present between the two points, the following condition should be satisfied:
\ba
\cos^2\theta_0+\sin^2\theta_0\cos(\phi_1-\phi_2)\geq 1-\frac{2}{\cosh^2 t_0}.\label{boundsp}
\ea
The boundary of this boundary looks like a past light-cone of the future infinity as depicted in figure  \ref{fig:lightconetd}.
We write the value of $\Delta\phi=\phi_1-\phi_2$ which saturates this bound as $\Delta\phi_{\mathrm{max}}$. The space-like geodesic exists only when $\Delta\phi\leq \Delta\phi_{\mathrm{max}}$.
For explicit construction of such a geodesic refer to the appendix \ref{sec:APgeo}.

\begin{figure}
	\centering
	\includegraphics[height=1.65in]{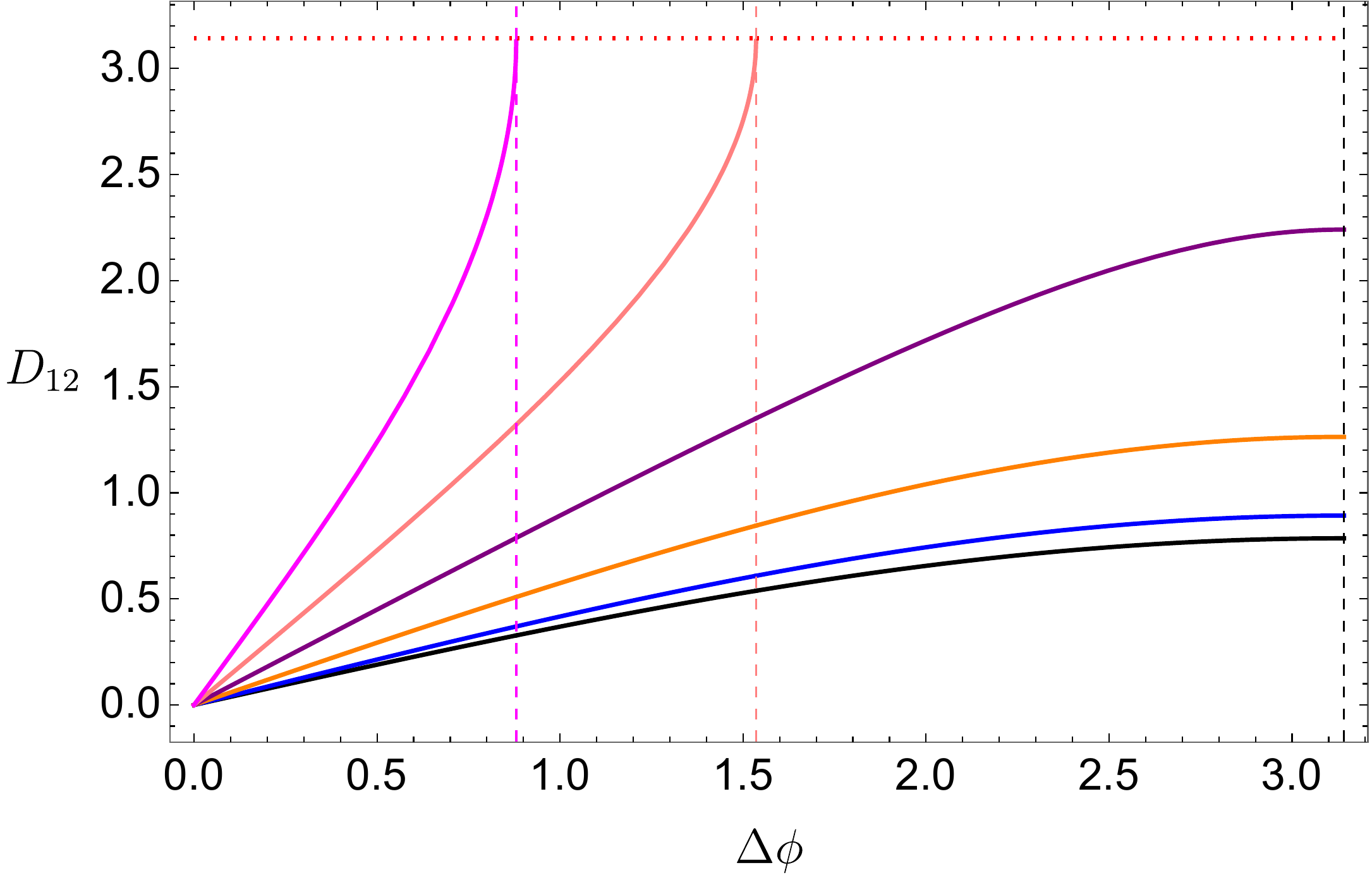}
	\includegraphics[height=1.65in]{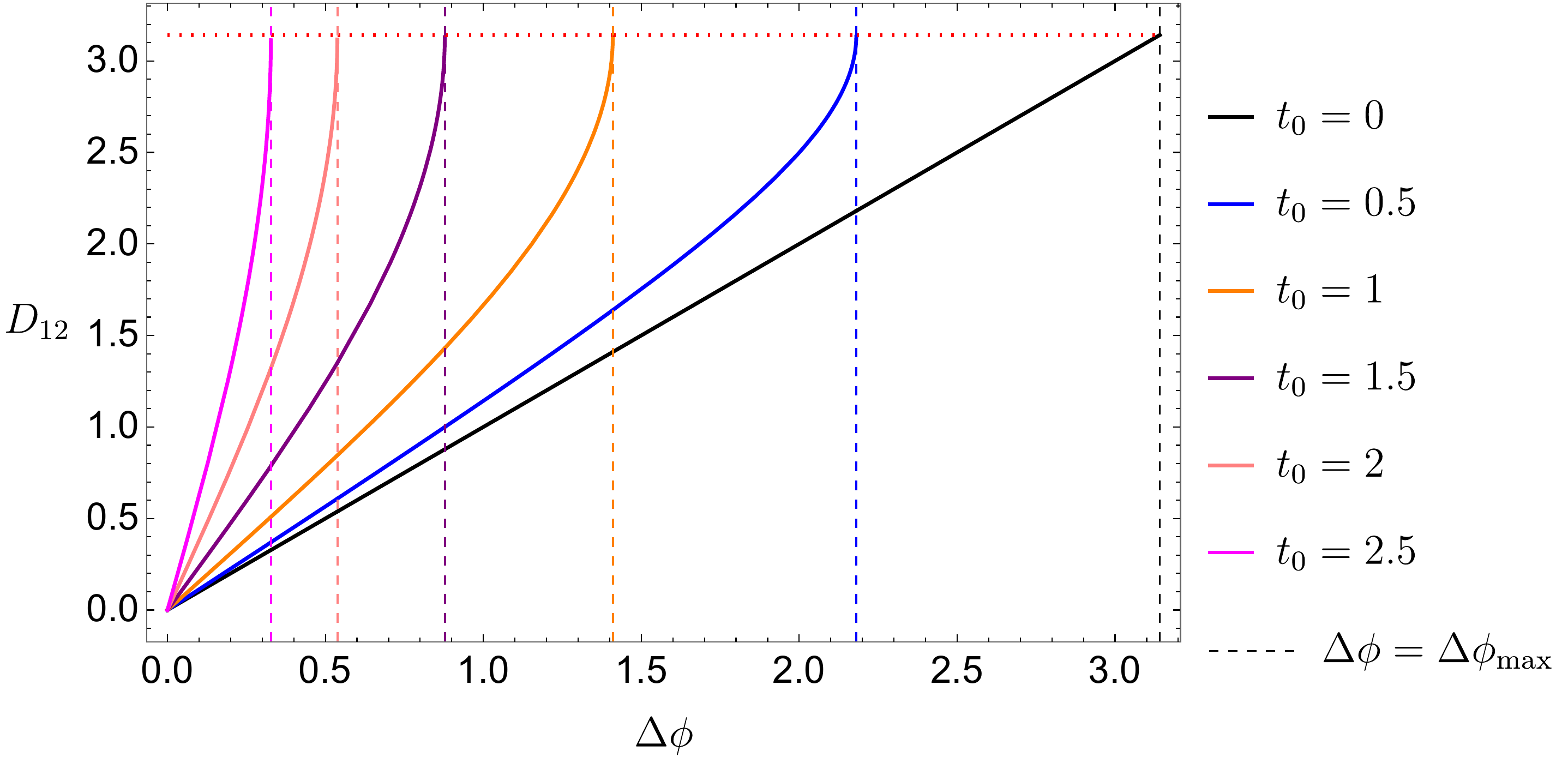}
	\caption{The behaviors of the length $D_{12}$ of space-like geodesics. The left panel shows $D_{12}$ as a function of $|\phi_1-\phi_2|$ for $\theta_0=\frac{\pi}{8}$ at various fixed times $t_0$. Notice that at the latter two times, the space-like geodesic ceases to excite for a large values of 
		$\Delta\phi=|\phi_1-\phi_2|$. The right panel shows  $D_{12}$ as a function of $|\phi_1-\phi_2|$ for the maximal half dS$_3$\ie $\theta_0=\frac{\pi}{2}$ at fixed times. The horizontal red dotted line describes $D_{12}=\pi$ and vertical dashed lines are referred as the critical size $\Delta \phi_{\mathrm{max}}$.}
	\label{fig:dSEEplot}
\end{figure}

It is clear that for two different points $\phi_1\neq \phi_2$, this inequality gets violated at enough later time. On the other hand, if $t_0$ and $\theta_0$ satisfy
\ba
\cos(2\theta_0)\geq 1-\frac{2}{\cosh^2 t_0}, \label{timeb}
\ea
then for any values of $\phi_1$ and $\phi_2$, the space-like geodesic which connects the two points does exist.
The geodesic length $D_{12}$ as a function of $\Delta\phi$ 
at $\theta_0=\frac{\pi}{8}$ and $\theta_0=\frac{\pi}{2}$
is depicted in figure \ref{fig:dSEEplot}.

When the bound (\ref{boundsp}) is violated, the geodesic length gets complex valued such that its real part is $\pi$:
\ba
D_{12}=\pi +i\arccosh\left[\sinh^2 t-\left(\cos^2\theta_0+\sin^2\theta_0\cos(\phi_1-\phi_2)\right)\cosh^2 t_0\right]. \label{geodetim}
\ea
The imaginary contribution comes from the time-like geodesic and the final real part $\pi$ does from the geodesic in an Euclidean space \ie a semi-sphere 
(see \cite{Hikida:2022ltr}). 

Consider the maximal case $\theta_0=\frac{\pi}{2}$. In this case, due to the Z$_2$ symmetry $\theta\to\pi-\theta$, the geodesics which connect two points on the boundary $\theta=\theta_0=\frac{\pi}{2}$ are all within the boundary dS$_2$, as shown in figure \ref{fig:horizon}. At $t_0=0$, there is always a geodesic which connects two points on the boundary and the geodesic length is simply found to be
\ba
D_{12}=\Delta\phi.
\ea
However for $t_0\neq 0$, a space-like geodesic does not exist for $\Delta\phi=|\phi_1-\phi_2|>\Delta\phi_{\mathrm{max}}$,
where the bound is explicitly given by
\begin{equation}\label{boundnu}
	\Delta\phi_{\mathrm{max}}=2 \( \frac{\pi}{2}- \arccos\( \frac{1}{\cosh t_0} \) \) =\pi-2\arctan\left[\sinh t_0\right]\,.
\end{equation}
This bound coincides with the boundary of the past light cone of a point at $t=\infty$ and is identical to the condition 
(\ref{tmaxt}) found in our previous analysis of AdS/CFT, shown in the left panel of figure \ref{fig:lightconetd}. These behaviors of geodesic length are plotted in the right panel of figure \ref{fig:dSEEplot}.

\begin{figure}
	\centering
	\includegraphics[width=2.5in]{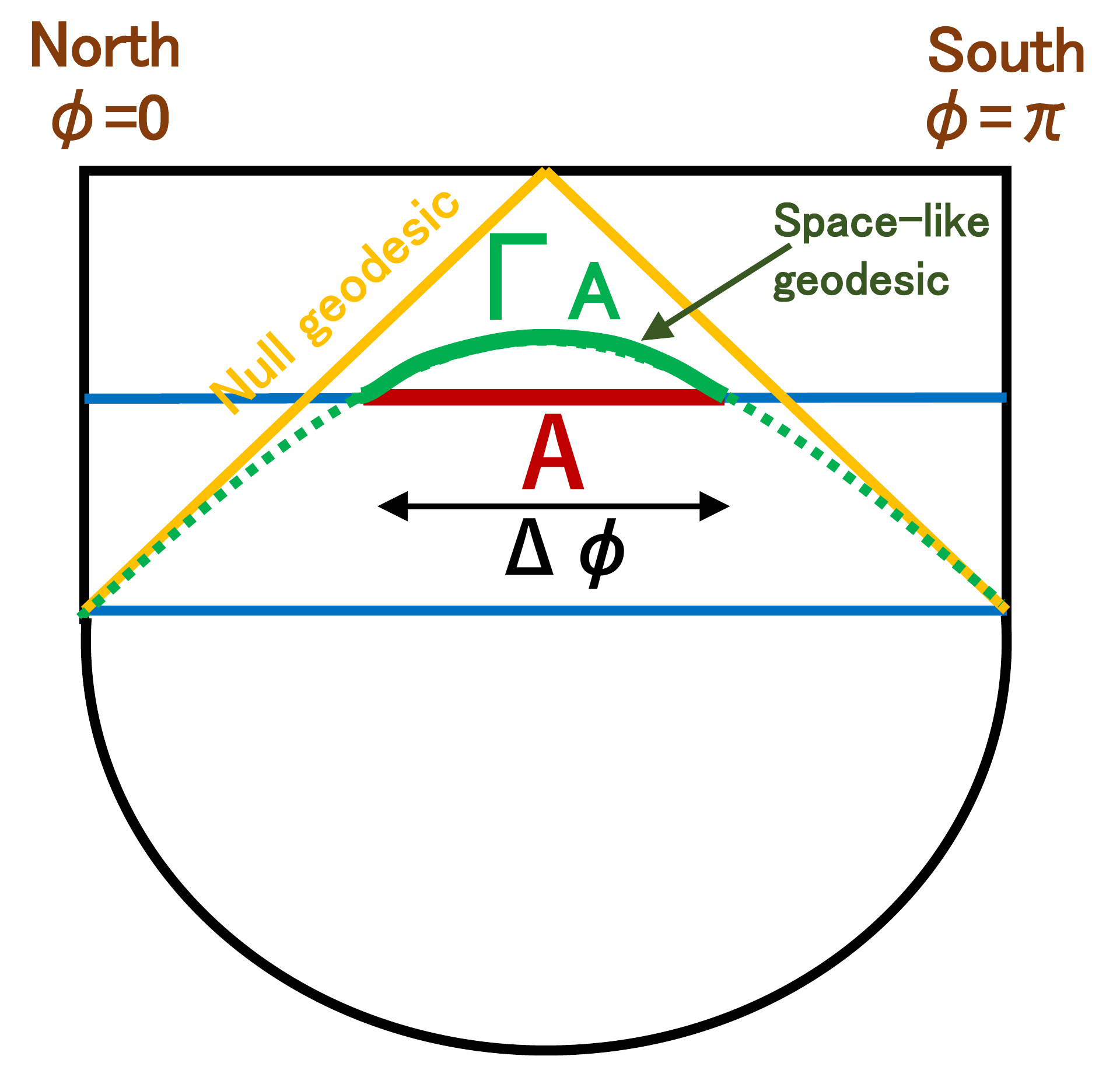}
	\includegraphics[width=2.5in]{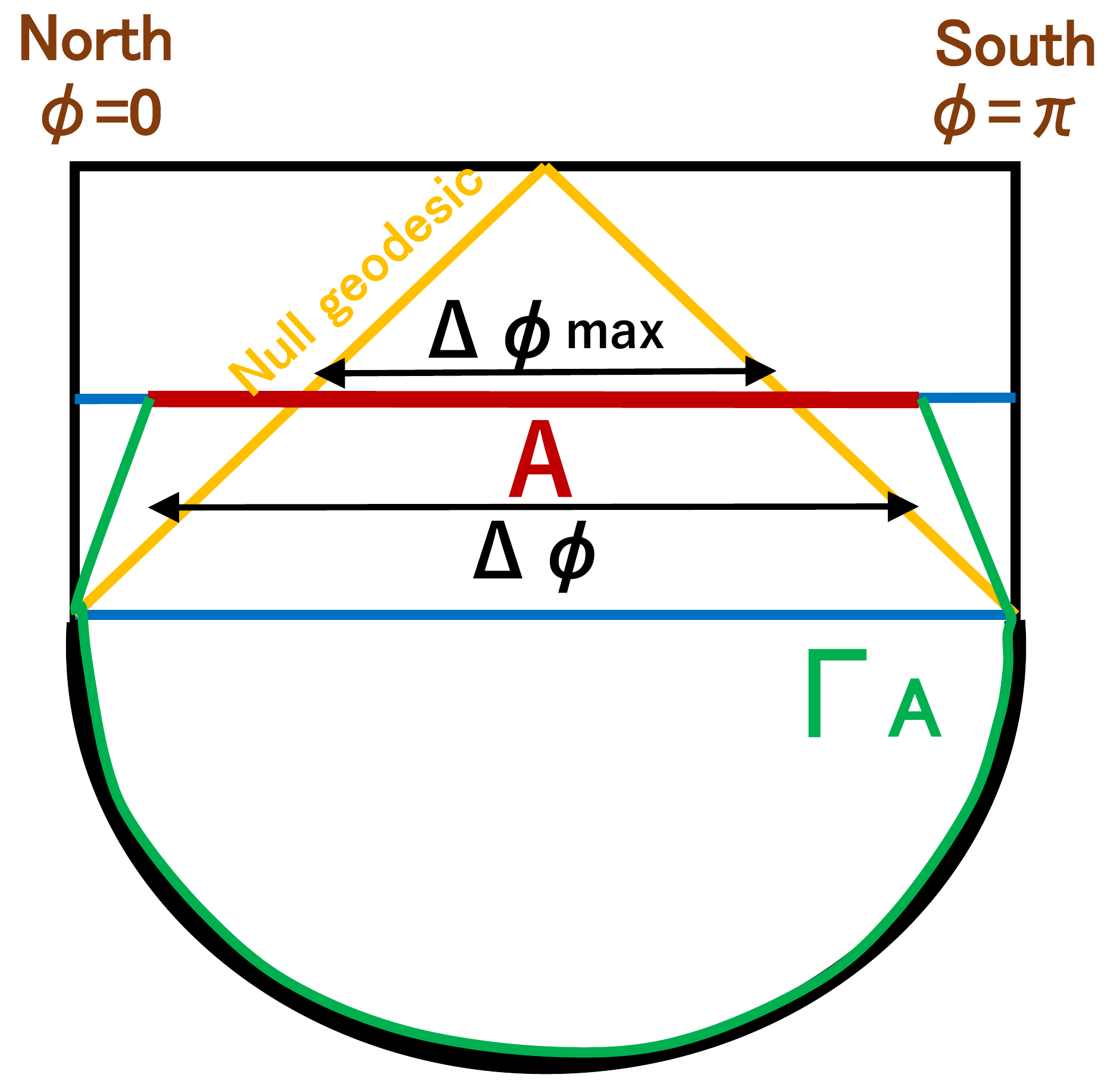}
	\caption{Sketches of connected geodesics $\Gamma_A$ in the boundary dS$_2$ for $\theta=\frac{\pi}{2}$. In the left panel, we present the space-like one with $\Delta\phi\leq \Delta\phi_{\mathrm{max}}$. In the right panel, we plotted what geodesic length computes for $\Delta\phi>\Delta\phi_{\mathrm{max}}$. It is a union of time-like and space-like geodesic.}
	\label{fig:setupp}
\end{figure}

The profile of the geodesic for $\Delta\phi\leq \Delta\phi_{\mathrm{max}}$ is sketched in 
the left panel of figure \ref{fig:setupp}. When $\Delta\phi>\Delta\phi_{\mathrm{max}}$, the geodesic length gets complex valued and can be interpreted as the union of time-like geodesic and space-like one as depicted in the right panel of figure \ref{fig:setupp}. Note that this transition at $\Delta\phi=\Delta\phi_{\mathrm{max}}$ is peculiar to our dS setup and can not be seen in the AdS setups.

\subsection{Holographic Entanglement Entropy and Violation of Subadditivity}

Now let us consider the calculation of  the holographic entanglement entropy (\ref{Area}) in Case 1 (figure \ref{fig:SCK}), namely Schwinger-Keldysh dS geometry, where there is no EOW brane. In this case, $S_A$ for an interval $A$, can be computed from the connected geodesic as
\ba
S_A=\frac{D_{12}}{4G_N}, \label{heeds}
\ea
whose length was already studied in the previous subsection.

Consider the holographic entanglement entropy at a fixed time $t=t_0$ for an interval $A$ defined by $\phi_1\leq\phi\leq \phi_2$. For a generic choice of the boundary $\theta_0$, the holographic entanglement entropy computed as the geodesic length, takes real and positive values only for $\Delta\phi\leq \Delta\phi_{\mathrm{max}}$, as depicted in figure \ref{fig:dSEEplot}. For the maximal value 
$\Delta\phi=\Delta\phi_{\mathrm{max}}$ we find $D_{12}=\pi$, which is a half of the length of the de Sitter horizon and thus $S_A$ becomes a half of de Sitter entropy $S_A=\frac{1}{2}S_{\rm dS}$. 

In terms of the time evolution, for an earlier time $0\leq t\leq t_{\mathrm{max}}$, $S_A$ is well-defined for any value of 
$\Delta\phi$, where $t_{\mathrm{max}}$ is the time when (\ref{timeb}) is saturated i.e. 
\begin{equation}
	t_{\mathrm{max}} = \arccosh \( \frac{1}{\sin \theta_0}  \) = \log \( \cot \(\frac{\theta_0}{2} \) \) \,.
\end{equation}

However, for $t>t_{\mathrm{max}}$, the behavior of holographic entanglement looks confusing partly because it takes a complex value for $\Delta\phi>\Delta\phi_{\mathrm{max}}$ and also because it is a convex function of $\Delta\phi$ even for $\Delta\phi\leq\Delta\phi_{\mathrm{max}}$
as can be seen from figure \ref{fig:dSEEplot}. The latter fact shows the violation of (strong) subadditivity. Indeed, the second derivative of the geodesic length 
\begin{equation}
	\frac{d^2 D_{12}}{d \Delta \phi d \Delta \phi} =\frac{\sqrt{2}\cosh t \sin \theta_0 \sin \(\frac{\theta_0}{2}\) \( \cosh^2 t \sin^2 \theta_0 -1 \) }{\(1 + \cos D_{12}
		\)^{3/2}} ,
\end{equation}
becomes positive when $t>t_{\mathrm{max}}$. In terms of the conformal time $T$ ($\cosh t \equiv  \frac{1}{\cos T}$), the critical time is rewritten as 
\begin{equation}
	T_{\mathrm{crt}} = \frac{\pi}{2} - \theta_0 \,, 
\end{equation}
which is nothing but the intersection between the boundary $\theta=\theta_0$ and the cosmological horizon located at $T=\frac{\pi}{2} \pm \theta$. 

\begin{figure}
	\centering
	\includegraphics[width=2.5in]{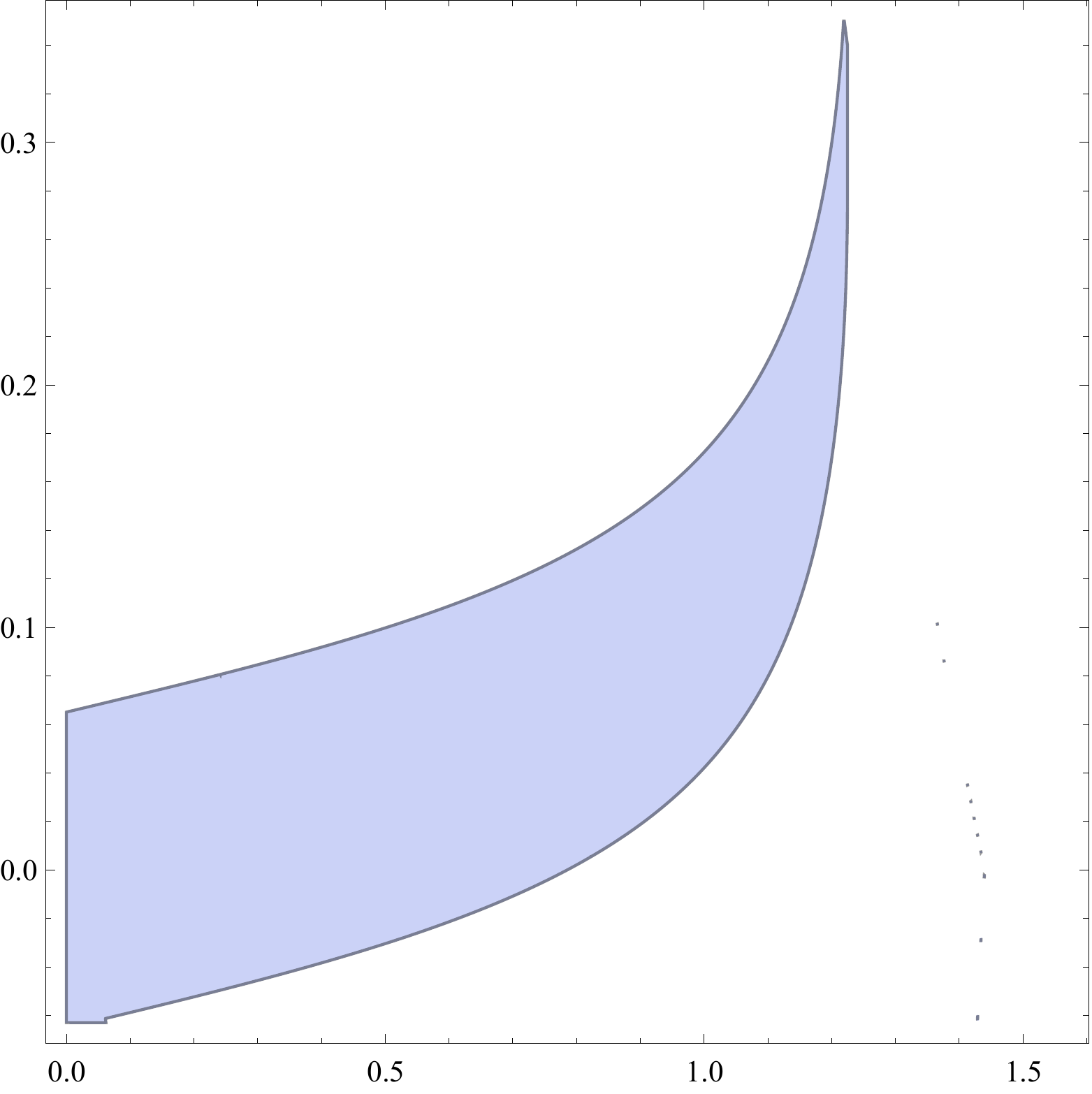}
	\includegraphics[width=2.5in]{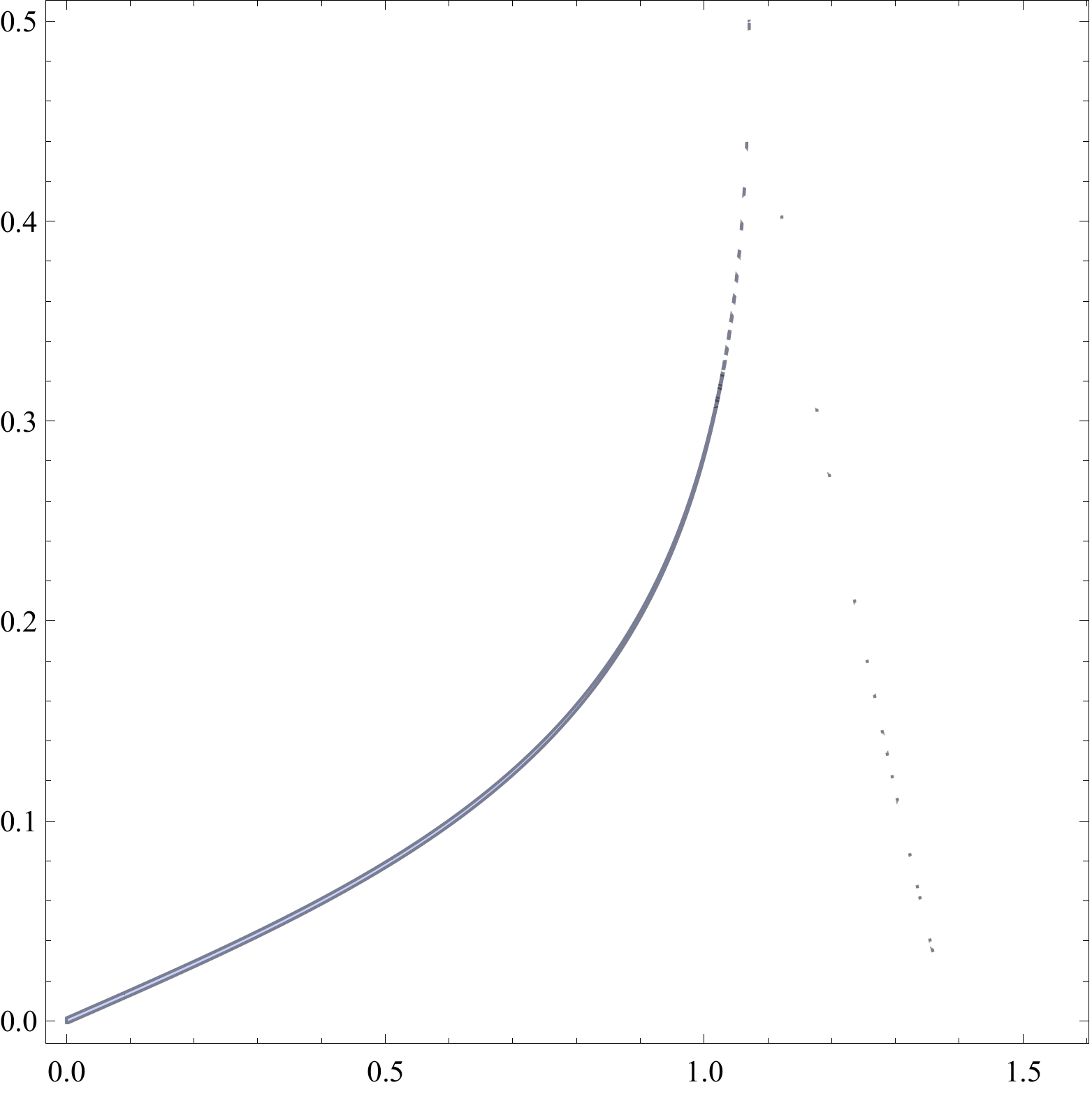}
	\caption{We plotted the behavior of subadditivity violation for $\theta_0=\frac{\pi}{4}$ (left) and $\theta_0=\frac{\pi}{2}-0.01$ (right). We consider the subsystem $A$ and $B$ such that $A$ connects the two points $(\phi,T)=(0,T_0)$  and $(1/2,T_0+\Delta T)$ and $B$ does the two points $(1/2,T_0+\Delta T)$ and $(1,T_0)$.
		$T$ is introduced by $\cos T=\frac{1}{\cosh t}$.
		The horizontal and vertical coordinates describe $T_0$ and $\Delta T$. We colored the region where the subadditivity holds i.e.  $S_A+S_B-S_{AB}\geq 0$.}
	\label{fig:lightregion}
\end{figure}

To see why the convex entropy function violates the subadditivity, first note that $S_A$ is a function of the length $y=L(A)$ of the interval $A$, owing to the translational invariance. It is obvious from the convex nature $\de^2_yS_A(y)>0$ that we have $S_A(2y)>2S_A(y)$, which violates the subadditivity $S_A+S_B\geq S_{AB}$. Of course, this shows that the strong subadditivity, which is expressed as $S_{AB}+S_{BC}\geq S_{ABC}+S_B$, is broken. Refer to figure \ref{fig:lightregion} for plots of the regions where the subadditivity is satisfied, whose further interpretation will be given later. This behavior is a complete contrast to that of AdS/CFT, where the strong subadditivity is always satisfied \cite{Headrick:2007km,Wall:2012uf}.

\subsection{Static Time Slices}

To study the nature of subadditvity violation, let us first focus on the maximal case $\theta_0=\frac{\pi}{2}$. In this case, it is useful to consider the static coordinate $(\chi,s)$ of dS$_2$  introduced as follows
\ba
\cos\chi=\cos\phi\cosh t,\ \ \ \ \ \cosh s=\frac{\sin\phi}{\s{\frac{1}{\cosh^2 t}-\cos^2\phi}}, 
\ea
which leads to the metric
\ba
ds^2=d\chi^2-\sin^2\chi ds^2.\label{staicc}
\ea
Then the geodesic length in the bulk dS$_3$ which connects between two points $(\chi_1,s_0)$ and $(\chi_2,s_0)$ on the boundary dS$_2$ at the same constant time $s_0$ is simply 
given by $D_{12}=|\chi_1-\chi_2|$. This means that the holographic entanglement entropy on the constant time slice in this static coordinate is linear about the subsystem size $|A|$ and saturates the subadditivity.  This clearly shows that if we consider time slices other than $s=$const. , including the constant $t$ slices, the strong subadditivity gets violated because the subsystem size $|A|$ gets shorter as
\ba
|A|=\int^{\chi_2}_{\chi_1}d\chi \s{1-\sin^2\chi \left(\frac{ds}{d\chi}\right)^2}\leq \chi_2-\chi_1=D_{12}.
\ea
This analysis shows that at $\theta_0=\frac{\pi}{2}$, the static slice $s=$const. is the only time slice which is consistent with the Hilbert space structure of dual unitary quantum systems.

\begin{figure}
	\centering
	\includegraphics[width=1.4in]{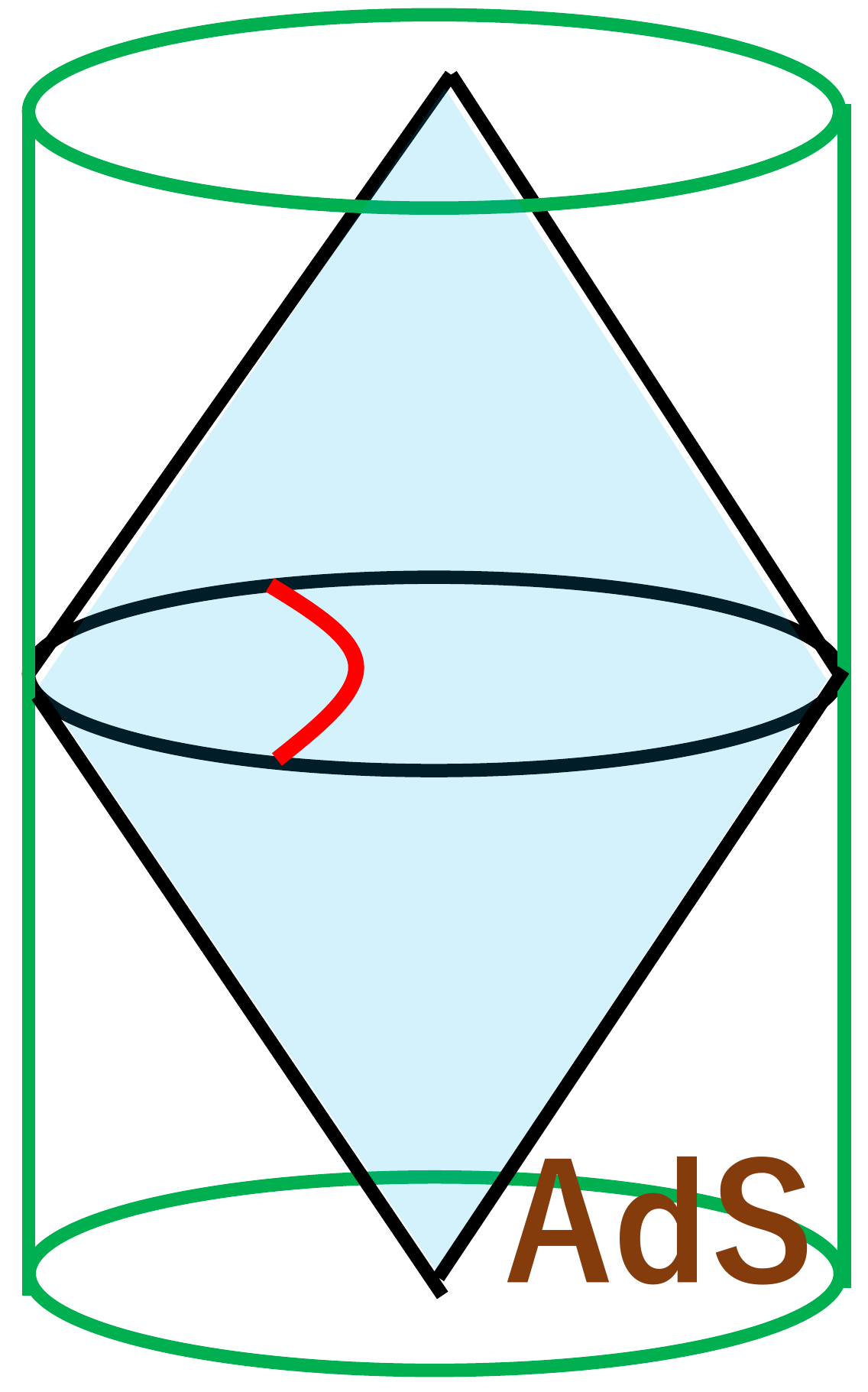}
	\includegraphics[width=4in]{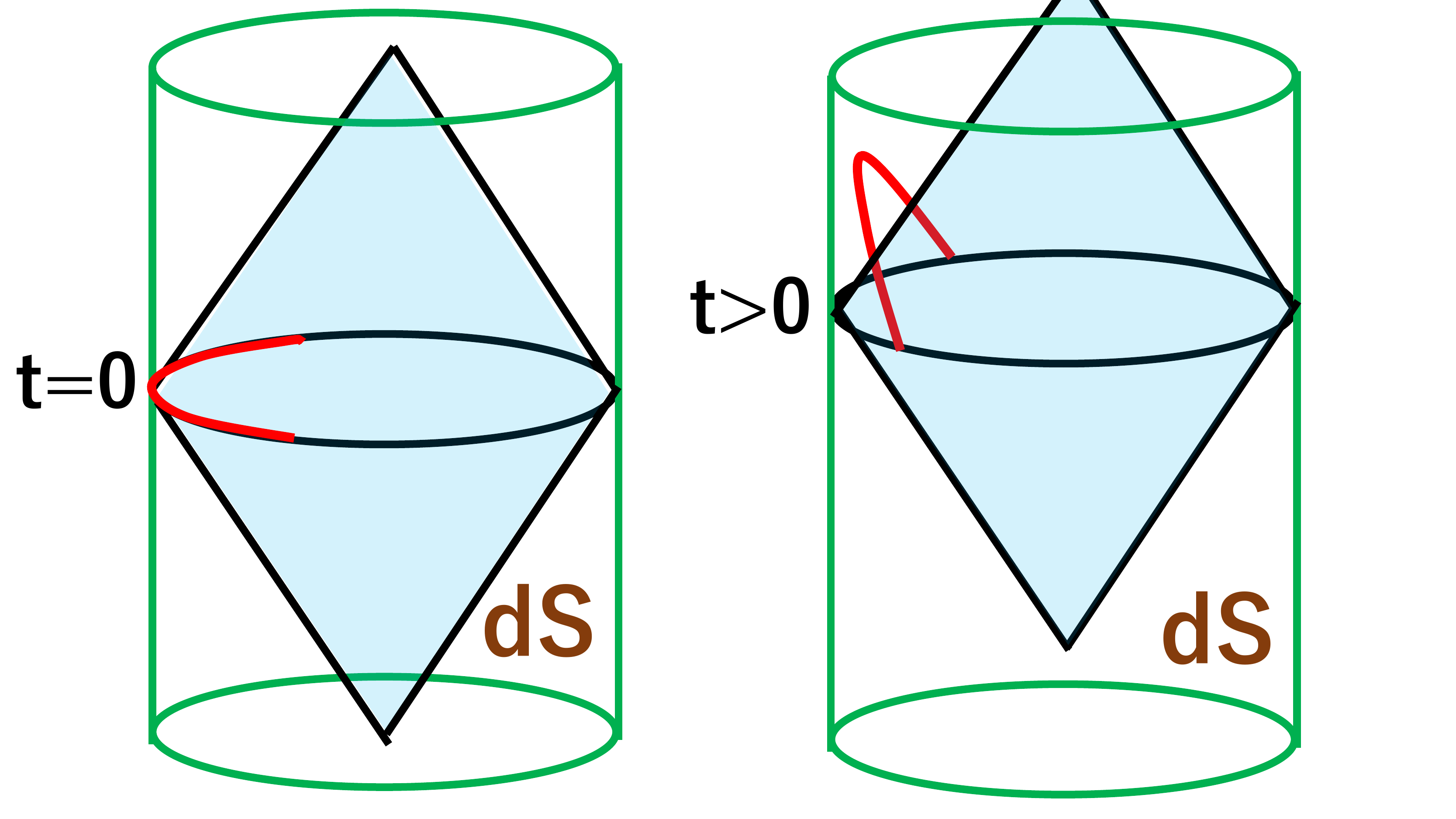}
	\caption{Sketches of Wheeler-DeWitt patches (blue regions) and the geodesics (red curves) in AdS$_3$ (left), in dS$_3$ at $t=0$ (middle) and in dS$_3$ at $t>0$ (right). }
	\label{fig:WDW}
\end{figure}

We can provide another feature of this restriction from the spacetime structure of de Sitter space. Consider a Wheeler-DeWitt patch in the bulk dS$_3$ for a generic time slice of the boundary dS$_2$. Since space-like geodesics which connects two boundary points are all on the boundary dS$_2$, the geodesics are outside of the Wheeler-DeWitt patch, except when the time slice is the constant $s$ slice, as depicted in figure \ref{fig:WDW} by comparing this with that in AdS. In a sensible holography we expect that any bulk counterpart dual to objects in the boundary theory at a specific time will be within its Wheeler-DeWitt patch. 

For generic values of $\theta_0$, the situation gets less sharp but looks qualitatively similar. This can be seen from the left panel of  figure \ref{fig:lightregion}. As the time evolves following constant $t$ (or equally $T$) slices, the region which satisfies the subadditivity gets squeezed into the future direction and eventually disappears. As $\theta_0$ gets closer to $\frac{\pi}{2}$, the region which satisfies the subadditivity becomes very narrow as in the right panel of figure \ref{fig:lightregion}.

\subsection{Time slices and dual Hilbert space}

\begin{figure}
	\centering
	\includegraphics[width=2.5in]{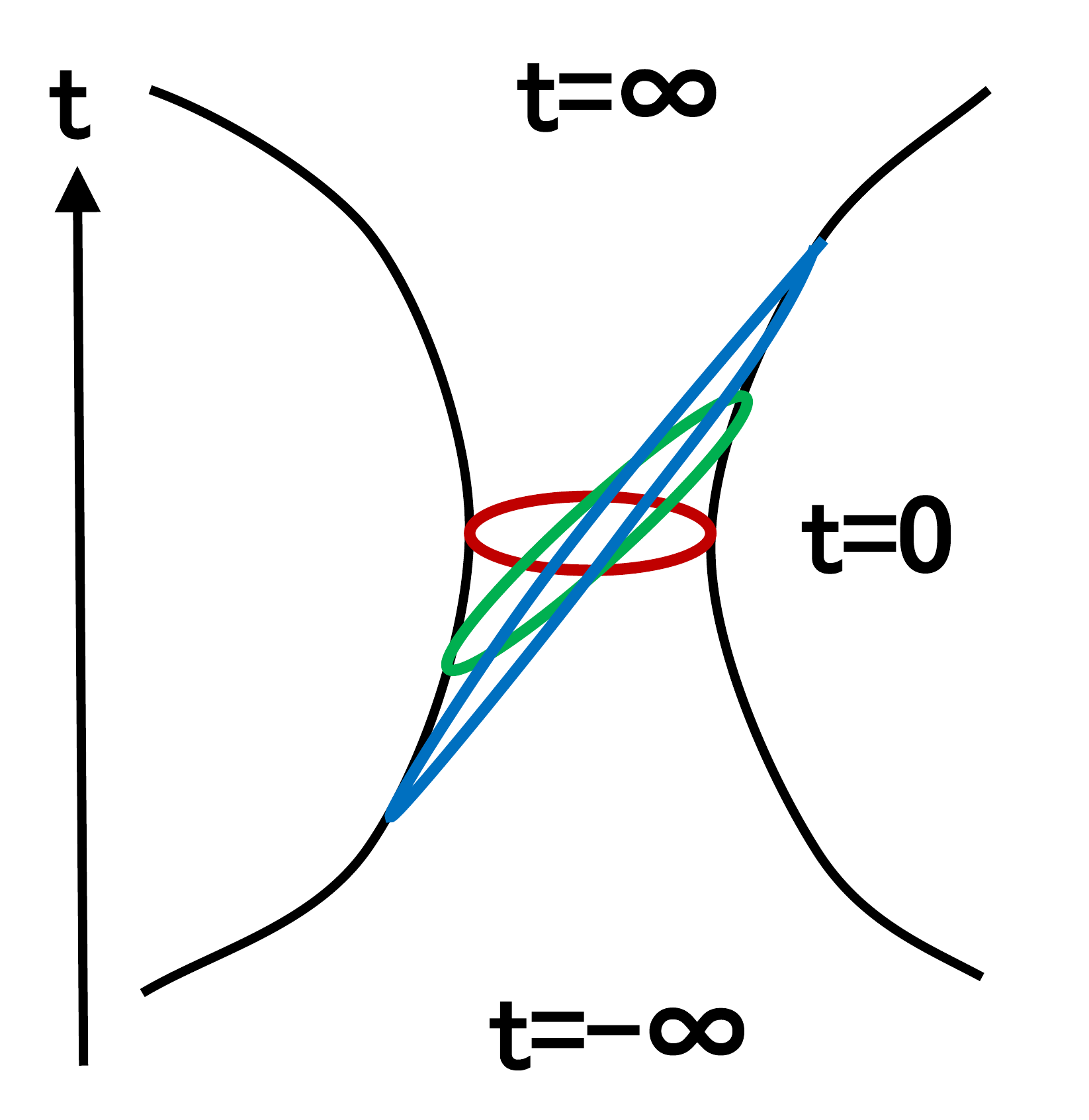}
	\includegraphics[width=2.5in]{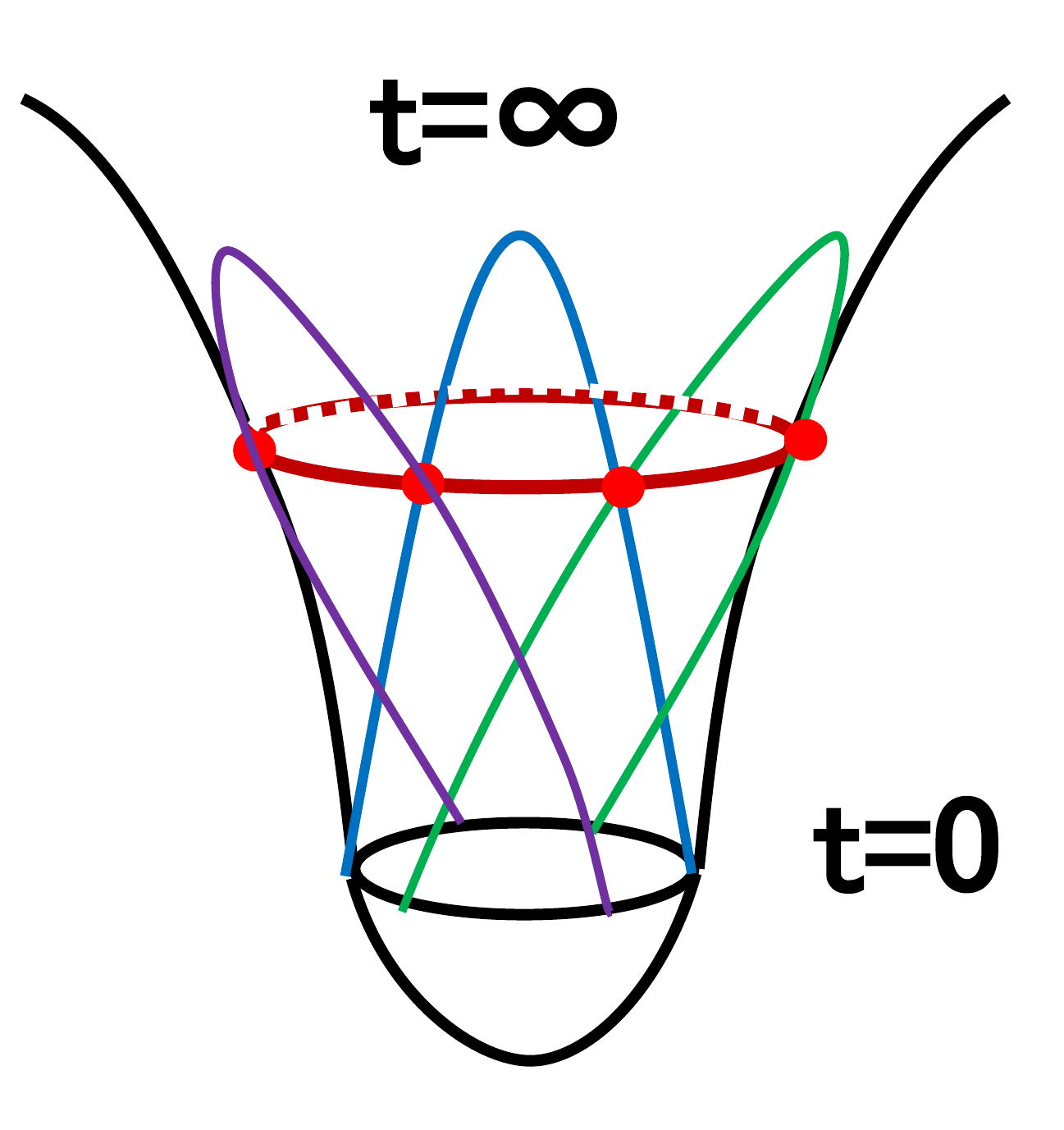}
	\caption{Sketches of nice time slices (constant $s$ slices) in the boundary dS$_2$ which describe the Hilbert space of dual quantum system. In the left panel we showed these slices in a global dS$_2$.  In the right panel, we showed that a constant $t$ slice (red curve) is decomposed into multiple constant $s$ slices (purple, blue and green curves).}
	\label{fig:Hilb}
\end{figure}

Having in mind the analysis of holographic entanglement entropy, we would like to consider the holographic dual interpretation. For simplicity let us focus on the case $\theta_0=\frac{\pi}{2}$. From the Hilbert space viewpoint of its dual quantum system, assuming that it is unitary, we can only allow the static time slice (constant $s$ one) on the boundary dS$_2$ as we have seen in the previous subsection.
Owing to the $SO(2,1)$ symmetry of dS$_2$, we can boost and rotate the canonical static time slice $t=0$ by this symmetry as depicted in the left panel of figure \ref{fig:Hilb}. Below we call these $SO(2,1)$ transformation of $t=0$ slice nice slices. Note that this family of nice time slices do not include $t=t_0>0$ slices but does the constant $s$ ones in the coordinate (\ref{staicc}). Therefore we expect that all nice slices correspond to an identical Hilbert space ${\cal H}_{\mathrm{dS}}$ of a quantum system dual to a half dS$_3$. Since the dual state looks maximally entangled due to the fact that the extremal surface $\Gamma_A$ is included in the boundary dS$_2$, the dimension of ${\cal H}_{\mathrm{dS}}$ should be given by 
$S_{dS}=\frac{2\pi}{4G_N}$, i.e. the de Sitter entropy of the dS$_3$ gravity. 

This argument looks very similar to the surface/state duality proposed in \cite{Miyaji:2015yva}, which argues that a codimension two space-like surface $\Sigma$ in general gravitational spacetimes is holographically dual to a certain quantum state $|\Phi_\Sigma\lb$. The difference is that in the present paper we put a real boundary by restricting the dS$_3$ to the half dS$_3$ so that it has a genuine boundary given by dS$_2$ at $\theta=\theta_0$. However, it is natural that there is a holographic duality even before we put the time-like boundary. In this context, our lesson from this present analysis is that in order for the surface/state duality works well we need to require that the extemal surface $\Gamma_A$ should be within the Wheeler-DeWitt patch of $\Sigma$.

We can decompose a constant $t=t_0>0$ slice into multiple nice time slices as depicted in the right panel of figure \ref{fig:Hilb}. As $t_0$ gets larger, we need more nice time slices to cover it as can be seen from that fact that the length of $t=t_0$ slice exponentially grows $\sim e^{t_0}$. 
This shows that the constant $t$ slice (except $t=0$ one) overestimates the dimension of Hilbert space. Indeed this is obvious from the fact that to cover the $t=t_0$ slice we need many nice slices whose Hilbert spaces clearly have overlaps as they cover the $t=0$ slice many times. Thus we cannot expect that such a slice describes a well-defined Hilbert space, which is consistent with our observation of the subadditivity violation. The division of the constant $t$ slice into subsystems does not correspond to the factorization of a Hilbert space. This behavior is quite different from that in local quantum field theories, where the choice of time slice on a given causal diamond  gives the identical subsystem of a Hilbert space. In our de Sitter holography, even though the segment $A$ on the constant $t$ slice and  $\Gamma_A$ on the nice slice both live on the same causal diamond as in the left panel of figure \ref{fig:setupp}, only the latter has a well-defined meaning as a proper subsystem of ${\cal H}_{\mathrm{dS}}$.

\subsection{Non-locality and Holographic Entanglement Entropy}
\label{EENon}

Then it is natural to ask why only nice slices (constant $s$ slices) can describe a proper Hilbert space. Even though we do not give a conclusive answer to this question, we would like to suggest that the non-local nature of dual quantum system plays a crucial role. First of all, it is obvious that the way we put the boundary of a half dS$_3$ leads to a finite cut off in the dual field theory. Also the volume law entanglement which we observe for nice slices is typical for vacuum states in highly non-local field theories \cite{Li:2010dr,Shiba:2013jja}.
Moreover, we can even find that a growth of entanglement entropy, defined by the replica method, can exceed the volume law i.e. $S_A\sim |A|^{p},\ \ p>1$ in highly non-local field theories, which shows the violation of subadditivity. Indeed, as shown in \cite{Li:2010dr} via the replica method, a non-local free scalar field theory in $d$ dimensions with the action 
\begin{equation}
	S=\int dx^d \phi(x)e^{(-\de_x^2)^q}\phi(x)\,,
\end{equation}
leads to the entanglement entropy whose UV divergence scales as 
\begin{equation}
	S_A\propto \left(\frac{L_A}{\ep}\right)^{d-2+2q}\,,
\end{equation}
where $L_A$ is the linear size of the subsystem $A$.
Thus, $S_A$ grows faster than the volume law if $q>\frac{1}{2}$. Here note that the entanglement entropy defined from the replica method based on Euclidean path-integral is not guaranteed to satisfy the subadditivity, though the entanglement entropy defined from a quantum state in a Hilbert space automatically satisfies the subadditivity. 

In our holographic dual of a half dS$_3$, it is natural to expect that a similar non-local field appears and this may explain the violation of subadditivity on generic time slices.
As the value of $\theta_0$ gets smaller, this non-local effects get milder and the range of time slices which satisfy the subadditivity, gets broader as depicted in figure \ref{fig:lightregion}.

\begin{figure}
	\centering
	\includegraphics[width=5in]{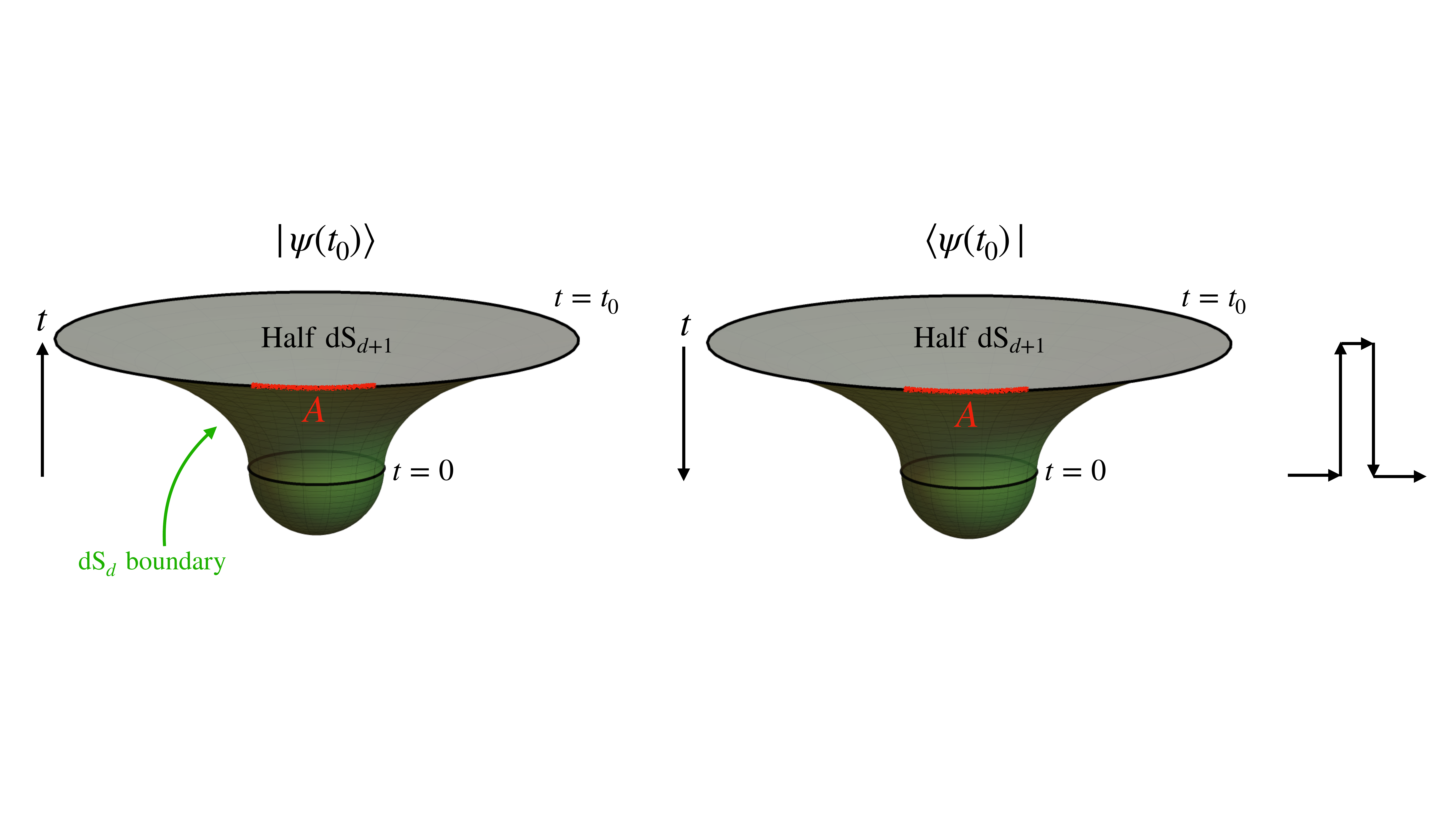}
	\caption{Sketches of calculation of holographic entanglement entropy in the Schwinger-Keldysh geometry. The sum of the contributions associated with two extremal surfaces $\Gamma^{(1)}_A, \Gamma^{(2)}_A$ gives rise to the real entanglement entropy.}
	\label{fig:braketEE}
\end{figure}

In this way, even though the proper Hilbert space interpretation becomes difficult for generic time slices including constant $t$ slices, we may be able to formally define the entanglement entropy by the replica method, which allows the violation of strong subadditivity. Notice that in such a non-local field theory, Hamiltonian cannot be defined in a standard way as the action involves derivatives with respect to the Euclidean time whose order is higher than two. If this interpretation is true, we should be able to define $S_A$ even when $A$ is so large that the two end points of $A$ cannot be connected by a space-like geodesic. If we apply the original holographic entanglement entropy formula (\ref{heeds}), we find that it is complex valued 
\begin{equation}
    D_{12}=\pi +i\arccosh\left[\sinh^2 t-\cos(2\theta_0)\cosh^2 t_0\right]\,.
\end{equation}
See \cite{Hikida:2022ltr,Chapman:2022mqd} for more studies about complex geodesics in dS space. As depicted in the right panel of figure \ref{fig:setupp}, the real part comes from the geodesic in the Euclidean instanton (semi-sphere) and the imaginary part comes from the time-like geodesic in the de Sitter space. However, if we consider the holographic calculation of entanglement entropy via the replica method (for the derivation of the covariant HEE in AdS/CFT see \cite{Dong:2016hjy}), there are two candidates of extremal surface: one is in the bra geometry and the other is the ket geometry as in figure \ref{fig:braketEE}.  This prescription was analyzed in \cite{Chen:2020tes} for a bra-ket wormhole in AdS/CFT.

The correct result is given by the sum of two contributions:
\ba
Z^{(n)}_{\mathrm{tot}}=e^{(1-n)S}+e^{(1-n)S^*},
\ea
where $n$ is the replica number. $S_A$ is given by 
\ba
S_A=\lim_{n\to 1} \frac{1}{1-n}\log Z^{(n)}_{\mathrm{tot}}\simeq \frac{S+S^*}{2}=\mbox{Re}[S].\label{preshde}
\ea
Thus we find that the actual $S_A$ is given by the real part:
\ba
S_A=\frac{\pi}{4G_N}=\frac{1}{2}S_{\mathrm{dS}},
\label{entreal}
\ea
where $S_{\mathrm{dS}}=\frac{2\pi}{4G_N}$ is the de Sitter entropy.

\begin{figure}
	\centering
	\includegraphics[width=3in]{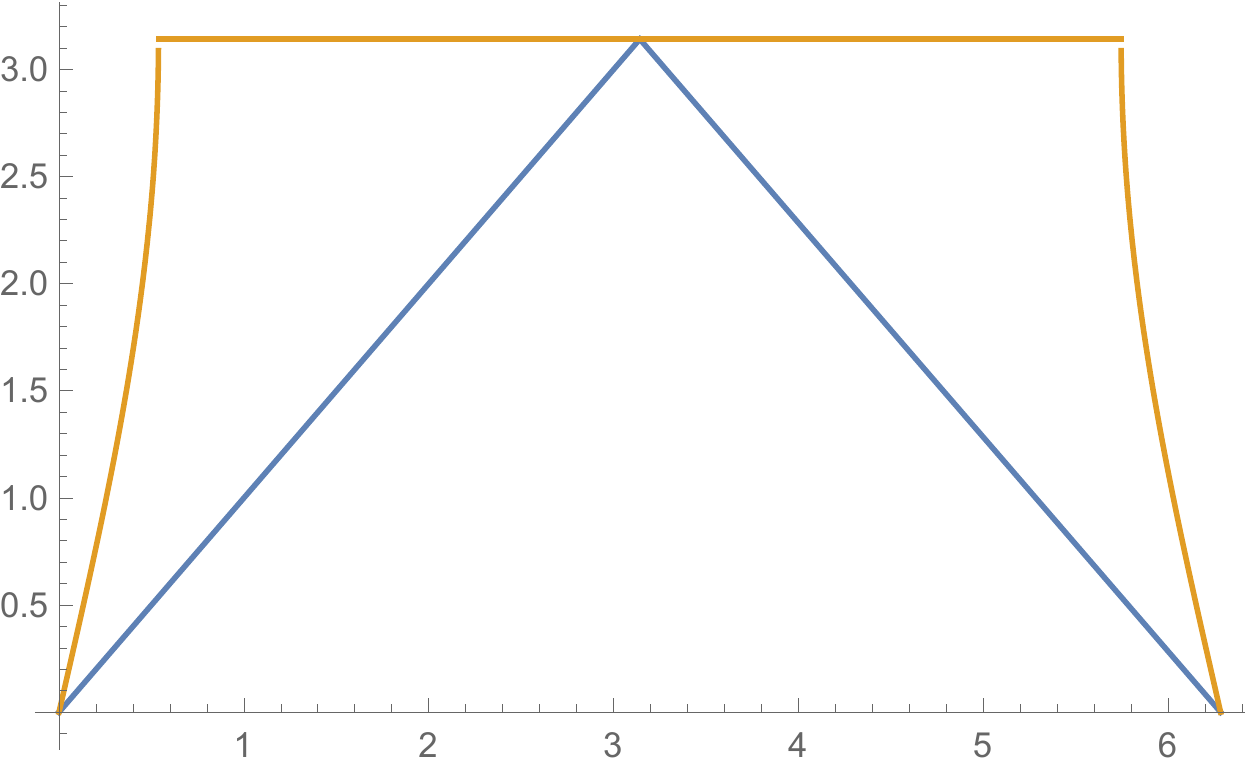}
	\includegraphics[width=3in]{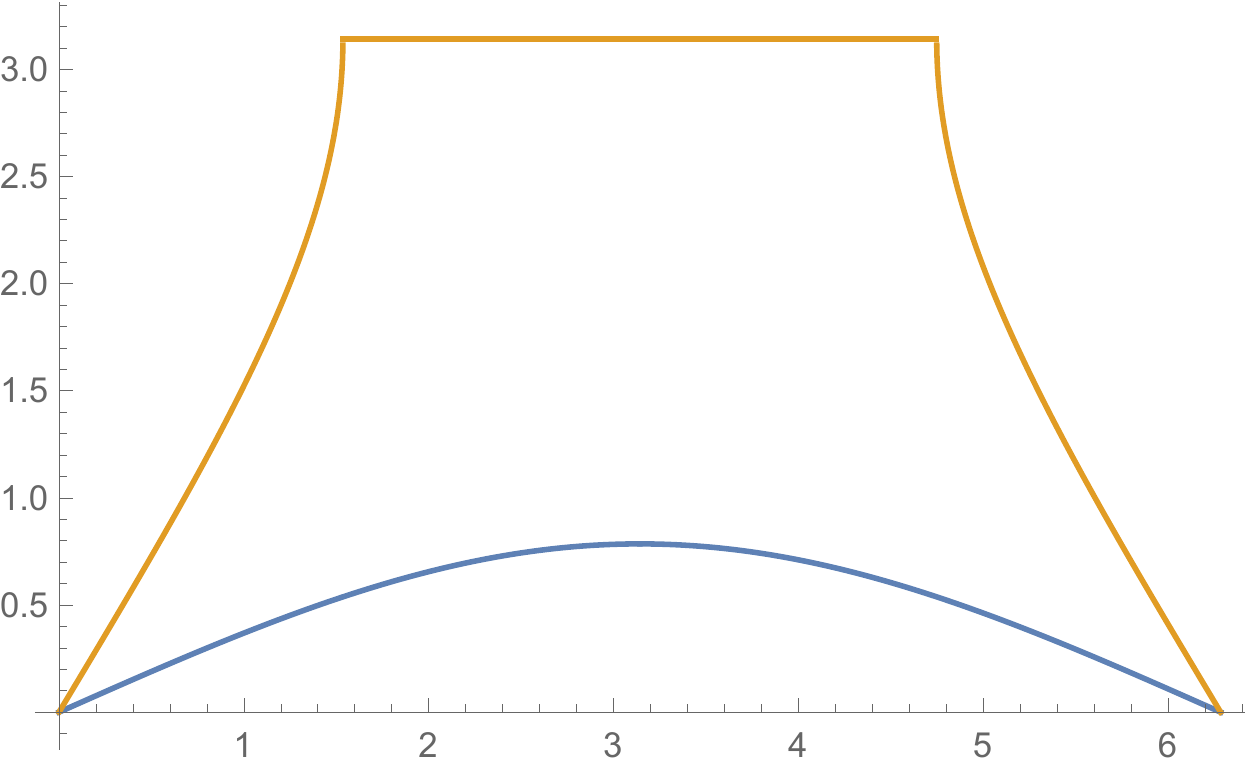}
	\caption{Plots of holographic entanglement entropy $S_A$ on a constant $t$ time slice based on the prescription (\ref{preshde}) for $d=2$. We plotted $S_A$ as a function of 
		$\Delta \phi$ in the range $0\leq \Delta\phi\leq 2\pi$ for 
		$\theta_0=\frac{\pi}{2}$ (left) and $\theta_0=\frac{\pi}{8}$ (right). The blue and orange curves correspond to 
		$t=0$ and $t=2$. Note that we have the critical time $t_{\mathrm{max}}=0$ for $\theta_0=\frac{\pi}{2}$ and $t_{\mathrm{max}}\simeq 1.61$ for $\theta_0=\frac{\pi}{8}$.}
	\label{fig:exHEE}
\end{figure}

In this way, we find that the saturation behavior i.e. $S_A$ grows monotonically for $0\leq \Delta\phi\leq \Delta \phi_{\mathrm{max}}$ and it takes the constant value $\frac{1}{2}S_{\mathrm{dS}}$ for $\Delta\phi\geq \Delta \phi_{\mathrm{max}}$. For $\Delta\phi>\pi$, the standard holographic prescription guarantees the relation $S_A=S_{A^c}$, where $A^c$ is the complement of $A$. Explicit 
plots are shown in figure \ref{fig:exHEE}.

\subsection{Overcounting of de Sitter Hilbert Space}
\begin{figure}
	\centering
	\includegraphics[width=5in]{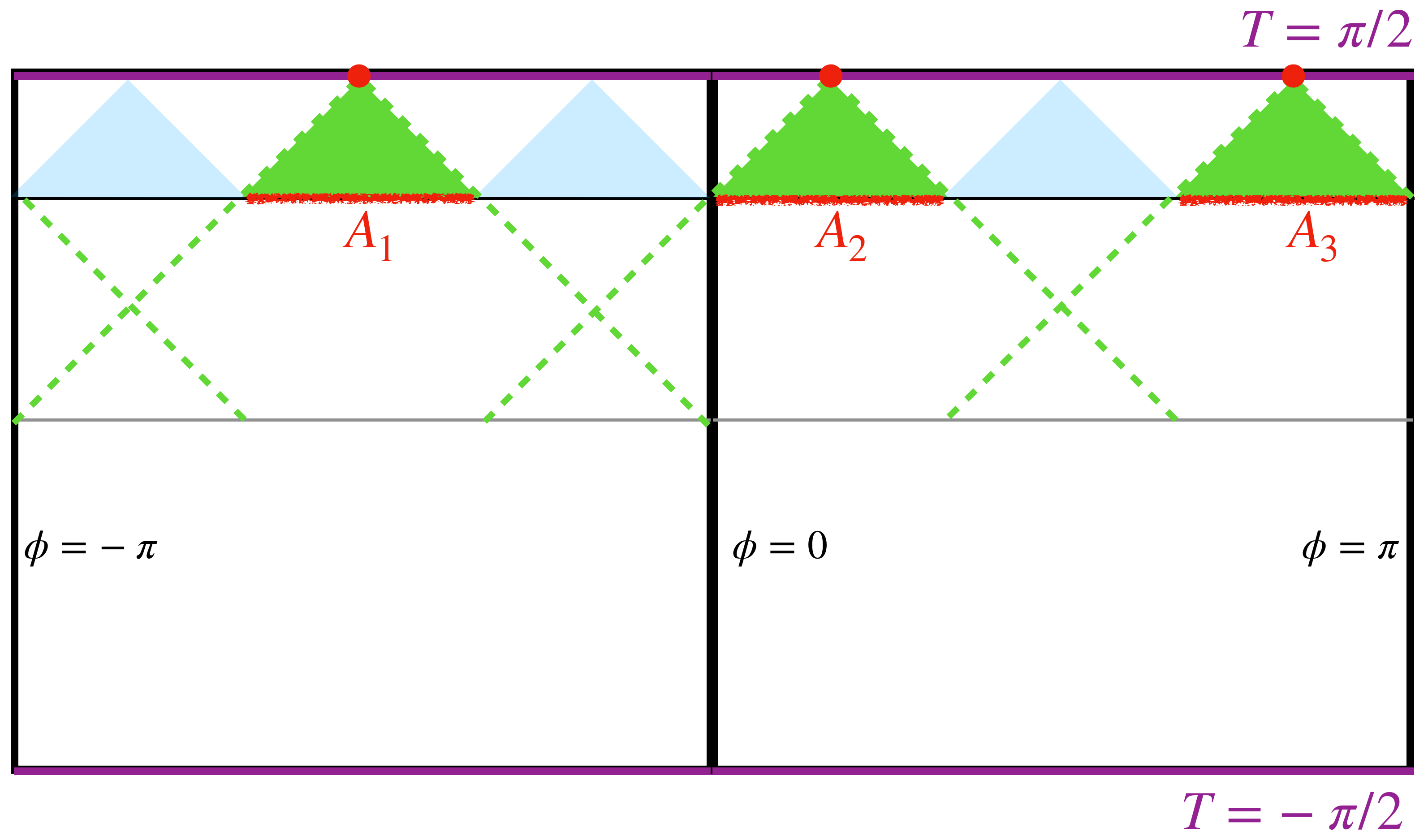}
	\caption{Penrose diagram of dS$_2$ spacetime which corresponds to the dS boundary located at $\theta=\theta_0=\frac{\pi}{2}$ in dS$_3$ bulk spacetime. The green dashed lines denote the null limit of geodesics. On the time slice $T=T_0$ or $t=t_0$, we divide the system into six parts.}
	\label{fig:dS2}
\end{figure}

In previous analyses regarding the holographic entanglement entropy of a single interval, it was observed that the entropy is bounded by half of the dS entropy, \ie 
\begin{equation}
	S_A \le \frac{1}{2} S_{\mathrm{dS}}\,,
\end{equation}
as illustrated in figure \ref{fig:exHEE}. This outcome is expected since the corresponding bulk dual represents only half of the dS spacetime. Generalizing this calculation to cases where subsystem $A$ comprises multiple disconnected intervals is also possible. Since this calculation follows the standard prescription \cite{Headrick:2010zt}, explicit results will not be presented in detail. However, it is important to note that the holographic entanglement entropy of a subregion consisting of multiple intervals is not bounded by the dS entropy $S_{\rm dS}$ at late times, as the sum of lengths of disconnected geodesics can be very large. Focusing on the case with the boundary dS space located at $\theta=\theta_0=\frac{\pi}{2}$, it has been derived in eq.~\eqref{boundnuA} that the entropy of a single interval reaches the maximal value $\frac{1}{2} S_{\mathrm{dS}}$ when its length corresponds to the critical size $\Delta \phi_{\rm max}$. It should be noted that the maximum size $\Delta \phi_{\rm max} (t_0)$ decreases to zero as the boundary time $t_0$ increases. Consequently, let us consider a time slice at $t=t_0$ which can be divided into $2N$ identical intervals, where $\frac{\pi}{N} \ge \Delta \phi_{\rm max} (t_0)$. Figure \ref{fig:dS2} provides an explicit example with $N=3$. By considering a subsystem with $N$ intervals, one can find that the holographic entanglement entropy is shown as 
\begin{equation}\label{eq:largeSA}
	S_{A_1\cup A_2 \cup \cdots \cup A_N} = N \times S_{A_i}= \frac{N}{2}S_{\rm dS} \ge S_{\rm dS}\,.
\end{equation}
This value can even approach infinity as $N$ tends to infinity, while $\lim\limits_{t_0\to \infty}\Delta \phi_{\rm max} \to 0$. It might be questioned whether this violates the entropy bound of de Sitter space, whose Hilbert space is expected to be finite. However, as observed from the violation of (strong) subadditivity, we expect that a generic time slice does not correspond to a pure state in a single de Sitter Hilbert space. Instead, the unbounded entropy of multiple intervals, as expressed in eq.~\eqref{eq:largeSA}, can be interpreted as the result of overcounting the dimension of the de Sitter Hilbert space. As shown in figure 
\ref{fig:dS2}, each critical interval $A_i$ at $t=t_0$ could be understood as a maximally entangled state in a single de Sitter Hilbert space. Back to the original time at $t=0$, those single Hilbert spaces are overlapping with each other along this time slice. The recent paper \cite{Cao:2023gkw} discusses the tensor network representation of dS space with overlapping qubits. However, we note that the studies in \cite{Cao:2023gkw} focus on bulk dS spacetime which is different from our dS boundary picture as shown in figure \ref{fig:dS2}.

\subsection{Higher dimensional half de Sitter space}\label{sec:highw}

It is straightforward to generalize our analysis for a half dS$_3$ to a half dS$_{d+1}$ for $d\geq 3$.
Below we will provide analysis for $\theta_0=\frac{\pi}{2}$ in the global dS$_{d+1}$ described by (\ref{dsg}) and (\ref{sth}). The time-like boundary of this space is given by dS$_{d}$ described by the global metric
\ba
ds^2=-dt^2+\cosh^2 t (d\phi^2+\sin^2\phi d\Omega^2_{d-2}),
\ea
where $d\Omega^2_{d-1}$ is the metric of the unit sphere S$^{d-2}$. As in the previous case of $d=2$, we would like to argue that gravity on a half dS$_{d+1}$ is dual to non-local field theory on the dS$_d$.

To examine this duality we would like to calculate the holographic entanglement entropy. This is given by the area of extremal surface $\Gamma_A$, which ends on the boundary of the subsystem $A$, by (\ref{Area}).

\begin{figure}[t]
	\centering
	\includegraphics[width=7cm]{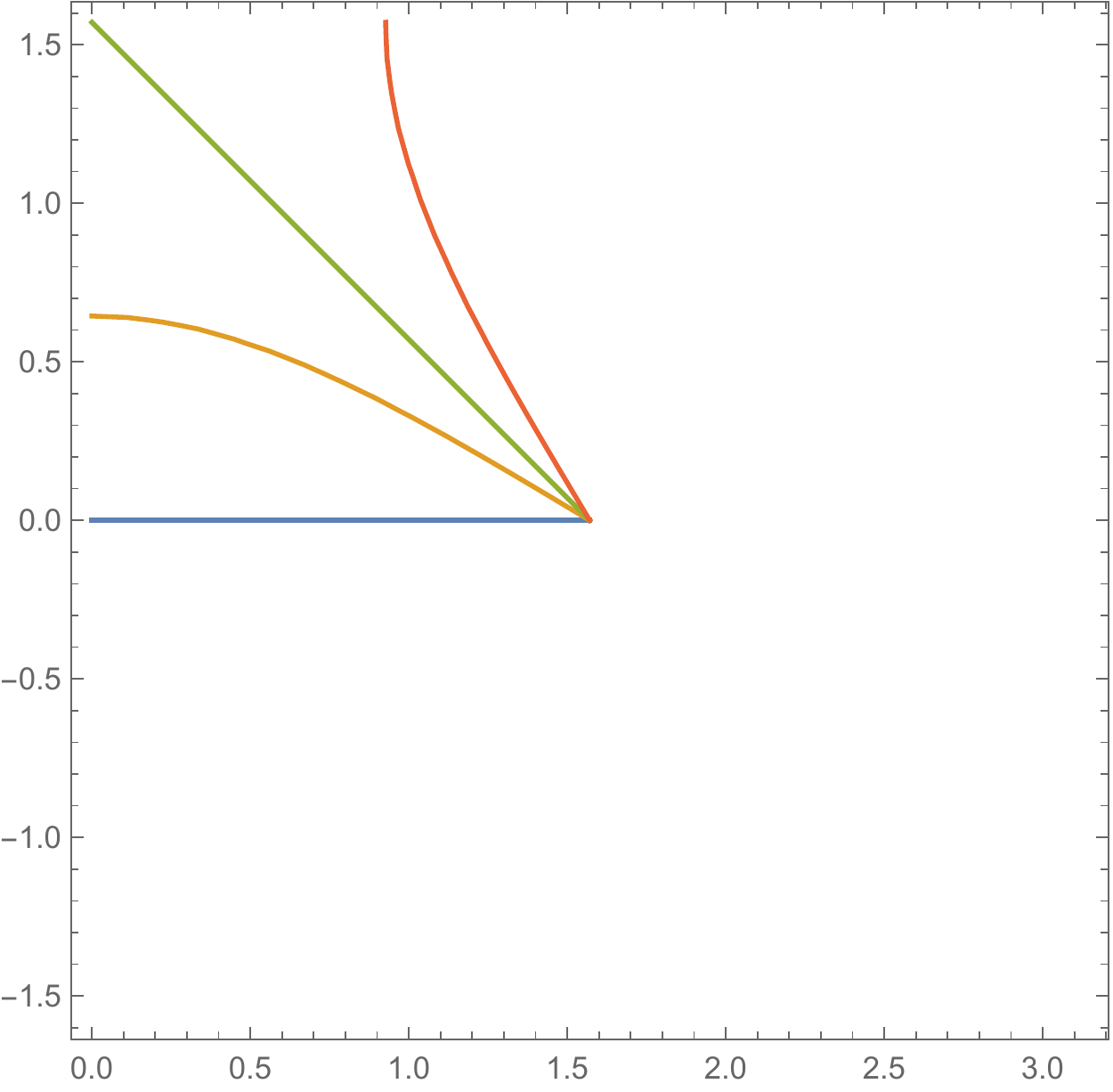}
	\caption{Profiles of extremal surfaces (\ref{extdsf}) in global dS. The horizontal and vertical coordinate are 
		$\phi$ and $T$, respectively. The blue, yellow, green ,and red curves describe the extremal surface for $L^2=1, 1/4, 0$ and $-1/4$, respectively. }
	\label{fig:dSEXT}
\end{figure}

For this we choose the subsystem $A$ to be a disk on the boundary dS$_d$:
\ba
t=t_0,\ \ 0\leq \phi\leq \phi_1.  \label{subads}
\ea
In this case, we can find the profile of the extremal surface $\Gamma_A$ which ends on the boundary $\phi=\phi_1$ of the subsystem $A$ (\ref{subads}) as
\ba
\frac{\cos\phi}{\tanh t}=\frac{1+L^2}{1-L^2}.\label{extdsf}
\ea
This surface can be simply obtained by mapping the extremal surface 
\be
t^2-\sum_{i=1}^{d-1}(x_i)^2=L^2,
\ee
in the Poincare dS$_{d}$ given by the metric 
\ba
ds^2=\frac{-dt^2+\Sigma_{i=1}^{d-1}dx_i^2}{t^2}.
\ea
The constant $L$ is related to the choice of the subsystem $A$ via
\ba
\frac{1+L^2}{1-L^2}=\frac{\cos\phi_1}{\tanh t_0}.
\ea
The profile of the extreme surface (\ref{extdsf}) is plotted in figure \ref{fig:dSEXT}.

This $d-1$ dimensional extremal surface $\Gamma_A$ is wrapped on S$^{d-2}$. The induced metric on $\Gamma_A$ reads 
\ba
ds^2=\frac{4L^2(1+L^2)^2}
{\left((1+L^2)^2-(1-L^2)^2\cos^2\phi\right)^2}d\phi^2
+\frac{(1+L^2)^2\sin^2\phi}{(1+L^2)^2-(1-L^2)^2\cos^2\phi}d\Omega^2_{d-2}.\no
\ea
This surface $\Gamma_A$ is space-like when $0< L^2\leq  1$.
It becomes light-like when $L^2=0$ and time-like when $L^2<0$.

\begin{figure}[t]
	\centering
	\includegraphics[width=7cm]{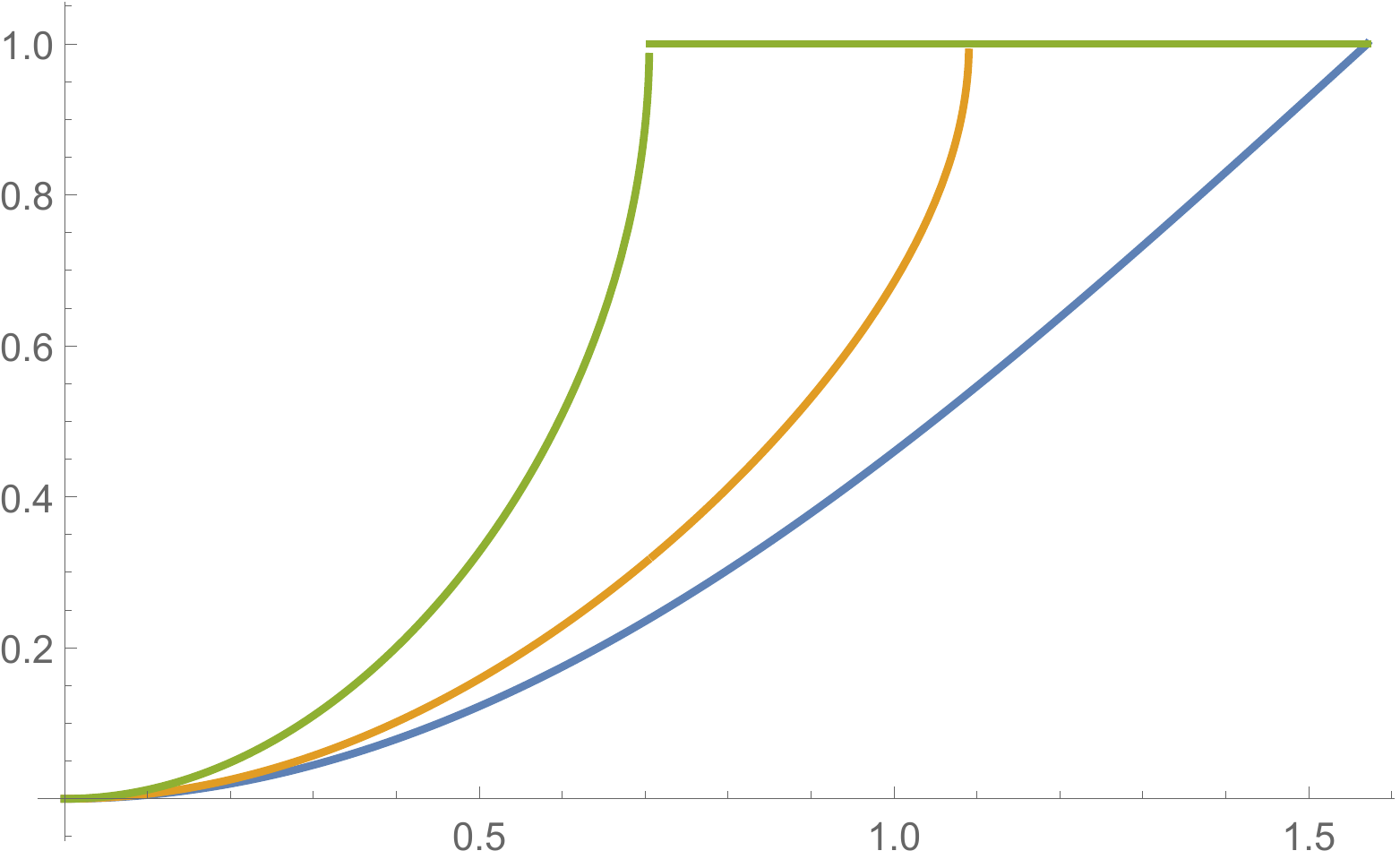}
	\includegraphics[width=7cm]{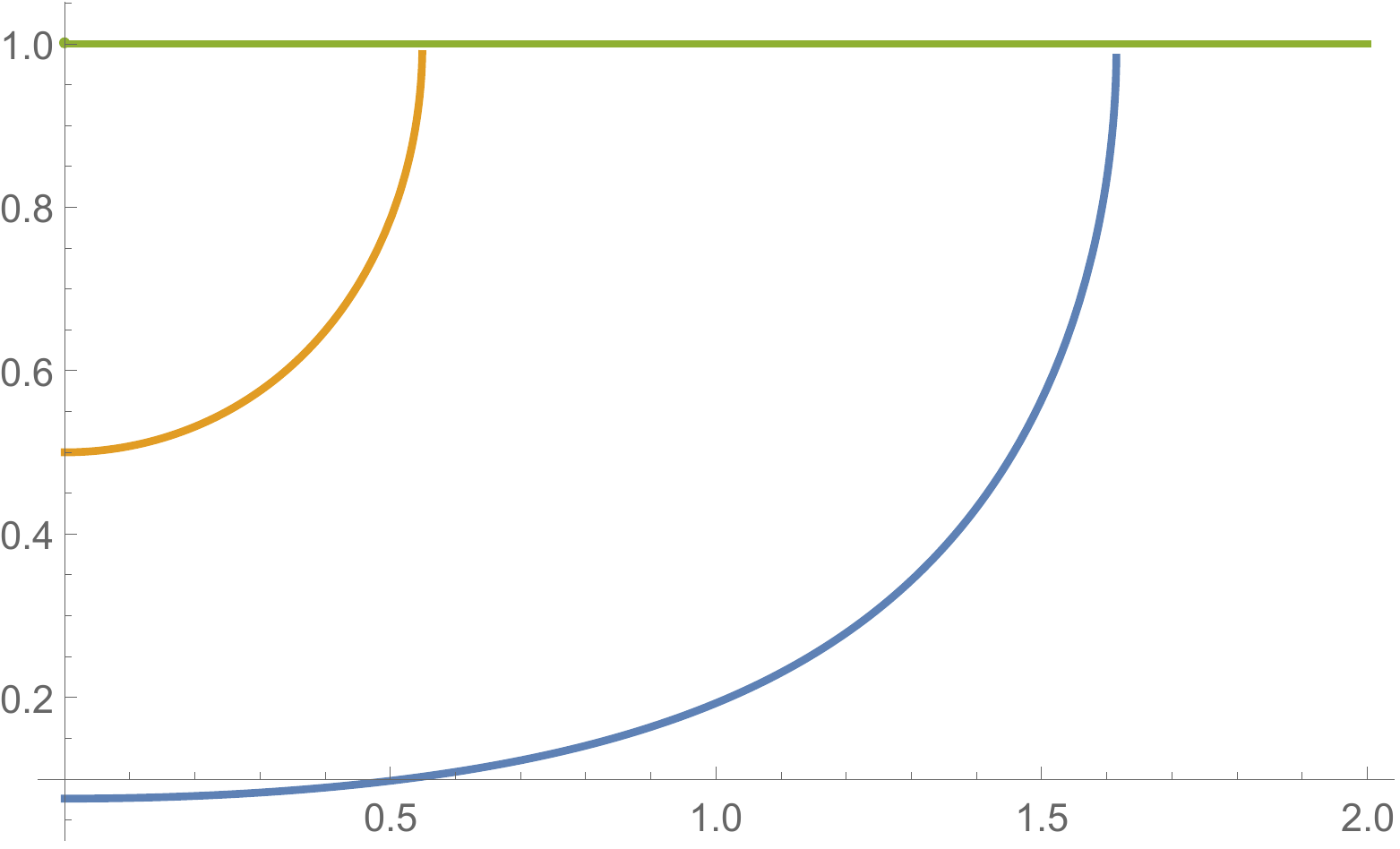}
	\caption{The behaviors of the holographic entanglement entropy for a half dS$_4$ ($d=3$) in case 1. In the left panel, we plotted $\frac{1}{2\pi}A(\Gamma_A)$ as a function of $\phi_1$ at $t=0$ (blue), $t=1/2$ (orange) and $t=1$ (green). In the right panel, we showed  $\frac{1}{2\pi}A(\phi_1,t_0)$ as a function of $t_0$ for $\phi_1=\pi/8$ (blue), $\phi_1=\pi/3$ (orange),
		$\phi_1=\pi/2$ (green).}
	\label{fig:dSarea}
\end{figure}

The area of $\Gamma_A$ is computed as 
\ba
A(\Gamma_A)=\mathrm{Vol}(\mathrm{S}^{d-2})\int^{\phi_1}_{0}d\phi
\frac{2L(1+L^2)^{d-1}(\sin\phi)^{d-2}}{\left((1+L^2)^2-(1-L^2)^2\cos^2\phi\right)^{\frac{d}{2}}}.
\ea
Since this is a function of $\phi_1$ and $t_0$, we write this as $A(\phi_1,t_0)$. It is straightforward to see that 
at $t_0=0$ and $\phi_1=\frac{\pi}{2}$ we have
\ba
A\left(\frac{\pi}{2},0\right)=\frac{1}{2}\mbox{Vol}(S^{d-1}).
\ea
This means that the corresponding holographic entanglement entropy coincides with the half of de Sitter entropy 
$\frac{1}{2}S_{dS}$. The behavior of  $A(\phi_1,t_0)$ is plotted in figure \ref{fig:dSarea}. In general it is a monotonically increasing function of $t_0$ and $\phi_1$ and gets finally saturated to the maximal value $\frac{1}{2}S_{dS}$. Note that when $L^2$ becomes negative i.e when $\Gamma_A$ gets time-like, we choose the other extremal surface  which goes past and which is wrapped on the semi-sphere in the Euclidean part as in right panel of figure\ref{fig:setupp}. Thus as in the same argument of 
(\ref{entreal}), the holographic entanglement entropy is given by the real part of $A(\Gamma_A)$, namely $\frac{1}{2}S_{\rm dS}$.

%%%%%%%%%%%%%%%%%%%%%%%%%%%%%%%%%%%%%%%%%%%%%%%%%%%%%%%%
%%%%%%%%%%%%%%%%%%%%%%%%%%%%%%%%%%%%%%%%%%%%%%%%%%%%%%%%
\section{Holography for a half dS with EOW brane (Case 2)}
\label{sec:HEEW}
%%%%%%%%%%%%%%%%%%%%%%%%%%%%%%%%%%%%%%%%%%%%%%%%%%%%%%%%
%%%%%%%%%%%%%%%%%%%%%%%%%%%%%%%%%%%%%%%%%%%%%%%%%%%%%%%%

Now we would like to study another setup of holography for a half de Sitter space in the presence of EOW brane (case 2).
We regard the future boundary $t=\infty$ of the de Sitter space as an EOW brane where we impose the Neumann boundary condition for the gravitational field. In the dual field theory, we have the final state projection as in the case of AdS/CFT analysis done in section \ref{catwadS}. 

To obtain the properties of the gravity dual, we would like to analyze the holographic entanglement entropy given by the area of extremal surface as in (\ref{Area}) and (\ref{heeds}). However, since the field theory sides has the final state projection, it should properly be regarded as the holographic pseudo entropy \cite{Nakata:2021ubr}. For simplicity focus on the maximal case $\theta_0=\frac{\pi}{2}$ in dS$_{d+1}$ for simplicity. We argue gravity in the half dS$_{d+1}$ with the EOW brane at $t=t_P$ is dual to a non-local field theory on its boundary dS$_d$ with a finial state projection at $t=t_p$. The new aspect here is that the space-like extremal surface  can end on the boundary of the EOW brane. 

Consider the case where the EOW brane is situated at the future infinity $t_p\to\infty$, which is dual to the final state projection at the future infinity.

\begin{figure}[t]
	\centering
	\includegraphics[width=2.5in]{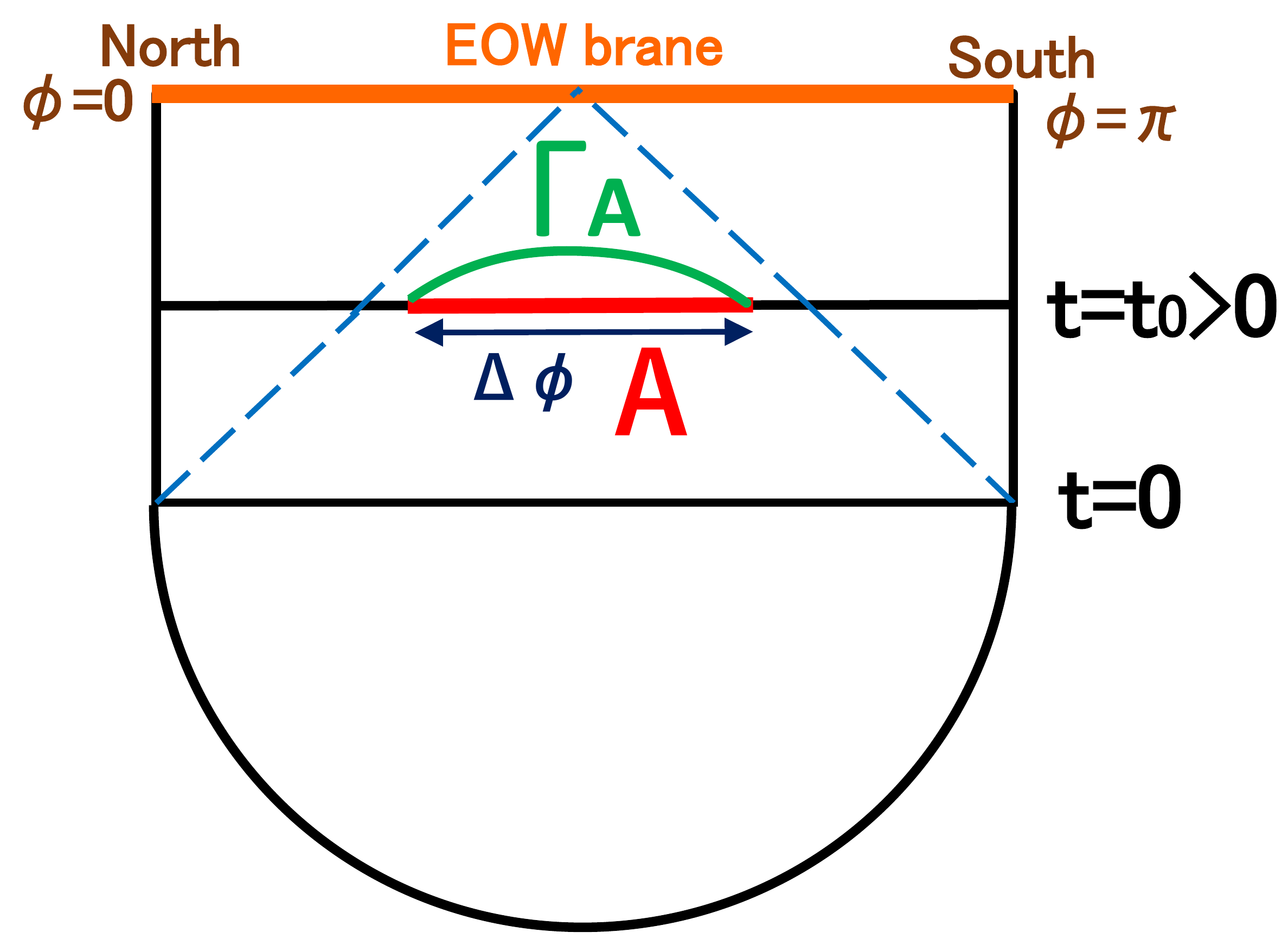}
	\includegraphics[width=2.5in]{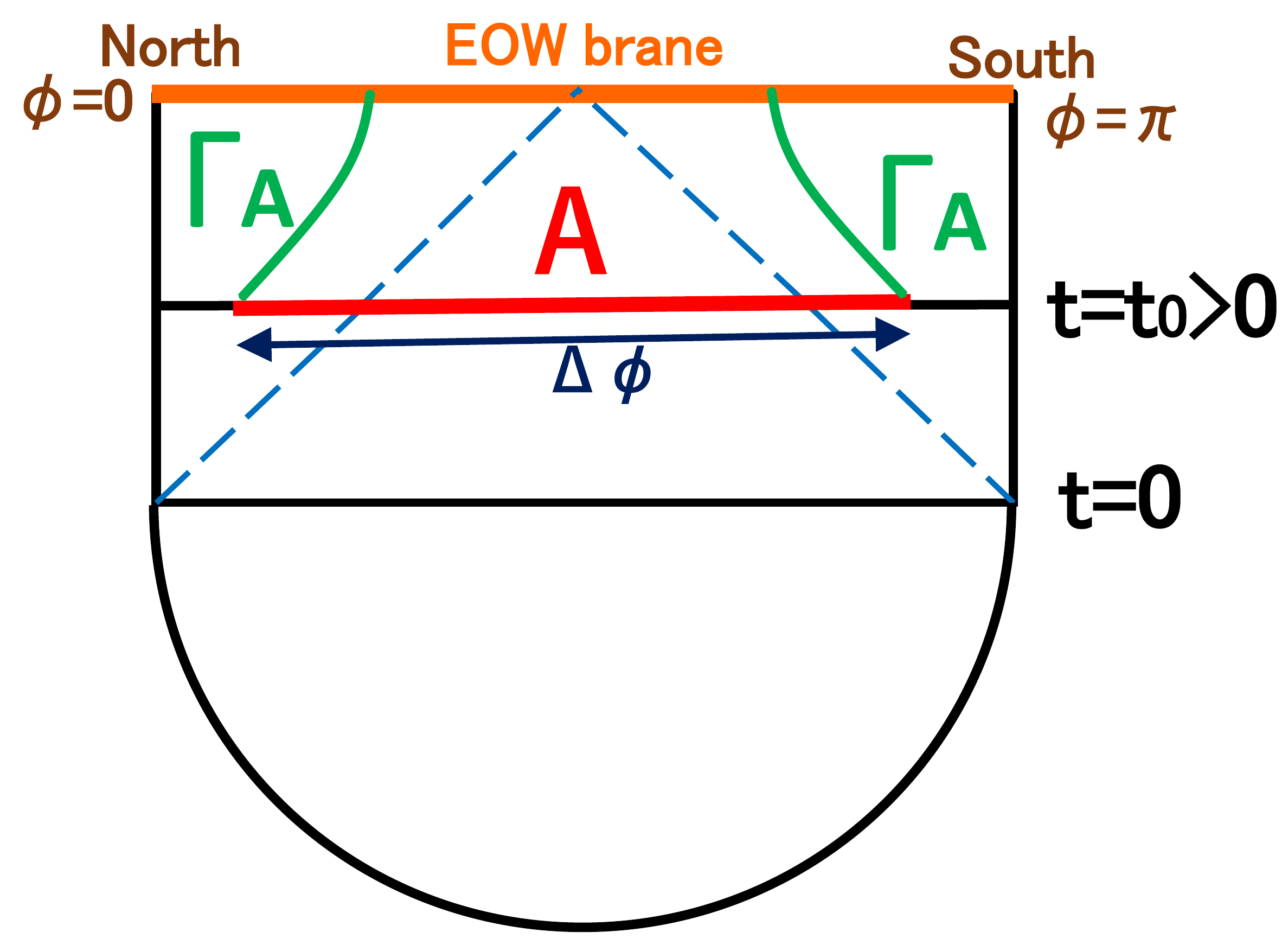}
	\caption{Sketches of geodesics in dS$_2$ which are used to calculation the holographic pseudo entropy in a half dS$_3$. When the subsystem $A$ is small $\Delta\phi\leq \Delta\phi_{\mathrm{max}}$, the geodesic $\Gamma_A$ is space-like and connected shown in the left panel. On the other hand when $A$ gets larger $\Delta\phi>\Delta\phi_{\mathrm{max}}$, $\Gamma_A$ consists of two disconnected time-like geodesics as in the right panel.}
	\label{fig:setudispp}
\end{figure}

First let us analyze the holographic pseudo entropy in this setup for $d=2$. When the interval $A$ is small such that $\Delta\phi<\Delta\phi_{\mathrm{max}}$, the holographic pseudo entropy $S_A$ is given by the length of connected space-like geodesic $D^{\mathrm{con}}_{12}$ as we had in the case 1. However, for $\Delta\phi> \Delta\phi_{\mathrm{max}}$, when a connected space-like geodesic does not exist, $S_A$ can be computed from the two disconnected geodesics which connect one of the end points of $A$ with a point on the EOW brane as depicted in figure \ref{fig:setudispp}. 

Note also that such a disconnected geodesic becomes time-like for  $\Delta\phi> \Delta\phi_{\mathrm{max}}$, which shows that the holographic pseudo entropy is pure imaginary. This is the crucial difference from the analysis of $S_A$ in the case 1. In the case 2, in the presence of EOW brane, the geodesics $\Gamma_A$ can end on it and the real part of $S_A$ for this contribution (see the right panel of figure \ref{fig:setudispp}) becomes smaller than the connected one (see the right panel of figure \ref{fig:setupp}) which has a positive real part.

In summary, the behavior of 
$S_A$ for $d=2$ is identical to that in section \ref{EENon} for $\Delta\phi<\Delta\phi_{\mathrm{max}}$, while we have $\mathrm{Re}[S_A]=0$
for $\Delta\phi>\Delta\phi_{\mathrm{max}}$:
\ba
&&\mbox{For}\ 0\leq \Delta \phi< \Delta\phi_{\mathrm{max}}:\ \ \ \ S_A=\frac{D^{\mathrm{con}}_{12}}{4G_N},\no
&& \mbox{For } \Delta \phi> \Delta\phi_{\mathrm{max}}:\ \ \ \ \mbox{Re}[S_A]=0.
\ea

\begin{figure}[t]
	\centering
	\includegraphics[width=7cm]{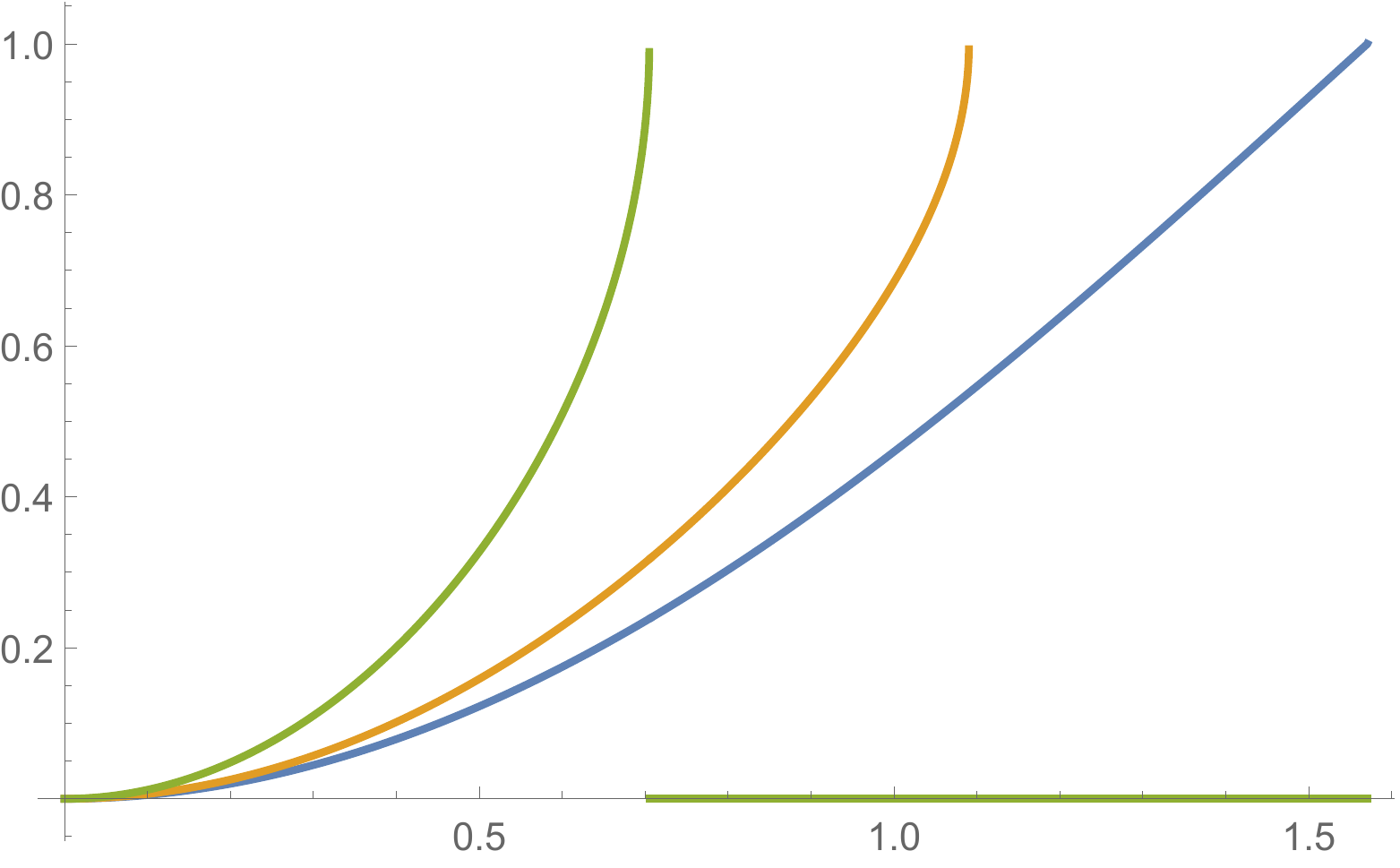}
	\includegraphics[width=7cm]{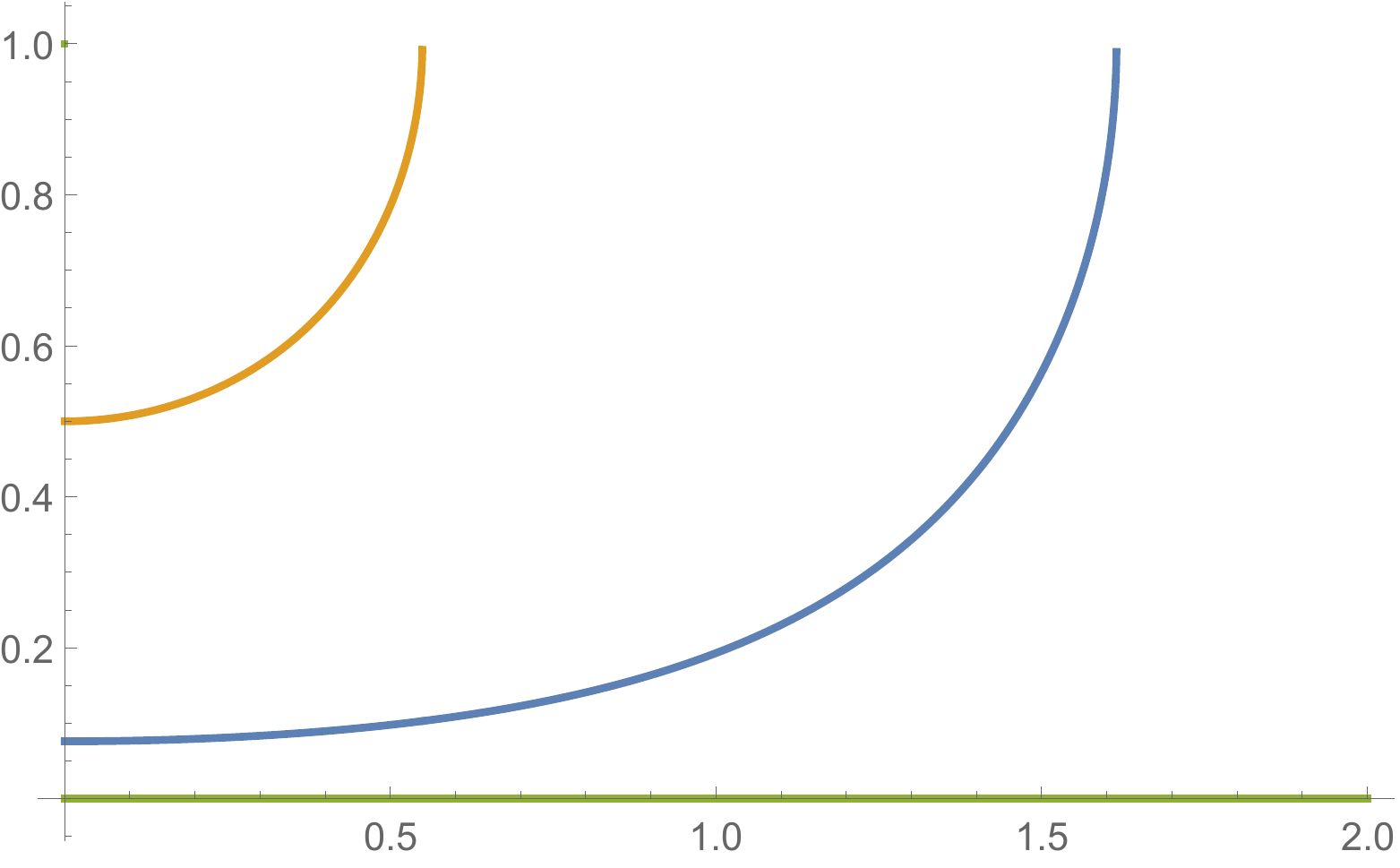}
	\caption{The behaviors of the holographic pseudo entropy for a half dS$_4$ ($d=3$) in case 2. In the left panel, we plotted $\frac{1}{2\pi}A(\Gamma_A)$ as a function of $\phi_1$ at $t=0$ (blue), $t=1/2$ (orange) and $t=1$ (green). In the right panel, we showed  $\frac{1}{2\pi}A(\Gamma_A)$ as a function of $t_0$ for $\phi_1=\pi/8$ (blue), $\phi_1=\pi/3$ (orange),
		$\phi_1=\pi/2$ (green).}
	\label{fig:dSarea2}
\end{figure}

It is straightforward to generalize this to higher dimensions by employing the results of $A(\phi_1,t_0)$ in case 1 presented in section \ref{sec:highw}.
When $\phi_1$ is smaller than the saturation value in case 1, $S_A$ takes the same value as that in case 1. For large values of $\phi_1$, we have Re$[S_A]=0$. An explicit plot for $d=3$ is shown in figure \ref{fig:dSarea2}.

Now let us interpret the behavior of $S_A$ in terms of dual field theory on dS$_d$. For $\Delta\phi<\Delta\phi_{\mathrm{max}}$, $S_A$ is identical to that in case 1, its interpretation is the same. Thus, we find the violation of subadditivity and this is expected to be due to the non-local nature of the dual field theory. For $\Delta\phi>\Delta\phi_{\mathrm{max}}$, we find a behavior special to case 2 that the real part of $S_A$ does vanish. 

This transition at $\Delta\phi=\Delta\phi_{\mathrm{max}}$ may be natural from the viewpoint of the dual field theory on dS$_d$. 
To see this, let us remember that the pseudo entropy is defined from the transition matrix (\ref{transm}), which depends on the two states $|\psi_1\lb$ and $|\psi_2\lb$. In our setup $|\psi_1\lb$ is the state which is obtained by the forward time evolution of Hartle-Hawking state, while $|\psi_2\lb$ is the one created by the backward time evolution of the final state. It is also useful to note that  the phenomenological observation \cite{Mollabashi:2020yie,Mollabashi:2021xsd,Akal:2021dqt}
implies that the real part of pseudo entropy typically measures the amount of entanglement in the intermediate state between the two states, though the interpretation of imaginary part is not well understood at present. 
Now the final state is defined by imposing a boundary condition at $t=t_p=\infty$ and the analysis of the space-like EOW brane \cite{Akal:2020wfl} suggests that the state has no real space entanglement as in the standard boundary state \cite{Miyaji:2014mca}. At the time slice $t=t_0$, we can have such an disentangling effect from a point at future infinity when two points are separated more than $2(t_p-t_0)$ assuming the light like propagation of physical signals on dS$_d$.
Indeed this border is $\Delta\phi_{\mathrm{max}}$. Therefore it is natural that the real part of pseudo entropy gets vanishing for $\Delta\phi>\Delta\phi_{\mathrm{max}}$. On the other hand, for $\Delta\phi<\Delta\phi_{\mathrm{max}}$, such two points do not feel the existence of the EOW brane and thus the result is the same as that in case 1.

\begin{figure}[t]
	\centering
	\includegraphics[width=5cm]{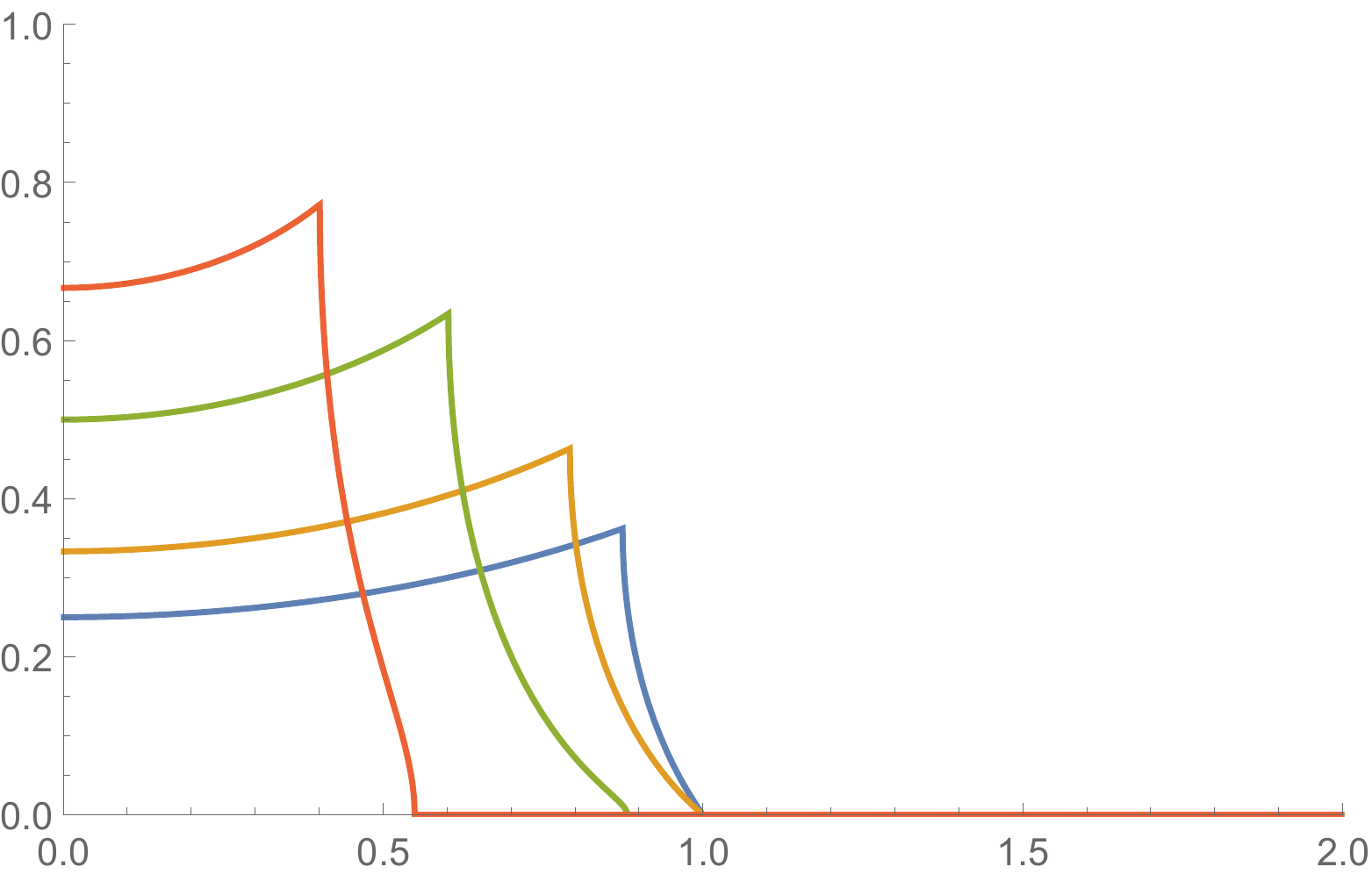}
	\includegraphics[width=5cm]{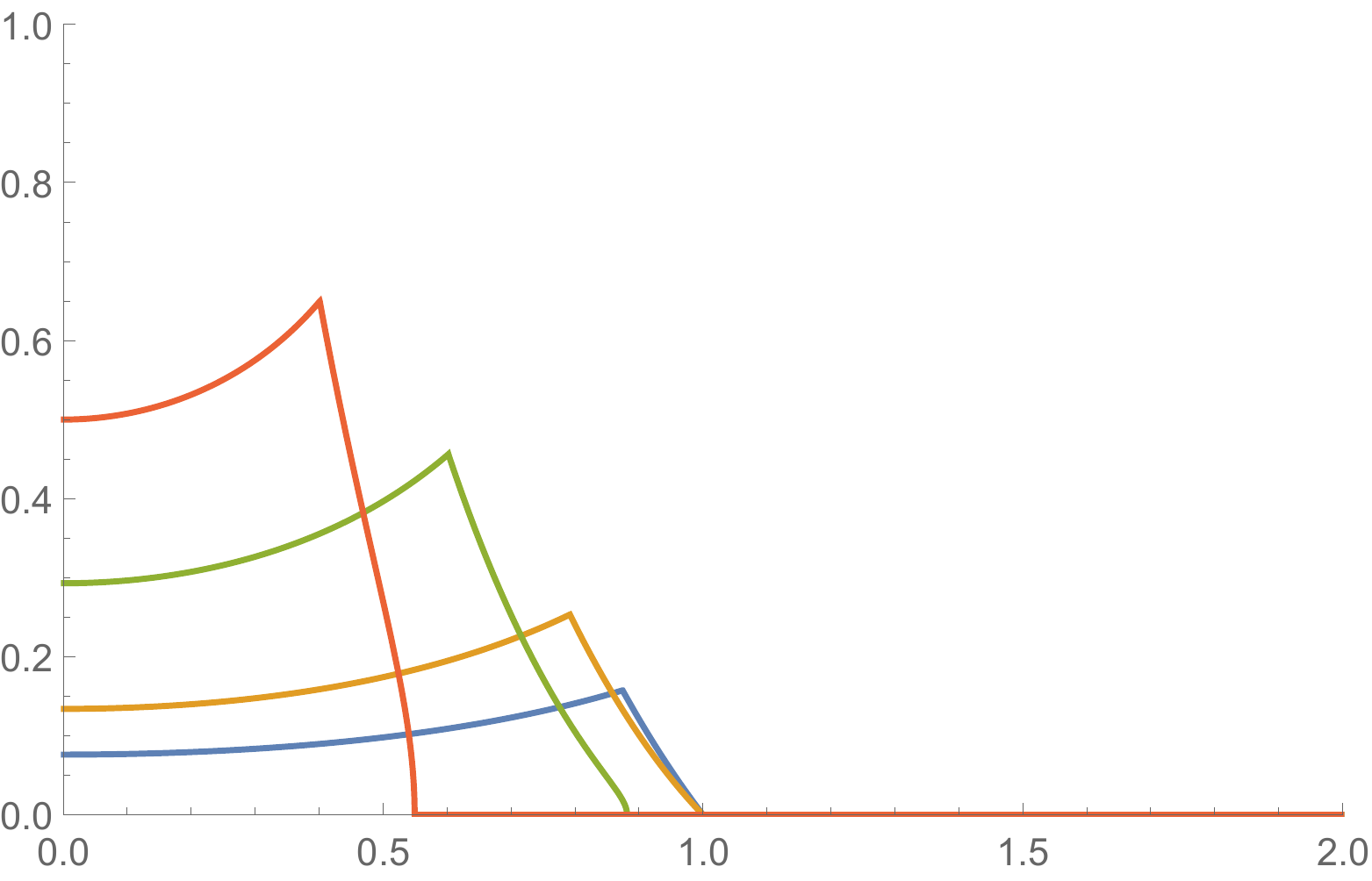}
	\includegraphics[width=5cm]{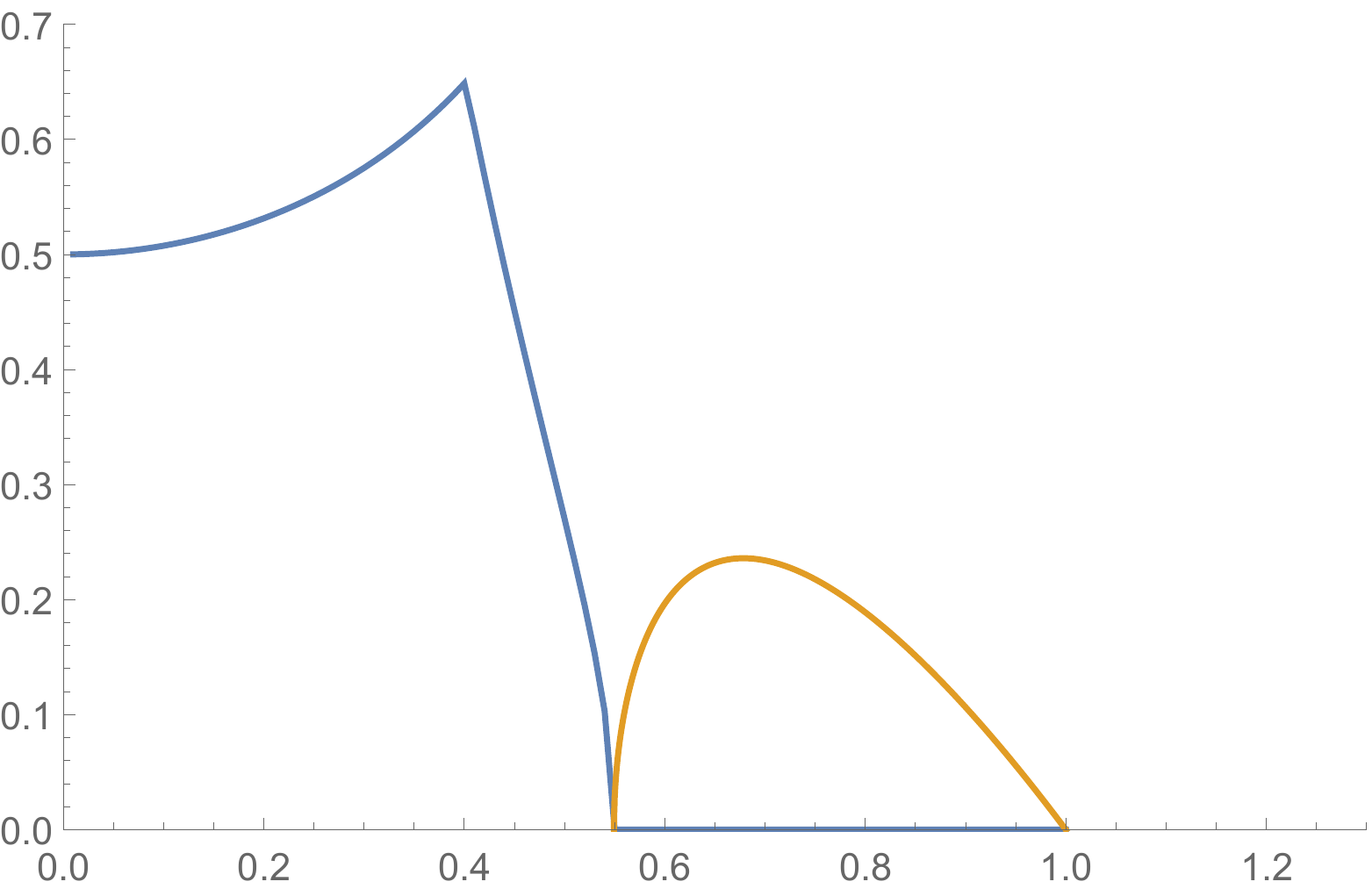}
	\caption{The behaviors of the holographic pseudo entropy for a half dS$_3$ ($d=2$, left) and a half dS$_4$ ($d=3$, middle and right) in case 2 with the EOW brane at $t_p=1$. We plotted $\frac{1}{\pi}A(\Gamma_A)$ (for $d=2$, left) and $\frac{1}{2\pi}A(\Gamma_A)$ (for $d=3$, middle) as a function of time $t_0$. We chose 
		$\phi_1=\frac{\pi}{8}$ (blue), $\phi_1=\frac{\pi}{6}$ (orange), $\phi_1=\frac{\pi}{4}$ (green), $\phi_1=\frac{\pi}{3}$ (red). In the right panel, we plotted Re$\left[\frac{1}{2\pi}A(\Gamma_A)\right]$ (blue) and Im$\left[\frac{1}{2\pi}A(\Gamma_A)\right]$ (orange) as a function of time for  $\phi_1=\frac{\pi}{3}$ for $d=3$. Note that  the real part starts vanishing when the geodesic becomes null and we have both the real and imaginary part vanishing for $t>t_p(=1)$.}
	\label{fig:dSareaPR4}
\end{figure}

In a similar way, we can analyze the holographic pseudo entropy when the final projection is inserted at a finite time $t=t_p$. Obviously at $t=t_p$, we have $S_A=0$. Moreover, the real part of $S_A$ gets vanishing when the geodesic $\Gamma_A$ becomes light-like. This can happen $t<t_p$ if the subsystem $A$ is large enough, in which case a imaginary part of $S_A$ starts to be non-zero until $t=t_p$ as $\Gamma_A$ becomes time-like, as depicted in figure \ref{fig:dSareaPR4} for plots. It is curious to note that this page curve like behavior of Re$S_A$ is qualitatively similar to the one found for the AdS/CFT setup in eq.~\eqref{HPED2}.

\section{Summary and Discussions}\label{sec:discussion}

The primary objective of this paper is to investigate the holographic duality involving gravity in de Sitter spaces. Unlike many other approaches, our focus centers on a half de Sitter space, achieved by introducing a timelike boundary in global dS spacetime. Within this framework, we propose that the gravity on a $d+1$-dimensional half de Sitter space is dual to a non-local field theory residing on its $d$-dimensional boundary.

Before delving into our investigation of de Sitter holography, we conducted an analysis of the holographic duality linking gravity in a $d+1$-dimensional AdS and a CFT living on a $d$-dimensional dS. This particular scenario can be regarded as a special case within the framework of the AdS/CFT correspondence. Of course, its validity and fundamental computational methods are well-established. We examined two distinct setups, referred to as the Case 1 and the Case 2, respectively. In the Case 1, the quantum state in the dual CFT is described by the Schwinger-Keldysh formalism, whereas in the Case 2, we consider the dual CFT incorporating a final state projection, as illustrated in figure \ref{fig:SCK} and figure \ref{fig:FSP}. In particular, the gravity dual in  the Case 2 is given by adding an end-of-the-world brane (EOW brane) on an AdS geometry.

In these setups, we evaluated holographic entanglement entropy, which is determined by the area of an extremal surface. In the Case 1, we observed that the holographic entanglement entropy for a subsystem of fixed size consistently increases as the size of de Sitter space is inflating with respect to the global time. Conversely, in the Case 2, it initially presents growth but eventually decreases to zero. At the time $t=t_{\mt P}$, corresponding to the implementation of the final state projection, the entanglement entropy vanishes. It is important to note that the extremal surface area we computed should be more appropriately interpreted as the pseudo entropy due to the presence of the final state projection. Additionally, we also confirmed that independent CFT calculations in CFT$_2$ reproduce the results which agree with that from gravity dual.

With the AdS/CFT results as our foundation, we proceeded to examine holography for gravity in a half de Sitter space. We focused on two setups, namely the Case 1 and the Case 2, and investigated the behaviour of holographic entanglement entropy. Note that we employed the standard calculation of holographic entanglement entropy \cite{Ryu:2006bv,Ryu:2006ef,Hubeny:2007xt}, where we minimize the area, as opposed to the prescription in \cite{Susskind:2021esx} where the area is maximized. Remarkably, we discovered that the properties of holographic entanglement entropy in a half de Sitter space diverge from those in the AdS/CFT correspondence. Notably, connecting two arbitrary points in a global de Sitter space using a spacelike geodesic is not always possible. Consequently, for a two-dimensional dS space, which serves as the timelike boundary of a three-dimensional half de Sitter space, there is typically no spacelike geodesic linking the endpoints of an interval. As a result, the definition of holographic entanglement entropy, denoted as $S_A$, in a conventional sense is problematic. Similarly, we observed the same limitation for extremal surfaces in higher dimensions ($d>2$).

However, a resolution to this issue arises when we consider both timelike and spacelike geodesics within a Hartle-Hawking contour, as illustrated in the right panel of figure \ref{fig:setupp}. The joint geodesics allow us to connect two endpoints beyond critical size. Consequently, the holographic entanglement entropy, denoted as $S_A$, acquires a complex value. In the Case 1, we contend that the appropriate holographic entanglement entropy can be obtained by taking its real part, which is nothing but half of the de Sitter entropy $\frac{1}{2}S_{dS}$. In the Case 2, with the presence of the EOW brane, we argue that the holographic pseudo entropy can be computed by utilizing timelike geodesics terminating on the EOW brane, as depicted in the right panel of figure \ref{fig:setudispp}.

Furthermore, even within parameter regions where a spacelike geodesic exists, the holographic entanglement entropy generally exhibits super-extensive behaviour relative to the subsystem size. Consequently, the corresponding function describing holographic entanglement entropy violates the (strong) subadditivity property. Notably, we have discovered that this issue is resolved solely by focusing on the time slices associated with the static coordinate in the case of the maximal half de Sitter space ($\theta_0=\frac{\pi}{2}$). On these particular time slices, the holographic entanglement entropy $S_A$ adheres to the volume law. In this regard, we anticipate that the quantum state represented by the half de Sitter space manifests maximal entanglement specifically on these special time slices. This implies that we can establish a well-defined Hilbert space solely for static time slices. Conversely, a generic time slice in the boundary de Sitter space spans the same Hilbert space as the static time slice multiple times, resulting in an overcounting of the genuine degrees of freedom (refer to figure \ref{fig:Hilb}). Additionally, it is worth noting that when subadditivity is violated, the extremal surface extends beyond the Wheeler-DeWitt patch, as depicted in figure \ref{fig:WDW}.

Although providing a Hilbert space interpretation for generic time slices, including the constant time slice of the global coordinate, presents challenges, we propose a potential understanding of the entanglement entropy by employing the replica method and defining the area of extremal surface $S_A$ as the entropy. Notably, in highly non-local field theories, the realization of super-extensive entanglement entropy can be realized. It is important to note that in such non-local field theories, a standard Hilbert space cannot be defined due to the appearance of infinitely many time derivatives in the action. Applying this perspective, the holographic entanglement entropy in the Case 1 exhibits initial growth, followed by saturation at $\frac{1}{2}S_{dS}$, as depicted in figure \ref{fig:exHEE} and figure \ref{fig:dSarea}. Note that this saturation was absent in the evolution of $S_A$ derived in the Case 1 in the framework of the AdS/CFT correspondence and that it shows that the entropy is bounded in de Sitter space.

In the Case 2, we observe that the real part of the holographic pseudo entropy exhibits initial growth and eventually vanishes at the critical time, while the imaginary part remains non-zero, as depicted in figure \ref{fig:dSarea2} and figure \ref{fig:dSareaPR4}. We identify that the real part of $S_A$ becomes zero when the subsystem $A$ surpasses the light cone (refer to the right panel of figure \ref{fig:setudispp}). This behaviour arises due to the influence of the EOW brane, where the reflection of two points on $A$ by the EOW brane boundary condition affects $S_A$. The EOW brane boundary state possesses vanishing quantum entanglement, thereby diminishing the overall quantum entanglement.

There are several promising avenues for future investigation. Firstly, it would be intriguing to explore alternative frameworks that provide a more manageable description of the ``overcounting" phenomenon in the Hilbert space dual to a half de Sitter space in global coordinates. One possible approach could involve qubit systems or tensor networks, offering a more controllable perspective. Additionally, it is of great interest to extend our analysis to holography in more generic asymptotically dS spacetimes, such as de Sitter black holes, as well as various cosmological models. Lastly, a pivotal and profound question lies in understanding how the creation and evolution of the universe can be described in terms of the dual field theory, utilizing the insights gained from the holographic duality.

%%%%%%%%%%%%%%%%%%%%%%%%%%%%%%%%%%%%%%%%%%%%%%%%%%%%%%%%
%%%%%%%%%%%%%%%%%%%%%%%%%%%%%%%%%%%%%%%%%%%%%%%%%%%%%%%%
\section*{Acknowledgements}

%%%%%%%%%%%%%%%%%%%%%%%%%%%%%%%%%%%%%%%%%%%%%%%%%%%%%%%%
%%%%%%%%%%%%%%%%%%%%%%%%%%%%%%%%%%%%%%%%%%%%%%%%%%%%%%%%

We are grateful to Takato Mori, Yusuke Taki and Zixia Wei  for useful discussions.
This work is supported by the Simons Foundation through the ``It from Qubit'' collaboration
and by MEXT KAKENHI Grant-in-Aid for Transformative Research Areas (A) through the ``Extreme Universe'' collaboration: Grant Number 21H05187. TT is also supported by Inamori Research Institute for Science, and by JSPS Grant-in-Aid for Scientific Research (A) No.~21H04469. SMR is also supported by JSPS KAKENHI Research Activity Start-up Grant NO. 22K20370. YS is supported by Grant-in-Aid for JSPS Fellows No.23KJ1337.
T. K. is supported by Grant-in-Aid for JSPS Fellows No. 23KJ1315.

\appendix

\section{Explicit Space-like Geodesics in dS\texorpdfstring{$_3$}{Lg}}
\label{sec:APgeo}

Here we study the connected geodesic anchored on the boundaries of an interval $A$ in dS$_3$. This is the geodesic which connects $(t,\theta,\phi)=(t_0,\frac{\pi}{2},-\phi_0+\frac{\pi}{2})$ and 
$(t,\theta,\phi)=(t_0,\frac{\pi}{2},\phi_0+\frac{\pi}{2})$ in the coordinate of 
dS$_3$ (\ref{dstheq}).
We assume $t_0\geq 0$ without losing generality.
The length $L$ of a curve is  
\ba
L=\int^{\phi_0+\frac{\pi}{2}}_{-\phi_0+\frac{\pi}{2}} d\phi\s{\cosh^2 t
	-\left(\frac{dt}{d\phi}\right)^2}.
\ea
This leads to the differential equation
\ba
\frac{dt}{d\phi}=\cosh t\s{1-\frac{\cosh^2 t}{\cosh^2 t_*}},
\ea
where the middle point $\phi=\frac{\pi}{2}$ is the turning point and we set $t=t_*$ at this point. 
By integrating this, we find
\ba
&& \phi_0=\frac{\pi}{2}-\arctan\left[\frac{\cosh t_*\sinh t_0}{\s{\cosh^2 t_*-\cosh^2 t_0}}\right],\no
&& L=\pi-2\arctan\left[\frac{\sinh t_0}{\s{\cosh^2 t_*-\cosh^2 t_0}}\right].
\ea

Note that assuming $t_0> 0$, there is an upper bound of $\phi_0$, reached at $t_*\to \infty$, which we call $\phi_{\max}$:
\ba
\phi_{\mathrm{max}}=\frac{\pi}{2}-\arctan\left[\sinh t_0\right].
\label{boundnuA}
\ea
In this maximal case, we find $L|_{\phi=\phi_*}=\pi$, which is a half of de Sitter horizon length. This maximal value $\phi=\phi_{\mathrm{max}}$ corresponds to the limit where the original space-like geodesic gets light-like as depicted in the left panel of figure \ref{fig:setupp}.
Therefore we cannot find an appropriate space-like geodesic when the subsystem $A$ is larger than this maximal size.

\bibliographystyle{JHEP}
\bibliography{dSHol}

%%%%%%%%%%%%%%%%%%%%%%%%%%%%%%%%%%%%%%%%%%%%%%%%%%%%%%%%
%%%%%%%%%%%%%%%%%%%%%%%%%%%%%%%%%%%%%%%%%%%%%%%%%%%%%%%%

\end{document}